\documentclass[reprint,showpacs,preprintnumbers,amsmath,amssymb]{revtex4-1}

\pdfoutput=1
\usepackage[pdftex]{graphicx}
\usepackage[unicode=true, bookmarks=false,bookmarksnumbered=true,bookmarksopen=true, breaklinks=false,backref=false,pagebackref=false, colorlinks=true]{hyperref}
\hypersetup{pdftitle={ },pdfauthor={ },linkcolor=black,citecolor=black, urlcolor=black, filecolor=black,pdfpagelayout=OneColumn,pdfnewwindow=true,pdfstartview=XYZ, plainpages=false}
\usepackage{color}

\usepackage{braket}
\usepackage[T1]{fontenc}
\usepackage{amsmath}
\usepackage{amsthm}
\usepackage{amssymb}
\usepackage{url}
\usepackage[english]{babel}
\usepackage{dcolumn}
\usepackage{bm}
\usepackage{subfig}
\usepackage{float}
\usepackage{url} 
\usepackage{grffile}

\usepackage[normalem]{ulem} 

\usepackage{tikz}
\usetikzlibrary{fadings}
\usetikzlibrary{shadings}

\tikzfading[name=fade out,
inner color=transparent!0,
outer color=transparent!100]

\tikzfading[name=middle,
top color=transparent!100,
bottom color=transparent!100,
middle color=transparent!0]

\tikzfading[name=middle rot,
right color=transparent!100,
left color=transparent!100,
middle color=transparent!0]

\providecommand{\tabularnewline}{\\}

\begin{document}


\title{Lattice QCD static potentials of the meson-meson and tetraquark systems
\\ computed with both quenched and full QCD}

\author{P. \surname{Bicudo}}
\email{bicudo@tecnico.ulisboa.pt}

\author{M. \surname{Cardoso}}
\email{mjdcc@cftp.ist.utl.pt}

\affiliation{CFTP, Instituto Superior T\'ecnico, Universidade de Lisboa}

\author{O. \surname{Oliveira}}
\email{orlando@fis.uc.pt}

\author{P. J. \surname{Silva}}
\email{psilva@teor.fis.uc.pt }

\affiliation{CFisUC, Department of Physics, University of Coimbra, P-3004 516 Coimbra, Portugal}

\pacs{12.38.Gc, 11.15.Ha}

\keywords{BLAH BLAH BLAH}

\begin{abstract}
We revisit the static potential for the $Q Q \bar Q \bar Q$ system using SU(3) lattice simulations, studying both the colour singlets groundstate and first excited state. We consider geometries where the two static quarks and the two anti-quarks are at the corners of rectangles of different sizes. 
We analyse the transition between a tetraquark system and a two meson system with a  two by two correlator matrix.
We compare the potentials computed with quenched QCD and with dynamical quarks. 
We also compare our simulations with the results of previous studies and
analyze quantitatively fits of our results with anzatse inspired in the string flip-flop model and in its possible colour excitations. 
\end{abstract}

\maketitle

\section{Introduction}

Our current understanding of strong interaction phenomenology, being the hadron spectrum or the form factors associated to
transitions between hadrons, relies on the description of the quark and gluon interaction within Quantum Chromodynamics. 
Despite the efforts of several decades, the non-perturbative nature of QCD still ensconce several properties of its fundamental particles. 
Indeed, we still do not understand the confinement mechanism,  which prevents the observation of free quarks and gluons in nature, 
and still do not have a satisfactory answer why the experimentally ~\cite{Olive:2016xmw} confirmed hadrons are composed of three valence quarks or a pair of quark and an anti-quark.

QCD is a gauge theory and physical observables should be gauge invariant objects. Gauge invariance implies
that only certain combinations of quarks and/or gluons can lead to observables particles. If one applies blindly such 
a simple rule, the observed hadrons are necessarily composite states involving multi-quarks and multi-gluon 
configurations. 
There is \textit{a priori} no reason why states with other valence composition than mesons or baryons, called in general exotic states, should not be observed. 
Exotic states can be pure glue states (glueballs), multi-quark states (tetraquark, pentaquarks, \textit{etc}) or hybrid states (mesons with a non-vanishing valence gluon content). Besides the hadron states compatible with the quark model, 
the particle data book ~\cite{Olive:2016xmw} also reports candidates for the different types of exotic states, see e.g. the reviews on pentaquarks and 
non-$q \overline q$ mesons. The masses of the experimental states listed as candidates to multi-quark/gluon hadrons cover the 
full range of energies of the particle spectrum. In particular the exotics with most observations are the tetraquarks.

In what concerns the experimental observation of exotic tetraquarks, the quarkonium sector of double-heavy tetraquarks including a $Q\bar Q$ pair is the most explored experimentally, see e.g. the recent reviews~\cite{Briceno:2015rlt,Lebed:2016hpi,Esposito20171}. In particular, the charged $Z_c^{\pm}$ and $Z_b^{\pm}$ are crypto-exotic, but technically they can be regarded as essentially exotic tetraquarks if we neglect $ c \bar c$ or $b \bar b$ annihilation.
There are two $Z_b^{\pm}$ observed only by the collaboration BELLE at KEK
\cite{Belle:2011aa}, 
slightly above $B \ \bar B^*$ and $B^* \ \bar B^*$ thresholds,
 the $Z_b(10610)^+$ and $Z_b(10650)^+$.
Their nature is possibly different from the  two $Z_c(3940)^{\pm}$ and  $Z_c(4430)^{\pm}$, whose mass is well above $DD$ threshold \cite{Choi:2007wga}.
The $Z_c^\pm$ has been observed with very high statistical significance
and has received a series of experimental observations by the BELLE collaboration \cite{Liu:2013dau,Chilikin:2014bkk}, the Cleo-C collaboration \cite{Xiao:2013iha}, the
BESIII collaboration \cite{Ablikim:2013mio,Ablikim:2013emm,Ablikim:2013wzq,Ablikim:2013xfr,Ablikim:2014dxl} and the LHCb
collaboration \cite{Aaij:2014jqa}. This family is possibly related to the closed-charm pentaquark recently observed at LHCb \cite{Aaij:2015tga}. 
Notice that, using na\"ive Resonant Group Method calculations, in 2008, some of us predicted
\cite{Cardoso:2008dd}
a partial decay width to $\pi \ J/\psi $ of the $Z_c(4430)^-$ consistent with the recently observed experimental value\cite{Aaij:2014jqa}.

On the other hand, in what concerns lattice QCD simulations, the most promising exotic tetraquark sector is also double-heavy, but it has a pair of heavy quarks $Q Q$ or antiquarks $\bar Q \bar Q$, and thus it differs from the quarkonium sector. Note that in lattice QCD, the study of exotics is presently even harder than in the laboratory, 
since the techniques and computer facilities necessary to study of resonances with many decay channels remain to be developed.
Lattice QCD searched for evidence of a large tetraquark component in the closed-charm $Z_c(3940)^-$ candidate but this resonance is well above threshold, and Ref. 
\cite{Prelovsek:2014swa,Leskovec:2014gxa}  concluded there is no robust lattice QCD evidence of a $Z_c^{\pm}$ tetraquark resonance.
Lattice QCD also searched for the expected boundstate in light-light-antiheavy-antiheavy channels \cite{Ikeda:2013vwa,Guerrieri:2014nxa}.
Using dynamical quarks, the only heavy quark presently accessible to Lattice QCD simulations is the charm quark. No evidence for boundstates in this possible family of tetraquarks, say for a $u d \bar c \bar c$ was found.
Moreover the potentials between two mesons, each composed of a light quark and a static (or infinitely heavy) antiquark , have been computed in lattice QCD \cite{Wagner:2010ad, Wagner:2011ev}.  A static antiquark constitutes a good approximation to a spin-averaged $\bar b$ bottom antiquark. The potential between the two light-static mesons can then be used, with the Born-Oppenheimer approximation \cite{Born:1927}, as a $B-B$ potential,
where the higher order $1/m_b$ terms including the spin-tensor terms are neglected.
From the potential of the channel with larger attraction, which occurs in the Isospin=0 and Spin=0 quark-quark system, 
the possible boundstates of the heavy antiquarks have been investigated with quantum mechanics techniques.
Recently, this approach indeed found evidence for a tetraquark $u d \bar b \bar b$ boundstate \cite{Bicudo:2012qt,Brown:2012tm}, while no boundstates have been found  for states where the heavy quarks are $\bar c \bar b$ or $\bar c \bar c$ (consistent with full lattice QCD computations \cite{Ikeda:2013vwa,Guerrieri:2014nxa}) or where the light quarks are $\bar s \bar s$ or $\bar c \bar c$ \cite{Bicudo:2012qt,Bicudo:2015kna,Peters:2015tra,Bicudo:2015vta,Bicudo:2016ooe,Bicudo:2016jwl,Peters:2016isf}. The $\bar b \bar b$ probability density in the only binding channel has also been computed in Ref.  \cite{Bicudo:2012qt,Bicudo:2015kna,Peters:2015tra,Bicudo:2015vta,Bicudo:2016ooe,Bicudo:2016jwl,Peters:2016isf}.

\begin{figure}[!t]
\includegraphics[width=1.1\columnwidth]{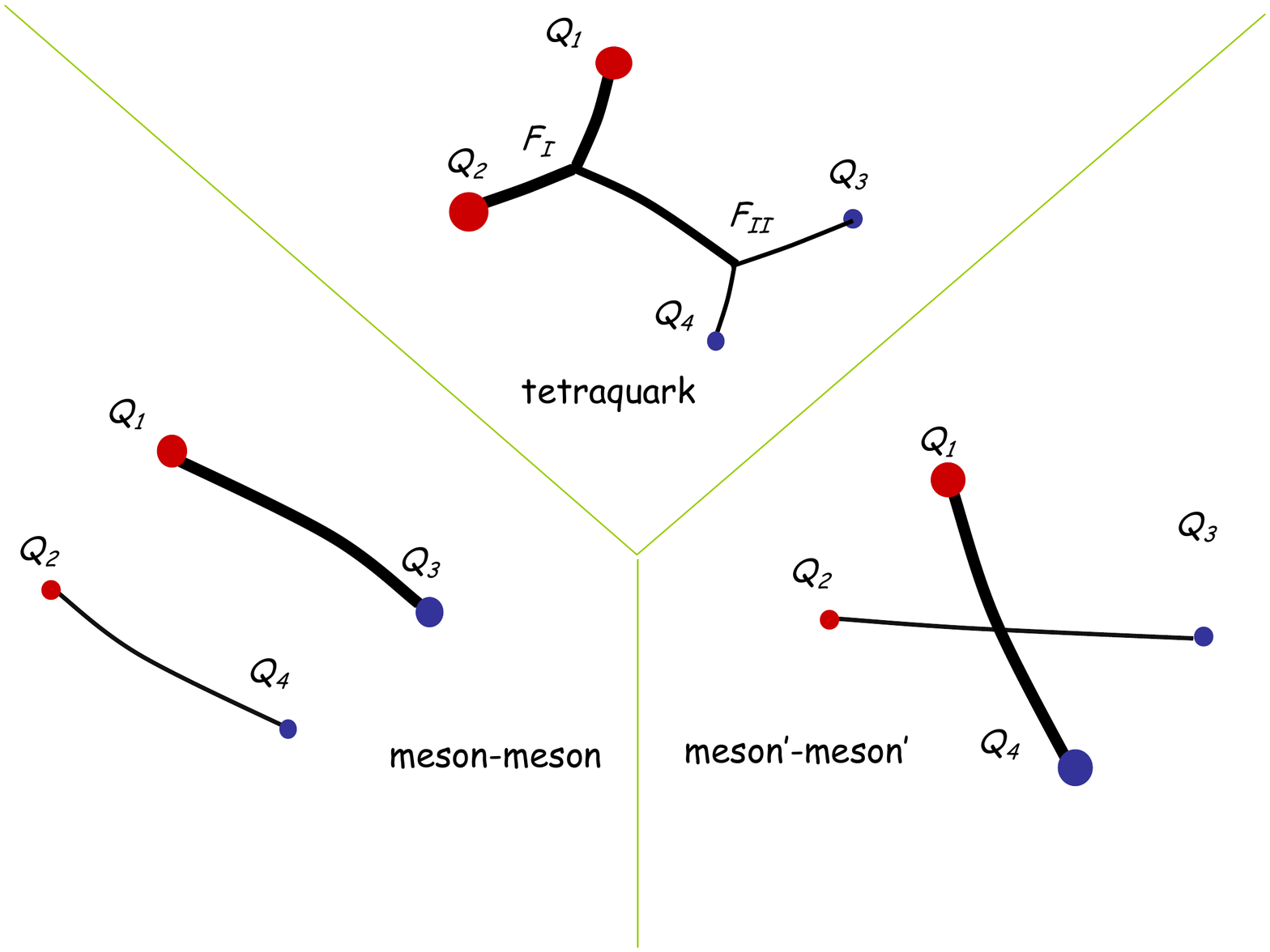}
\caption{ (Colour online.) The paradigm of the string flip flop model, with three possible arrangements of quark colours in the groundstate for a $QQ \bar Q \bar Q$ colour singlet system: meson-meson, tetraquark and meson-meson'.\label{fig:tripleflipflop}}
\end{figure}

The quark models for tetraquarks with the most sophisticated description of confinement are the string flip-flop models. Clearly, tetraquarks are always coupled to meson-meson systems, and we must be able to address correctly the meson-meson interactions. The first quark models had confining two-body potentials proportional to the SU(3) colour Casimir invariant 
$ \vec \lambda_i \cdot \vec \lambda_j \, V(r_{ij})$ suggested by the One-Gluon-Exchange type of potential. However this would lead to an additional Van der Waals potential $ V_\text{Van der Waals}  = { V'(r) \over r}  \times T, $
where $T$ is a polarization tensor. The resulting Van der Waals 
\cite{Fishbane:1977ay,Appelquist:1978rt,Willey:1978fm,Matsuyama:1978hf,Gavela:1979zu,Feinberg:1983zz}
force between mesons, or baryons would be extremely large and this is clearly not compatible with observations.
The string flip-flop potential for the meson-meson interaction was developed in Refs. 
\cite{Miyazawa:1979vx,Miyazawa:1980ft,Oka:1984yx,Oka:1985vg,Karliner:2003dt},
to solve the problem of the Van der Waals forces produced by the two-body
confining potentials.
The first considered string flip-flop potential was the one
minimizing the energy of the possible two different meson-meson configurations, say $M_{13} \, M_{24}$ or $M_{14} \, M_{23}$.
This removes the inter-meson potential, and thus solves the problem of the Van der Waals force.
An upgrade of the string flip-flop potential includes a third possible
configuration \cite{Carlson:1991zt}, in the tetraquark channel, say $T_{12 \, , \, 34}$,
where the four constituents are linked
by a connected string
\cite{Vijande:2007ix,Vijande:2009xx}. 
The three confining string configurations differ in the strings linking the quarks and antiquarks, 
this is illustrated in 
Fig. \ref{fig:tripleflipflop}. 
When the diquarks $qq$ and $\bar q \bar q$ distances are small, the tetraquark configuration minimizes the string energy.
When the quark-antiquark pairs $q \bar q$ and $q \bar q$ are close, the meson-meson configuration minimizes
the string energy. 
With a triple string flip-flop potential, 
boundstates below the threshold for hadronic coupled channels have been found
\cite{Beinker:1995qe,Zouzou:1986qh,Vijande:2007ix,Vijande:2009xx,Bicudo:2010mv,Bicudo:2015bra}.
On the other hand, the string flip-flop potentials allow fully unitarized studies of resonances
\cite{Lenz:1985jk,Oka:1984yx,Oka:1985vg,Bicudo:2010mv,Bicudo:2015bra}.
Analytical calculations with a double flip-flop harmonic oscillator potential,
\cite{Lenz:1985jk},
using the resonating group method again with a double flip-flop confining harmonic oscillator potential,
\cite{Oka:1984yx,Oka:1985vg},
anfd with the triple string flip-flop potential \cite{Bicudo:2010mv,Bicudo:2015bra}
have already predicted resonances and boundstates.

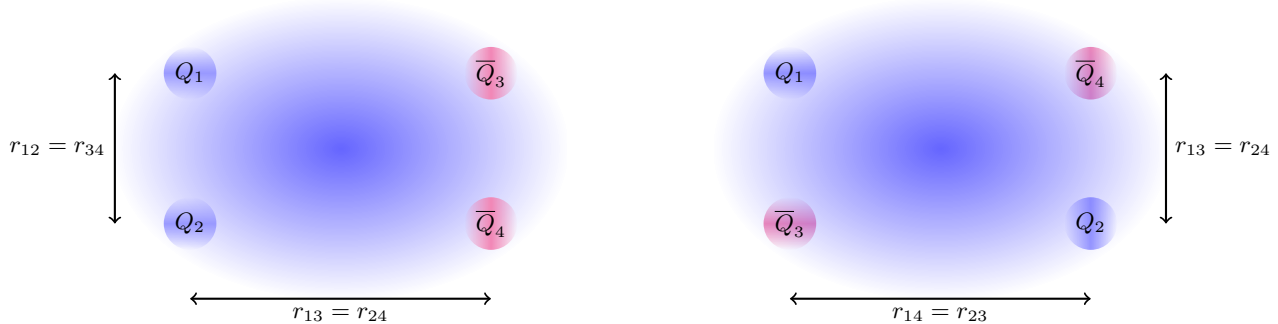
\begin{figure*}[!t]
\centering
\begin{tikzpicture}
\draw[<->,thick] (-2,0) -- (2,0) node [pos=0.5,below] {$r_{13} = r_{24}$};
\draw[<->,thick] (-3,1) -- (-3,3) node [pos=0.5,left] {$r_{12} = r_{34}$}; 
%
%
%
%
\draw[fill=white!40!blue, path fading=fade out, draw=none] (-3,0) rectangle (3,4);
%
%
   \fill[fill=blue!40!white, path fading=middle] (-2,1) circle (0.35);
   \fill[fill=blue!40!white, path fading=middle] (-2,3) circle (0.35);
   \fill[fill=white!40!magenta, path fading=middle rot](2,1) circle (0.35);
   \fill[fill=white!40!magenta, path fading=middle rot](2,3) circle (0.35);
\node at (-2,1) {\textbf{$Q_2$}};
\node at (-2,3) {\textbf{$Q_1$}};
\node at (2,1) {\textbf{$\overline Q_4$}};
\node at (2,3) {\textbf{$\overline Q_3$}};
\end{tikzpicture}
\hspace{50pt}
\begin{tikzpicture}
\draw[<->,thick] (-2,0) -- (2,0) node [pos=0.5,below] {$r_{14} = r_{23}$};
\draw[<->,thick] (3,1) -- (3,3) node [pos=0.5,right] {$r_{13} = r_{24}$}; 
%
   \fill[fill=white!40!magenta, path fading=middle] (-2,1) circle (0.35);
   \fill[fill=blue!40!white, path fading=middle] (-2,3) circle (0.35);
   \fill[fill=blue!40!white, path fading=middle rot](2,1) circle (0.35);
   \fill[fill=white!40!magenta, path fading=middle rot](2,3) circle (0.35);
\draw[fill=white!40!blue, path fading=fade out, draw=none] (-3,0) rectangle (3,4);
\node at (-2,1) {\textbf{$\overline Q_3$}};
\node at (-2,3) {\textbf{$Q_1$}};
\node at (2,1) {\textbf{$Q_2$}};
\node at (2,3) {\textbf{$\overline Q_4$}};
\end{tikzpicture}
\caption{ (Colour online.) Our two different planar geometries for the static tetraquark potential: the parallel geometry (left) and the anti-parallel geometry (right)
considered in our simulations.}
\label{Fig:geometrias}
\end{figure*}

So far. the theoretical and experimental interpretations of the observed states that can possibly be exotics is not clear crystal and, certainly, a better understanding of the colour force helps to elucidate our present view of the hadronic spectrum. 
For heavy quark systems its dynamics can be represented by a potential which, in general, is a function of the geometry of the hadrons, 
of the spin orientation of its components and of the quark flavours. In the limit of infinite quark mass one can compute the so-called
static potential using first principle lattice QCD techniques via the evaluation of Wilson loops. The static potential provides an important
input to the modelling of hadrons and it gives a simple realisation of the confinement mechanism. Moreover it can be applied to study tetraquarks $Q  Q \bar Q \bar Q$ with two heavy quarks and two heavy antiquarks, see for instance a Dyson-Schwinger study in Ref. 
\cite{Heupel:2012ua}, at the intersection of the two sectors most studied experimentally and theoretically. 

The static potential has been computed using lattice QCD for mesons, tetraquark, pentaquarks and hybrid systems see 
\cite{Bali:2000gf,Alexandrou:2004ak,Bornyakov:2005kn,Okiharu:2004ve,Okiharu:2004wy,Bicudo:2007xp,Cardoso:2008sb}.
For a quark and an anti-quark system, the static potential $V_{Q\overline Q}$ is a landmark calculation in lattice QCD and it is 
used to set the scale of the simulations. $V_{Q\overline Q}$ has been computed both in the quenched theory and 
in full QCD with the lattice data being well described by a one-gluon exchange potential (a Coulomb like potential) at short distances and a linear rising function of the quark distances at large separations. The behaviour at large interquark distances
provides a nice explanation of the confinement mechanism. 
Moreover, for other hadronic systems and for large separations of its constituents a similar pattern of the corresponding
static potentials has been observed in lattice simulations, i.e. a linear rising potential which, once more, is a simple realisation of
quark confinement.

In the current work we revisit the static potential for tetraquarks using lattice simulations. The static potential for tetraquarks was 
computed for the gauge group SU(3) and in the quenched approximation in~\cite{Alexandrou:2004ak,Bornyakov:2005kn,Okiharu:2004ve}. 
The hybrid potential defined and measured in~\cite{Bicudo:2007xp} can also be viewed as a particular limit of the tetraquark potential. 
Herein, of all the possible geometries for the $QQ \bar{Q} \bar{Q}$ system we consider the case where quarks and anti-quarks are
at the corners of a rectangle, see Fig.~\ref{Fig:geometrias}, and recompute the static potential of the system both in the quenched 
approximation and in full QCD. We focus our analysis in the comparison of the quenched and full QCD and also in the transition between
a tetraquark system and a two meson system. Thus we go beyond the triple string flip-flop paradigm of Fig. \ref{fig:tripleflipflop} and analyse, in the transition region, the mixing between the meson-meson and tetraquark string configurations. Moreover we explore not only the groundstate but also the first excited state.

The current work is organised as follows. In Sec.~\ref{Sec:colourstructure} we discuss the possible colour structures for a 
$QQ \bar{Q} \bar{Q}$ system and introduce the $Q \bar{Q}$ potentials used to compare the results of the static potentials
for the tetraquark. In Sec.~\ref{Sec:geometries} we revisit the geometries used to compute the
static potentials and discuss the expected configurations at large separations. In Sec.~\ref{Sec:calculopotencial} the method used to 
evaluate the static potentials is described. In Sec.~\ref{Sec:lattice} we report on the parameters used in the lattice simulations and how we set the 
scale of the simulations. The results for the static tetraquark potential for the two geometries are described in Sec~\ref{Sec:resultados}.
In Sec.~\ref{Sec:final} we resume and conclude. In the Appendix, the reader can find various tables with all our numerical results.

\section{The Color structure of a $QQ\bar{Q}\bar{Q}$ system \label{Sec:colourstructure}}

The colour-spin-spatial wave function of a $QQ\bar{Q}\bar{Q}$ system has multiple combinations, relevant for the computation of static potentials. In this section, we analyse the possible colour wave functions associated with a tetraquark system.

The quarks belong to the fundamental $3$ representation of SU(3), while anti-quarks are in a $\overline 3$ representation of the group. 
The  space built from the direct product $3\otimes3\otimes\overline 3\otimes\overline 3$ includes two independent colour singlet states. 

In a $QQ\bar{Q}\bar{Q}$ system, quarks and anti-quarks can combine into colour singlet meson-like states, 
leading naturally to the two meson states,
\begin{eqnarray}
|\mathbf{1}_{13}\mathbf{1}_{24}\rangle  &=&  \frac{1}{3}\delta_{ik}\delta_{jl}|Q_{i}Q_{j}\bar{Q}_{k}\bar{Q}_{l}\rangle \ ,
\nonumber
\\
|\mathbf{1}_{14}\mathbf{1}_{23}\rangle  &=&  \frac{1}{3}\delta_{il}\delta_{jk}|Q_{i}Q_{j}\bar{Q}_{k}\bar{Q}_{l}\rangle \ ,
\label{eq:singleto2}
\end{eqnarray}
where only the colour indices are written explicitly and $\mathbf{1}_{ij}$ refers to the meson-like colour singlet state built
combining quark $i$ and anti-quark $j$. The two colour singlet states in Eq. (\ref{eq:singleto2}) are not orthogonal to each other and a straightforward algebra gives,
\begin{equation}
\langle\mathbf{1}_{13}\mathbf{1}_{24}|\mathbf{1}_{14}\mathbf{1}_{23}\rangle=\frac{1}{3}\label{eq:mm_nonortho} \ .
\end{equation}

Moreover, a quark and anti-quark pair, besides a colour singlet state, can also form a colour octet state. With two colour octets it is again possible to build a colour singlet state.
For the $QQ\bar{Q}\bar{Q}$  system the colour singlet states
built from the octets read,
\begin{eqnarray}
|\mathbf{8}_{13}\mathbf{8}_{24}\rangle  =  \frac{1}{4\sqrt{2}}\lambda_{ik}^{a}\lambda_{jl}^{a}|Q_{i}Q_{j}\bar{Q}_{k}\bar{Q}_{l} \rangle \ ,
\nonumber
\\
|\mathbf{8}_{14}\mathbf{8}_{23}\rangle  =  \frac{1}{4\sqrt{2}}\lambda_{il}^{a}\lambda_{jk}^{a}|Q_{i}Q_{j}\bar{Q}_{k}\bar{Q}_{l}\rangle \ ,
 \label{eq:octeto2}
\end{eqnarray}
where the factors comply with the normalization condition,
\begin{equation}
\langle \mathbf{8}_{13}\mathbf{8}_{24} \, |\,\mathbf{8}_{13}\mathbf{8}_{24}\rangle = 
\langle \mathbf{8}_{14}\mathbf{8}_{23} \, | \, \mathbf{8}_{14}\mathbf{8}_{23}\rangle = 1 \ . 
\end{equation}
The colour octet-octet states in Eq. (\ref{eq:octeto2}) can be written in terms of the meson-meson states defined in Eq. (\ref{eq:singleto2}),
\begin{eqnarray}
|\mathbf{8}_{13}\mathbf{8}_{24}\rangle  &=&  \frac{3|\mathbf{1}_{14}\mathbf{1}_{23}\rangle-|\mathbf{1}_{13}\mathbf{1}_{24}\rangle}{2\sqrt{2}} \ ,
\nonumber
\\
|\mathbf{8}_{14}\mathbf{8}_{23}\rangle  &=&  \frac{3|\mathbf{1}_{13}\mathbf{1}_{24}\rangle-|\mathbf{1}_{14}\mathbf{1}_{23}\rangle}{2\sqrt{2}} \ .
\end{eqnarray}
A simple calculation shows that the colour octet states (\ref{eq:octeto2}) are not orthogonal (in colour space) to each other.
However, each of the octet-octet states is orthogonal to the corresponding meson-meson state, i.e.
\begin{eqnarray}
\langle\mathbf{1}_{13}\mathbf{1}_{24}|\mathbf{8}_{13}\mathbf{8}_{24}\rangle  &=&  0 \ ,
\nonumber
\\
\langle\mathbf{1}_{14}\mathbf{1}_{23}|\mathbf{8}_{14}\mathbf{8}_{23}\rangle  &=&  0 \ .
\end{eqnarray}

The states in Eqs. (\ref{eq:singleto2})  and (\ref{eq:octeto2}) do not represent all the possible colour singlet states that can be associated to a 
$QQ\bar{Q}\bar{Q}$ system.   We can also consider diquark-antidiquark configurations.
For the group SU(3) it follows that  $\mathbf{3}\otimes\mathbf{3}=\bar{\mathbf{3}}\oplus\mathbf{6}$, 
$\bar{\mathbf{3}}\otimes\bar{\mathbf{3}}=\mathbf{3}\oplus\bar{\mathbf{6}}$ and the two colour singlet states belong to the space spanned by
$\mathbf{3}\otimes\bar{\mathbf{3}}$ and $\mathbf{6}\otimes\bar{\mathbf{6}}$,
\begin{eqnarray}
|\mathbf{\bar{\mathbf{3}}}_{12}\mathbf{3}_{34}\rangle 
   &=&  \frac{1}{2\sqrt{3}}~ \epsilon_{ijm} ~ \epsilon_{klm} ~ |Q_{i}Q_{j}\bar{Q}_{k}\bar{Q}_{l}\rangle
\nonumber 
\\
  &=&  \sqrt{\frac{3}{4}}\bigg(|\mathbf{1}_{13}\mathbf{1}_{24}\rangle-|\mathbf{1}_{14}\mathbf{1}_{23}\rangle\bigg) \ ,
  \label{eq:antitrip_trip_def}
\\\
|\mathbf{6}_{12}\bar{\mathbf{6}}_{34}\rangle  
&=&  \sqrt{\frac{3}{8}}\bigg(|\mathbf{1}_{13}\mathbf{1}_{24}\rangle+|\mathbf{1}_{14}\mathbf{1}_{23}\rangle\bigg)
\ .
\label{eq:sext_antisext_def}
\end{eqnarray}
The states in Eqs. (\ref{eq:antitrip_trip_def}) and (\ref{eq:sext_antisext_def}) are orthogonal to each other in colour space,  i.e.
$\langle\bar{\mathbf{3}}_{12}\mathbf{3}_{34}|\mathbf{6}_{12}\mathbf{\bar{6}}_{34}\rangle  =  0$. Furthermore,
they are eigenstates of the exchange operators of quarks or anti-quarks, and verify the following relations,
\begin{eqnarray}
P_{12}|\bar{\mathbf{3}}_{12}\mathbf{3}_{34}\rangle &= P_{34}|\bar{\mathbf{3}}_{12}\mathbf{3}_{34}\rangle  =& - ~
|\bar{\mathbf{3}}_{12}\mathbf{3}_{34}\rangle \ ,
\nonumber
\\
P_{12}|\mathbf{6}_{12}\mathbf{\bar{6}}_{34}\rangle &= P_{34}|\mathbf{6}_{12}\mathbf{\bar{6}}_{34}\rangle  =&  + ~
|\mathbf{6}_{12}\mathbf{\bar{6}}_{34}\rangle  ,
\end{eqnarray}
where $P_{ij}$ is the exchange operator of (anti)quark $i$ with (anti)quark $j$. 
Eqs.  (\ref{eq:antitrip_trip_def}) and (\ref{eq:sext_antisext_def}) can be inverted, giving,
\begin{eqnarray}
|\mathbf{1}_{13}\mathbf{1}_{24}\rangle  =  \sqrt{\frac{2}{3}}|\mathbf{6}_{12}\mathbf{\bar{6}}_{34}\rangle+\frac{1}{\sqrt{3}}|\bar{\mathbf{3}}_{12}\mathbf{3}_{34}\rangle \ ,
\nonumber
\\
|\mathbf{1}_{14}\mathbf{1}_{23}\rangle  =  \sqrt{\frac{2}{3}}|\mathbf{6}_{12}\mathbf{\bar{6}}_{34}\rangle-\frac{1}{\sqrt{3}}|\bar{\mathbf{3}}_{12}\mathbf{3}_{34}\rangle \ ,
\end{eqnarray}
which shows that the meson-meson states of Eq. (\ref{eq:singleto2}) are not eigenstates of the quark and of the anti-quark
exchange operators $P_{12}$ and $P_{34}$.

The static potential $V$ for a $QQ\bar{Q}\bar{Q}$ system is a complicated object which may involve, two, three and four body interactions.
In general, $V$ also depends on the allowed quantum numbers of the constituents of the multiquark state. The static potential should allow, when combined with quantum mechanics, for the groundstates to be the ones of Fig. \ref{fig:tripleflipflop}.
For example, the static potential should allow for the formation of two-meson states when the quark-anti-quark distances are small compared to the quark-quark and anti-quark-anti-quark distances, or possibly for the formation of a tetraquark at other particular distances.

As an approximate model to understand the results of the lattice simulations for the static potential in terms of overlaps with the various colour singlets, one can consider the two-body potential given by the Casimir scaling,
\begin{equation}
V_{CS} = \sum_{i<j} C_{ij} ~ V_{M} \ ,
\label{eq:VCasimir}
\end{equation}
where $V_M$ is the mesonic static $Q \bar Q$ potential
with $C_{ij} = \frac{\lambda_{i}^{a}\cdot\lambda_{j}^{a}}{-16/3}$ and compare the results of the simulations with the one of any of the colour singlet states and the Casimir potential given by,
\begin{equation}
V_\Psi = \langle \Psi | V_{CS} | \Psi \rangle \ .
\end{equation}
Note, for a two body system, the one gluon exchange predicts a static potential proportional to $\lambda^a_i \cdot \lambda^a_j$.

\begin{table}[!t]
\begin{tabular}{|@{\hspace{0.5cm}}c@{\hspace{0.5cm}}|@{\hspace{0.5cm}}c@{\hspace{0.5cm}}|@{\hspace{0.5cm}}c@{\hspace{0.5cm}}|@{\hspace{0.5cm}}c@{\hspace{0.5cm}}|}
\hline 
$|\Psi\rangle$ & $C_{12}$ & $C_{13}$ & $C_{14}$\tabularnewline
\hline 
\hline 
$|\mathbf{1}_{13}\mathbf{1}_{24}\rangle$ & 0 & 1 & 0\tabularnewline
\hline 
$|\mathbf{8}_{13}\mathbf{8}_{24}\rangle$ & 1/4 & -1/8 & 7/8\tabularnewline
\hline 
$|\mathbf{1}_{14}\mathbf{1}_{23}\rangle$ & 0 & 0 & 1\tabularnewline
\hline 
$|\mathbf{8}_{14}\mathbf{8}_{23}\rangle$ & 1/4 & 7/8 & -1/8\tabularnewline
\hline 
$|\bar{\mathbf{3}}_{12}\mathbf{3}_{34}\rangle$ & 1/2 & 1/4 & 1/4\tabularnewline
\hline 
$|\mathbf{6}_{12}\mathbf{\bar{6}}_{34}\rangle$ & -1/4 & 5/8 & 5/8\tabularnewline
\hline 
\end{tabular}\caption{Normalized mean values of the Casimir invariant operators $\langle\Psi|C_{ij}|\Psi\rangle$. The indices $i$ and $j$ refer to the quarks and anti-quarks. 
\label{tab:NormCas}}
\end{table}

\begin{figure*}[!t]
\begin{centering}
\includegraphics[scale=0.8]{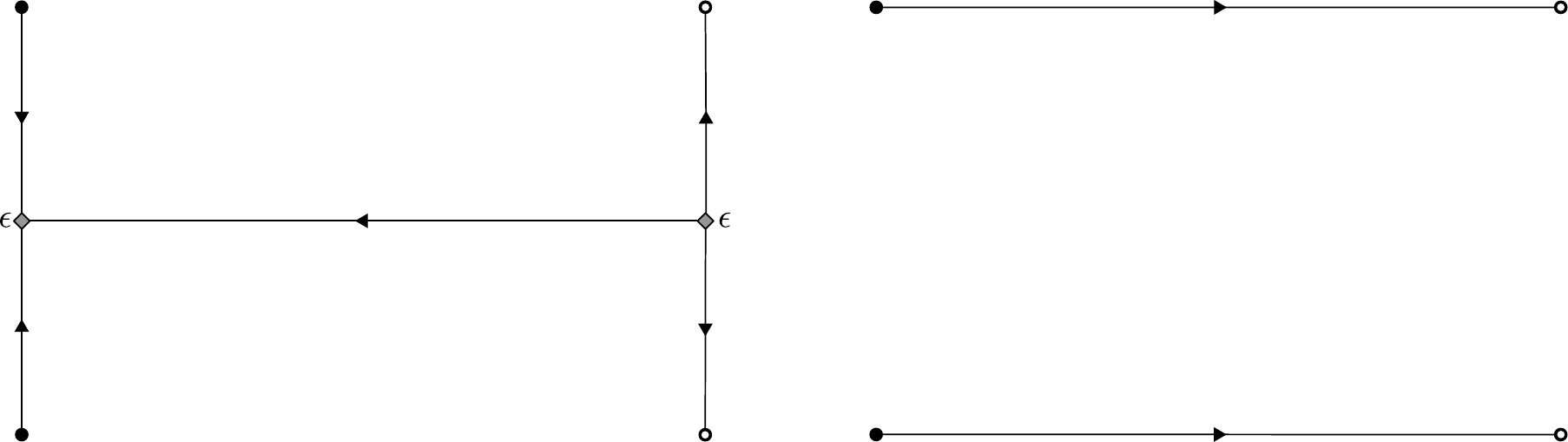}
\par\end{centering}
\caption{The two operators used in the computation of the static potential for the parallel alignment geometry. 
The left operator $\mathcal{O}_{YY}$ (where $\epsilon$ stands for a Levi-Civita symbol , see text for details), creates a $|\mathbf{\bar{3}}_{12}\mathbf{3}_{34}\rangle$ state, 
while the right one creates a two-meson state.
 \label{fig:parallel_ops}}
\end{figure*}

\begin{figure*}[!t]
\begin{centering}
\includegraphics[scale=0.8]{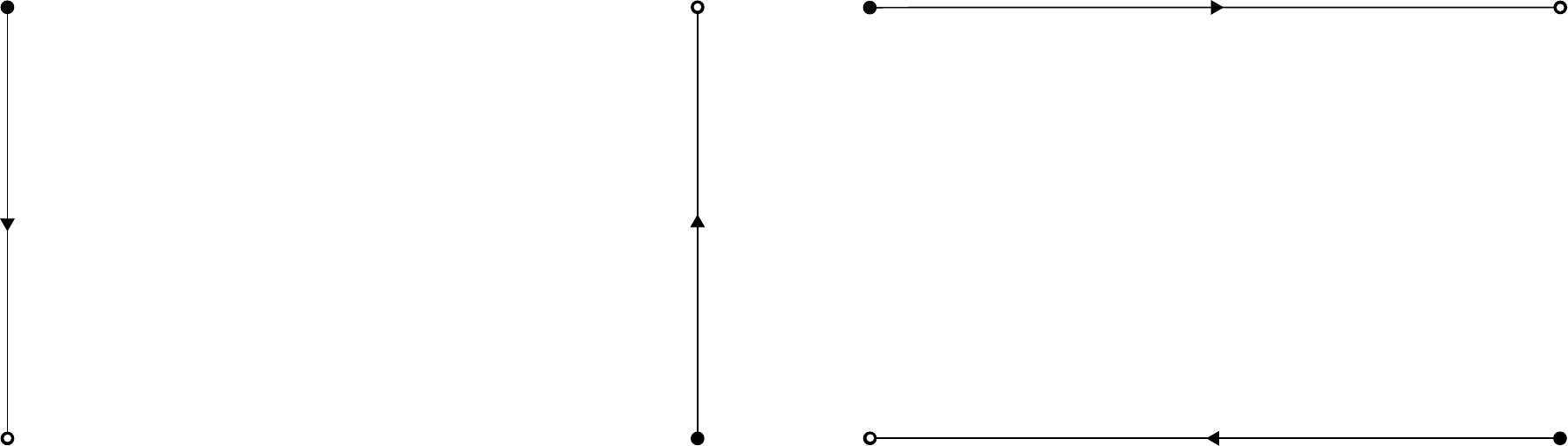}
\par\end{centering}
\caption{The two operators used in the computation of the static potential for the anti-parallel alignment case. Note that
both Wilson lines describe two-meson operators. 
\label{fig:anti-parallel_ops}}
\end{figure*}

The expectation values $\langle \Psi | C_{ij} | \Psi \rangle$ for the possible colour singlet states associated to the $QQ\bar{Q}\bar{Q}$ system
are reported in Table \ref{tab:NormCas}. These numbers are important to obtain a qualitative insight into the result of the simulations. 
For instance, if for a given state $C_{ij} < 0$, we don't expect that the lattice result would give us a strong attraction between the 
particles $i$ and $j$ and, therefore, one can expected significant deviations of the static potential relative to the potential $V_{CS}$ associated to
the corresponding colour singlet state.

Moreover we consider as well the first excitation of the $QQ\bar{Q}\bar{Q}$, which also depends in the particular distances of the system.
Based in the orthogonality conditions and in a crude Casimir scaling where $V_M$ would be a spatial independent potential,
we would expect the pairs of colour singlet states,
($ |1_{13} 1_{24} \rangle$, $ |8_{13} 8_{24} \rangle$) ,
($ |1_{14} 1_{23} \rangle$, $ |8_{14} 8_{23} \rangle$) and 
($ | \overline 3_{12} 3_{34} \rangle$, $ |6_{13} \overline 6_{24} \rangle$)
to form possible (groundstate, first excited state) pairs. This already goes beyond the simple paradigm of Fig. \ref{fig:tripleflipflop}.

Nevertheless, Eq. (\ref{eq:VCasimir}) is clearly an approximation, and our aim is to compute more rigorous potentials.
Previous lattice studies 
\cite{Alexandrou:2004ak,Bornyakov:2005kn,Okiharu:2004ve,Okiharu:2004wy}
show that the static potential for a tetraquark system is not described entirely by a function proportional
to this potential. 
An example of such a kind of potentials is the two-meson potential,
\begin{eqnarray}
V_{33} 
&=& 
\langle \mathbf{1}_{13} \mathbf{1}_{24} | V_{CS} | \mathbf{1}_{13} \mathbf{1}_{24} \rangle \ ,
\nonumber
\\
&=&
V_M(r_{13}) + V_M(r_{24}) \ ,
\end{eqnarray}
which we expect to saturate the ground state when the quark-quark and anti-quark-anti-quark distances are large.


\section{Geometrical Setup \label{Sec:geometries}}


We aim to measure the static potential for the $QQ\bar{Q}\bar{Q}$ system but also to investigate the
transition between the  tetraquark and a two meson state,  and the transition between the two two-meson states. This computation within lattice QCD simulations requires choosing a particular geometrical setup of the 
quark system under investigation. In principle, one could choose any of the available geometrical configurations allowed 
by the hypercubic lattice. In order to study in detail the transitions between the different states, in the current work we opt for restricting our study to the case where the four particles are at the corners of a 
rectangle and look at 
%
%
%
%
%
%
two particular alignments.  In the so-called parallel alignment, see Fig.~\ref{Fig:geometrias} (left), the two quarks (anti-quarks) are at adjacent
corners of the rectangle. In the anti-parallel alignment, see Fig.~\ref{Fig:geometrias} (right), the quarks (anti-quarks) are at the opposite corners of the 
rectangle.


\begin{figure}[!t]
\begin{centering}
\includegraphics[scale=0.7]{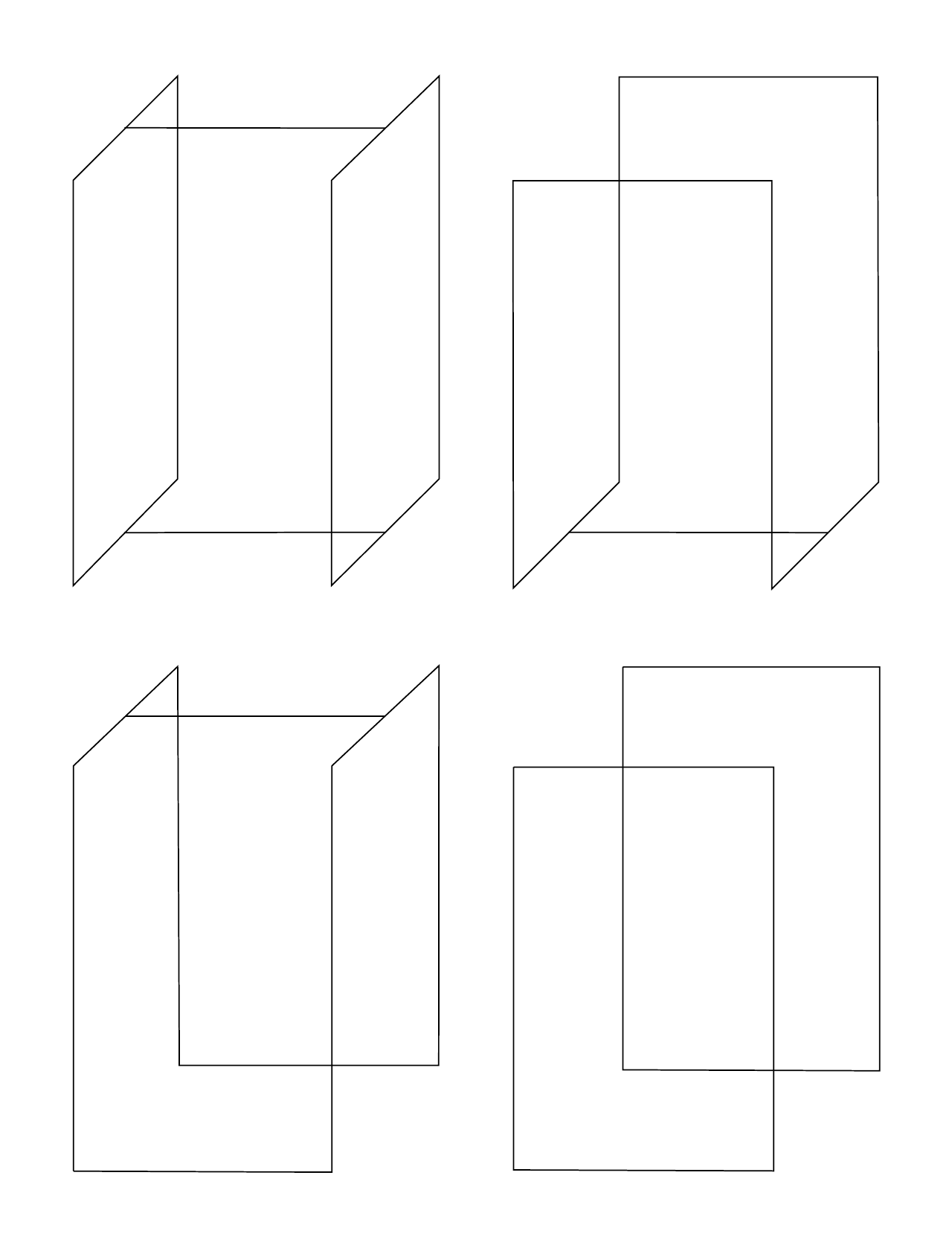}
\par\end{centering}
\caption{Correlation matrix, defined with Wilson lines, for the parallel alignment case.
\label{fig:parallel_loops}}
\end{figure}

\begin{figure}[!t]
\begin{centering}
\includegraphics[scale=0.7]{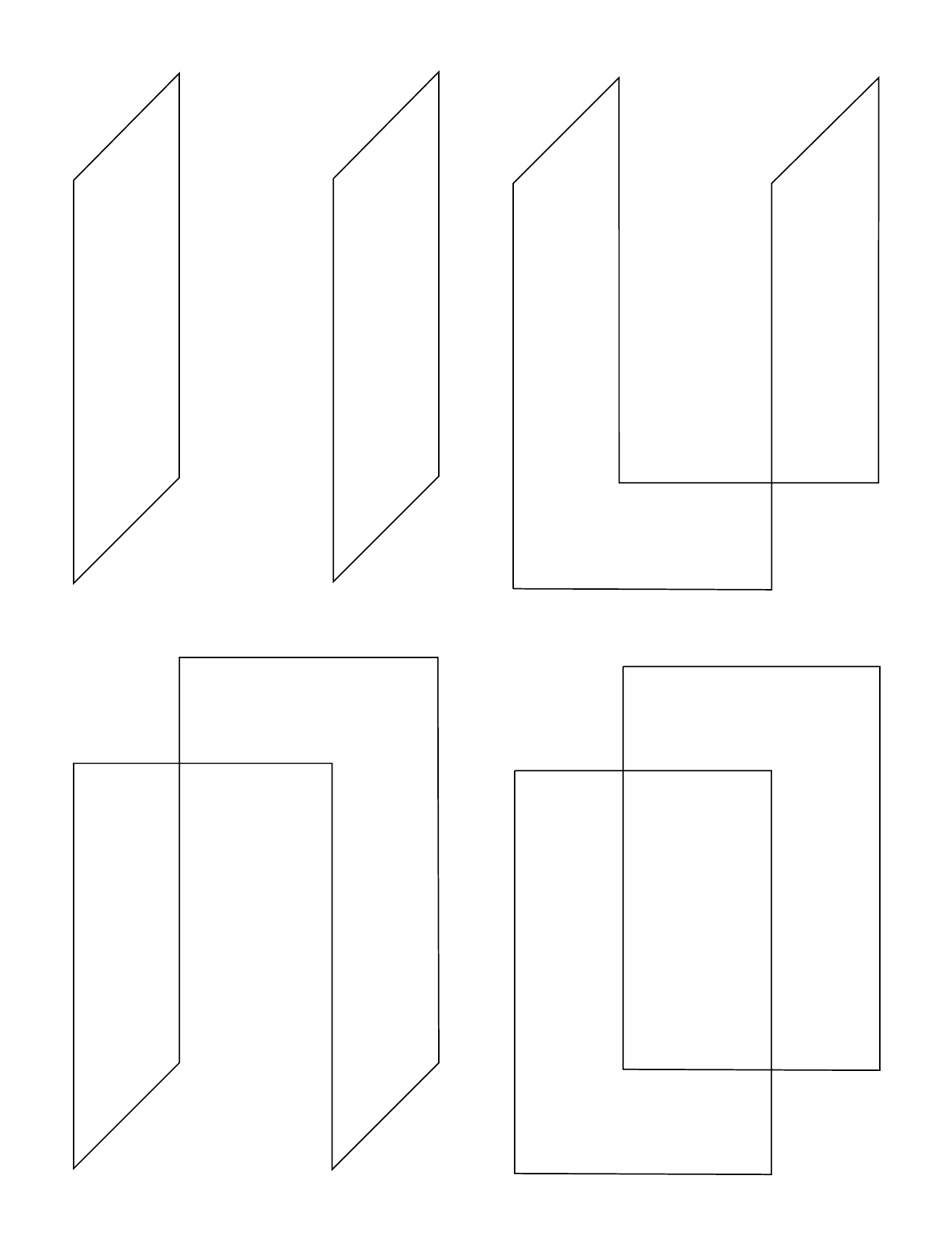}
\par\end{centering}
\caption{Correlation matrix, defined with Wilson lines, for the anti-parallel alignment of quarks. \label{fig:anti-parallel_loops}}
\end{figure}

\subsection{Parallel Alignment of Quarks}

For this geometry, where the two quarks are at neighbour corners of the rectangle, we can describe the system via
the intra-diquark distances,
\begin{equation}
r_{12}=|\mathbf{x}_{1}-\mathbf{x}_{2}|=|\mathbf{x}_{3}-\mathbf{x}_{4}| \ ,
\label{eq:r12_def}
\end{equation}
and the inter-diquark distances,
\begin{equation}
r_{13}=|\mathbf{x}_{1}-\mathbf{x}_{3}|=|\mathbf{x}_{2}-\mathbf{x}_{4}| \ .
\label{eq:r13p_def}
\end{equation}
Note that for both cases the second equality  holds only due to the particular geometrical configuration considered.

If one assumes that quarks are confined within colourless states, this geometrical setup has two limits which allow to study the transition
between a tetraquark state and a two meson system. Indeed, when $r_{12}\ll r_{13}$ one expects the ground state of the 
$QQ\bar{Q}\bar{Q}$  system to be that of a tetraquark,  while for the opposite case, i.e. for $r_{13}\ll r_{12}$, one expects the system, i.e.
its potential, to behave as a two meson system.

For this geometrical setup, in the evaluation of the static potential we consider the basis of operators shown in Fig. \ref{fig:parallel_ops}. 
They are associated with a tetraquark operator (left in the figure) and a two-meson operator (right in the figure), the two ground state configurations 
expected for this particular geometry.

\subsection{Anti-parallel Alignment of Quarks}

For the anti-parallel alignment of quarks described in Fig.~\ref{Fig:geometrias} (right), we take as distance variables,
\begin{eqnarray}
r_{13} &=&|\mathbf{x}_{1}-\mathbf{x}_{3}|=|\mathbf{x}_{2}-\mathbf{x}_{4}| \ ,
\label{eq:r13a_def} 
\nonumber 
\\
r_{14} &=&|\mathbf{x}_{1}-\mathbf{x}_{4}|=|\mathbf{x}_{2}-\mathbf{x}_{3}|  \ , 
\label{eq:r14_def} 
\end{eqnarray}
where, again, the second equalities are valid due to the particular characteristics of the geometrical distribution
of quarks and anti-quarks.

For this geometrical setup, one expects the ground state of the system when $r_{13}\ll r_{14}$ and $r_{14}\ll r_{13}$
to be dominated by the two possible independent two-meson states. For the computation of the static
potential we use the basis of operators shown in Fig. \ref{fig:anti-parallel_ops} that are associated with the
two two-meson operators.

\section{Computing the Static Potential \label{Sec:calculopotencial}}

\begin{table*}[!t]
\begin{tabular}{|c|c|c|c|c|c|c|c|}
\hline 
$r_\text{min}$ & $r_\text{max}$ & $\chi^{2}/\mbox{d.o.f.}$ & $Ka$ & $\gamma$ & $\sigma a^{2}$ & $a$ (fm) & $a^{-1}(\mbox{GeV})$\tabularnewline
\hline 
\hline 
5 & 12 & 0.98 & 0.6406(21) & 0.3078(77) & 0.02490(14) & 0.0681 & 2.898\tabularnewline
\hline 
6 & 12 & 0.62 & 0.6382(49) & 0.2987(199) & 0.02506(29) & 0.0681 & 2.898\tabularnewline
\hline 
5 & 11 & 1.08 & 0.6409(21) & 0.3085(75) & 0.02488(14) & 0.0681 & 2.898\tabularnewline
\hline 
6 & 11 & 0.79 & 0.6385(55) & 0.2996(224) & 0.02504(34) & 0.0681 & 2.898\tabularnewline
\hline 
\end{tabular}
\caption{Fits of the  static $Q \bar Q$ meson potential (Wilson loop) in quenched QCD, for different intervals $r \in [r_\text{min},r_\text{max}]$, to the Cornell potential model of Eq. (\ref{eq:cornell}). 
\label{tab:fits_meson_q}}
\end{table*}

\begin{table*}[!t]
\begin{tabular}{|c|c|c|c|c|c|c|c|}
\hline 
$r_\text{min}$ & $r_\text{max}$ & $\chi^{2}/\mbox{d.o.f.}$ & $Ka$ & $\gamma$ & $\sigma a^{2}$ & $a$ (fm) & $a^{-1}(\mbox{GeV})$\tabularnewline
\hline 
\hline 
3 & 12 & 0.43 & 0.2995(23) & 0.3625(49) & 0.03092(25) & 0.0775 & 2.546\tabularnewline
\hline 
4 & 12 & 0.50 & 0.3005(84) & 0.3654(270) & 0.03085(64) & 0.0775 & 2.546\tabularnewline
\hline 
5 & 12 & 0.43 & 0.2931(169) & 0.3386(638) & 0.03129(109) & 0.0772 & 2.546\tabularnewline
\hline 
3 & 11 & 0.19 & 0.3017(42) & 0.3666(117) & 0.03065(33) & 0.0773 & 2.553\tabularnewline
\hline 
4 & 11 & 0.04 & 0.3063(25) & 0.3799(76) & 0.03030(21) & 0.0772 & 2.546\tabularnewline
\hline 
5 & 11 & 0.05 & 0.3042(58) & 0.3728(204) & 0.03044(40) & 0.0772 & 2.546\tabularnewline
\hline 
\end{tabular}
\caption{Fits of the  static $Q \bar Q$ meson potential (Wilson loop) in full QCD, for different intervals $r \in [r_\text{min},r_\text{max}]$, to the Cornell potential
model of Eq. (\ref{eq:cornell}). 
\label{tab:fits_meson_dyn}}
\end{table*}

For the computation of the static potential, including the groundstate and the first excited state, we rely on a basis of two operators $\mathcal{O}_{i}$ for each of the geometrical setups 
discussed in Sec. \ref{Sec:geometries}.  Defining the correlation matrix,
\begin{eqnarray}
   M_{ij}
   &=&\langle\mathcal{O}_{i}(0)^{\dagger} \, \mathcal{O}_{j}(t)\rangle 
   \nonumber \\
   &=& \sum_{n}c_{in}^{*} \, c_{jn} \, e^{-V_{n}t} \ ,
\end{eqnarray}
where $\langle \cdots \rangle$ stands for vacuum expectation value, $c_{in}=\langle n|\mathcal{O}_{i}|0\rangle$
and $|n \rangle$ are the eigenstates of the Hamiltonian of the system,
the determination of the potential requires the knowledge of the solutions of the generalized  eigenvalue problem 
\begin{equation}
M_{ij}(t)\, a_{j}(t) = \lambda_{k}(t) \, M_{ij}(t_{0}) \, a_{j}(t)\label{eq:gen_eigen_wilson} \ .
\end{equation}
In our calculation, we assume that the creation of an excited state out of the vacuum occurs at $t = 0$. 
From the generalized eigenvalues $\lambda_{k}$, the energy levels of system $V_k$ can be estimated from the plateaux on
the effective mass given by,
\begin{eqnarray}
M_{eff}(t) &=& \log\frac{\lambda_{k}(t)}{\lambda_{k}(t+1)}
\nonumber \\
&=& V_{k}+\mathcal{O}(e^{-(V_{k+1}-V_{k})t}) \ .
\end{eqnarray}
In practice, the effective mass plateaux are identified fitting to a constant  both generalized eigenvalues.
In this way, one is able to compute both the static potential for the ground state and the first excited state of the system.

As described above, the basis of operators chosen to compute $V$ depends on the geometry of the system 
and on the expected ground states. For the anti-parallel alignment, we use two meson-meson operators, while
for the parallel alignment a meson-meson operator and a diquark-antidiquark 
operator, i.e. a $\bar{\mathbf{3}}_{12}\mathbf{3}_{34}$ colour configuration, are used to compute the correlation matrix.

In the case where the quarks are in the anti-parallel alignment  the operators used to compute the potential are,
\begin{eqnarray}
\mathcal{O}_{13,24}  &=&  \frac{1}{3}Q_{1}^{i}L_{13}^{ij}\bar{Q}_{3}^{j}\,Q_{2}^{k}L_{24}^{kl}\bar{Q}_{4}^{l} \ ,
\nonumber
\\
\mathcal{O}_{14,23}  &=&  \frac{1}{3}Q_{1}^{i}L_{14}^{ij}\bar{Q}_{4}^{j}\,Q_{2}^{k}L_{23}^{kl}\bar{Q}_{3}^{l} \ ,
\label{eq:anti-parallel_ops_def}
\end{eqnarray}
where $L$ are Wilson lines connecting the quark. Its representation in terms of closed Wilson loops is given in 
Fig. \ref{fig:anti-parallel_loops}. The corresponding correlation matrix reads,
\begin{equation}
M = 
\left(\begin{array}{cc}
W_{13}W_{24} & \frac{1}{3}W_{1324}\\
\frac{1}{3}W_{1423} & W_{14}W_{23}
\end{array}\right) \ ,
\end{equation}
where $W_{i}$ are normalized mesonic Wilson loops $W=\mbox{\ensuremath{\frac{1}{3}}Tr}[U]$.

On the other hand, for the parallel alignment the two operators we consider are, 
\begin{eqnarray}
&\mathcal{O}_{YY}  =&  \frac{1}{2\sqrt{3}}Q_{1}^{i}Q_{2}^{j}\epsilon_{i'j'k}L_{1a}^{ii'}L_{2a}^{jj'}L_{ab}^{kk'}\epsilon_{k'l'm'}L_{b3}^{l'l}L_{b4}^{m'm}\bar{Q}_{3}^{l}\bar{Q}_{4}^{m} \ , \nonumber \\
&\mathcal{O}_{13,24}&  =  \frac{1}{3}Q_{1}^{i}L_{13}^{ij}\bar{Q}_{3}^{j}\,Q_{2}^{k}L_{24}^{kl}\bar{Q}_{4}^{l} \ .  
\label{eq:parallel_ops_def}
\end{eqnarray}
The closed Wilson loops associated to $\mathcal{O}_{YY}$ and $\mathcal{O}_{13,24}$ are represented in Fig. \ref{fig:parallel_loops} and
the corresponding correlation matrix is given by,
\begin{equation}
M =
\left(\begin{array}{cc}
W_{YY} & \frac{1}{2\sqrt{3}}W_{YY,1324}\\
\frac{1}{2\sqrt{3}}W_{1324,YY} & W_{13}W_{24}
\end{array}\right) \ .
\end{equation}

%

\section{Lattice Setup \label{Sec:lattice}}

From the static potential we aim to understand the transition between possible configurations of a
$QQ\bar{Q}\bar{Q}$ system. Furthermore, we also want to glimpse any possible differences due to the quark dynamics.
Therefore, for the computation of $V_k$ we consider two different simulations.  

Our quenched simulation uses an ensemble of 1199 configurations provided by the PtQCD collaboration \cite{Cardoso:2011xu,Cardoso:2010di,PtQCD},  generated using the Wilson action in a
$24^3 \times 48$ lattice for a value of $\beta = 6.2$. The quenched configurations were generated using GPU's and a combination of  Cabbibo-Marinari, pseudo-heatbath and over-relaxation algorithms, and computed in the GPU servers of the PtQCD collaboration.

Our full QCD simulation uses a Wilson fermion dynamical ensemble of 156 configurations generated in a $24^3 \times 48$ lattice and a $\beta = 5.6$. 
In the dynamical ensemble we take $\kappa=0.15825$ for hopping parameter, which corresponds to a pion mass of
$m_\pi = 383$ MeV. 
For Wilson fermions the deviations from continuum physics are of order $\mathcal{O} (a)$
in the lattice spacing and, therefore, one can expect relative large systematic errors. However, we expect that the static
potential as measured from the full QCD simulation away from the physical point to be more realistic when compared
to the quenched simulation.
The full QCD configuration generation has been performed in the Centaurus cluster \cite{LCA} using the Chroma library \cite{Edwards:2004sx}. The Hybrid Monte Carlo integrator scheme has been tuned using the methods described in 
\cite{Clark:2011ir,Kennedy:2012gk}.

Then, with both the quenched and full QCD ensembles of configurations, we perform our correlation matrix computations at the PC cluster ANIMAL of the PtQCD collaboration.

%
%

The Wilson loops at large Euclidean time are decaying exponential functions of the static potential times the Euclidean time and, therefore, 
for large Euclidean times the Wilson loops are dominated by the statistical noise of the Monte Carlo. 
A reliable measurement of the static potential requires techniques which reduce the contribution of the noise to the correlation functions
used in the evaluation of $V$.

The quality of the measurement of the effective masses depends strongly on the overlap with the ground state of the system.
In order to improve the ground state overlap we applied  50 iterations of APE smearing \cite{Albanese:1987ds} 
with $w = 0.2$ to the spatial links in both configuration ensembles. 
Furthermore, for the quenched ensemble, to further improve the signal to noise ratio, we used the extended multihit 
technique \cite{Cardoso:2013lla}. This procedure generalizes the multihit as described in \cite{Parisi:1983hm}
by fixing the $n^{th}$ neighbouring links instead of the first ones when performing the averages of the links. 
However, this technique has the inconvenient of changing the short distance behaviour of the correlators and, therefore,
one should not consider the points with $r<r_\text{min}$. In previous studies with the multihit, $r_\text{min}=2$ was sufficient, but in our study we consider $r_\text{min}=4$.
For the dynamical configurations the multihit technique can not be applied and, therefore, we resorted on
hypercubic blocking \cite{Hasenfratz:2001hp} with the parameters $\alpha_1 = 0.75$, $\alpha_2 = 0.60$ and
$\alpha_3 = 0.30$ to improve the signal to noise ratio.


\begin{figure}[!t]
\includegraphics[width=0.75\columnwidth]{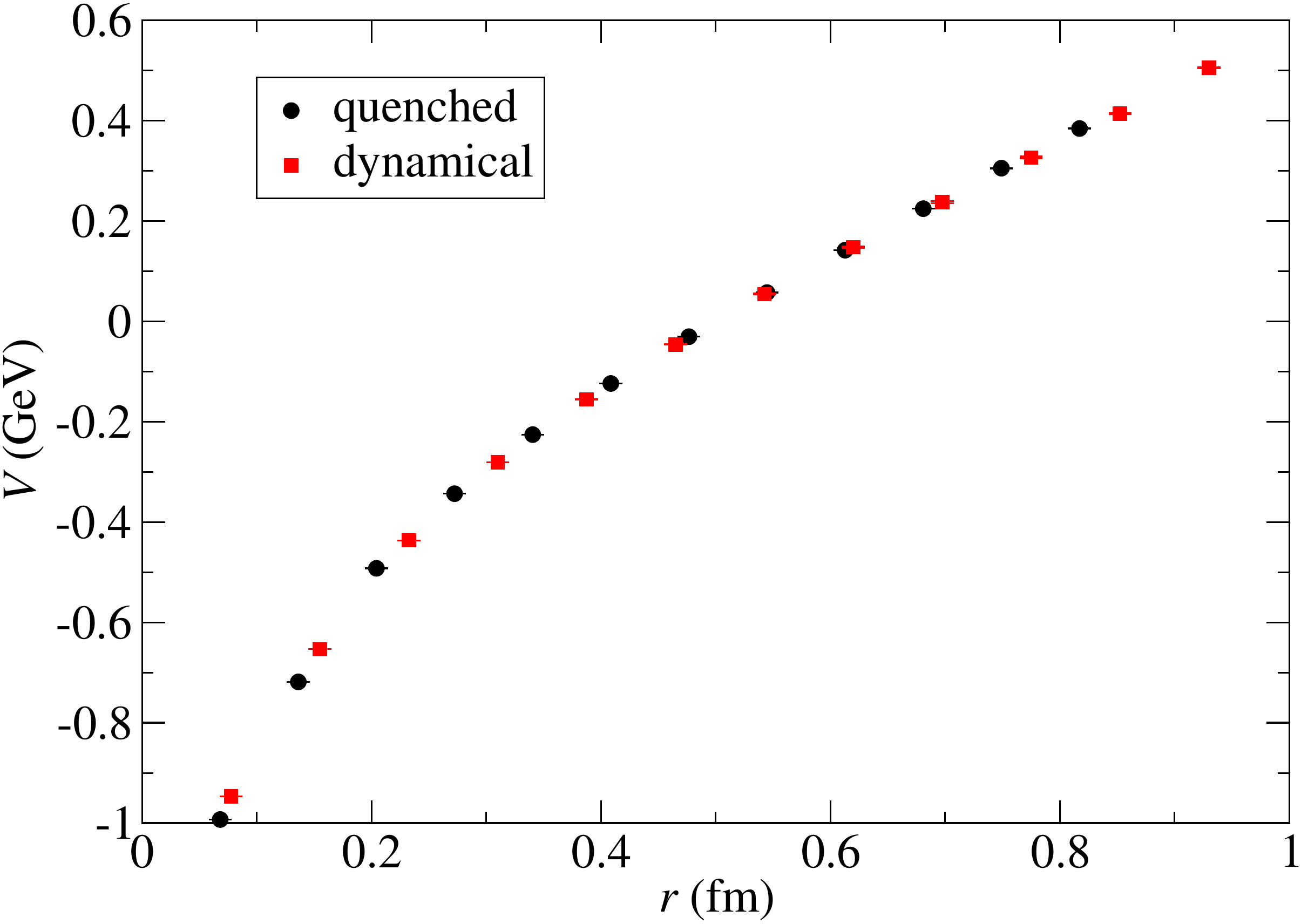}
\caption{ (Colour online.) Meson $Q \bar Q$ static potential for the two sets of data. Both potentials are
shifted by a constant, for $V(r)$ to vanish at $r_{0} = 0.5$ fm. \label{fig:meson_staticpot}}
\end{figure}

For the conversion into physical units we first evaluate Wilson loops to access the ground state meson static potential on a single
axis. In this calculation, we use a variational basis built using four different smearing levels to access the ground state meson
static potential. The lattice data for the static meson potential is then fitted to the Cornell potential functional form,
\begin{equation}
V_M(r)=K-\frac{\gamma}{r}+\sigma r \ .
\label{eq:cornell}
\end{equation}
The fits for different fitting ranges are reported in Tables \ref{tab:fits_meson_q} and \ref{tab:fits_meson_dyn} for
the quenched and the dynamical ensembles, respectively. 
The fits allows for the evaluation of the physical scale associated to the two ensembles through the Sommer method \cite{Sommer:1993ce}.
Indeed, by demanding that,
\begin{equation}
r_{0}^{2}\frac{dV_M}{dr}(r_{0}) = 1.65 \ ,
\end{equation}
where $r_{0}=0.5$ fm, the lattice spacing $a$ is measured and we present it in 
Tables \ref{tab:fits_meson_q} and \ref{tab:fits_meson_dyn} for various fitting ranges. The results show that $a$ is fairly independent of
the fitting intervals and, in the following,  we take $a \simeq 0.0681$ fm for the quenched data ensemble and $a \simeq 0.0775$ fm
for the dynamical data set. 
Our QCD lattice spacing is essentially similar to the one obtained with different techniques.
It follows that the lattice volumes used in the simulation are
$( 1.63 \, $fm$)^3 \times 3.27 \, $fm$ $ for the quenched case and 
$( 1.86 \, $fm$)^3 \times 3.72 \, $fm$ $ for the dynamical simulation.
For completeness, in  Fig. \ref{fig:meson_staticpot} we show the ground state meson potentials for the two ensembles in physical units.

\section{Results for the static $QQ\bar{Q}\bar{Q}$ potential \label{Sec:resultados}}

In this section, we report on the results for the static potential with the two different geometries mentioned  in Sec. \ref{Sec:geometries}, and we apply fits with ansatze bases in the string flip-flop potential and in the Casimir scaling.

%

\begin{figure*}[!t]
\begin{centering}
\includegraphics[scale=0.28]{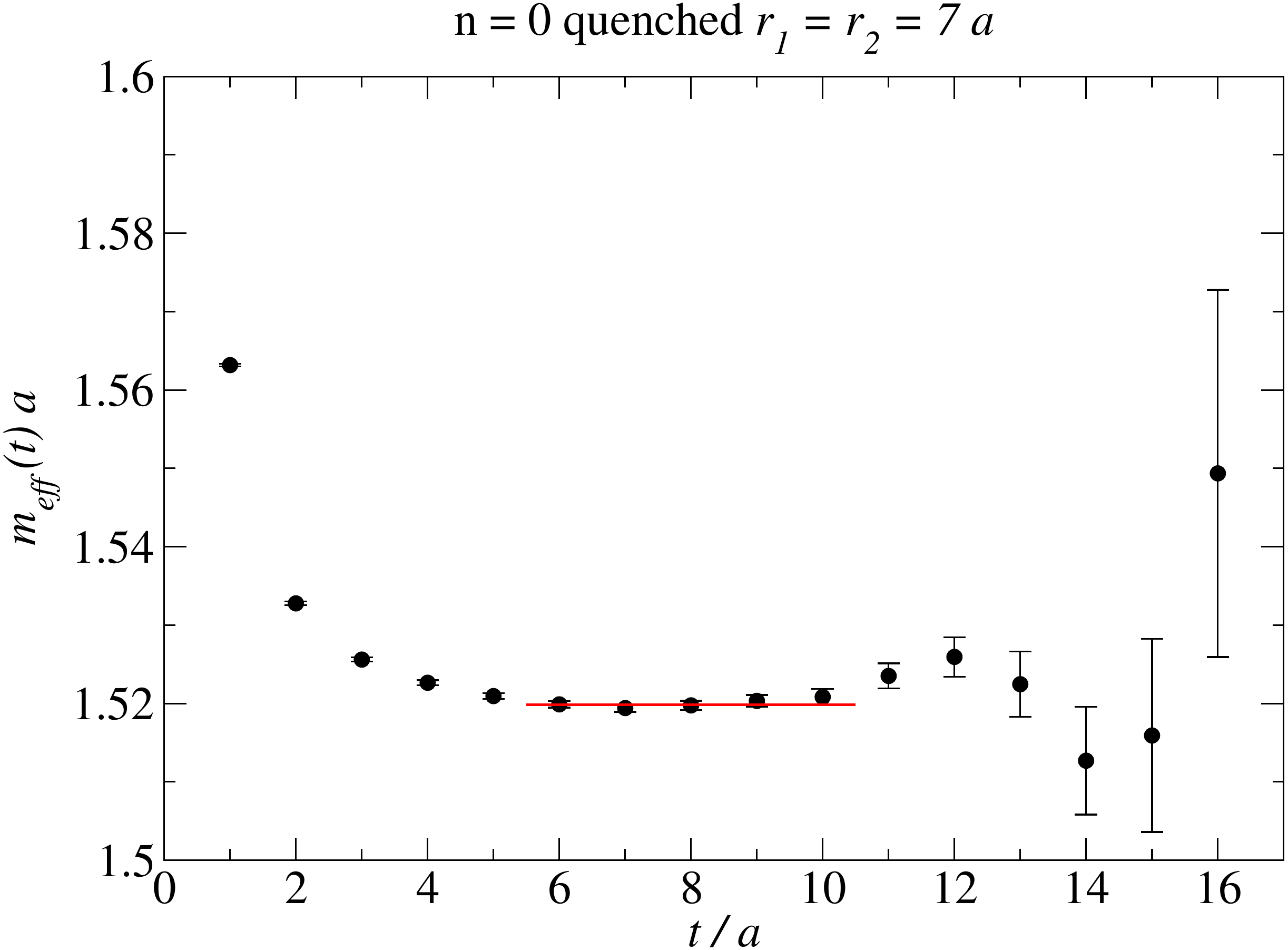}
~
\includegraphics[scale=0.28]{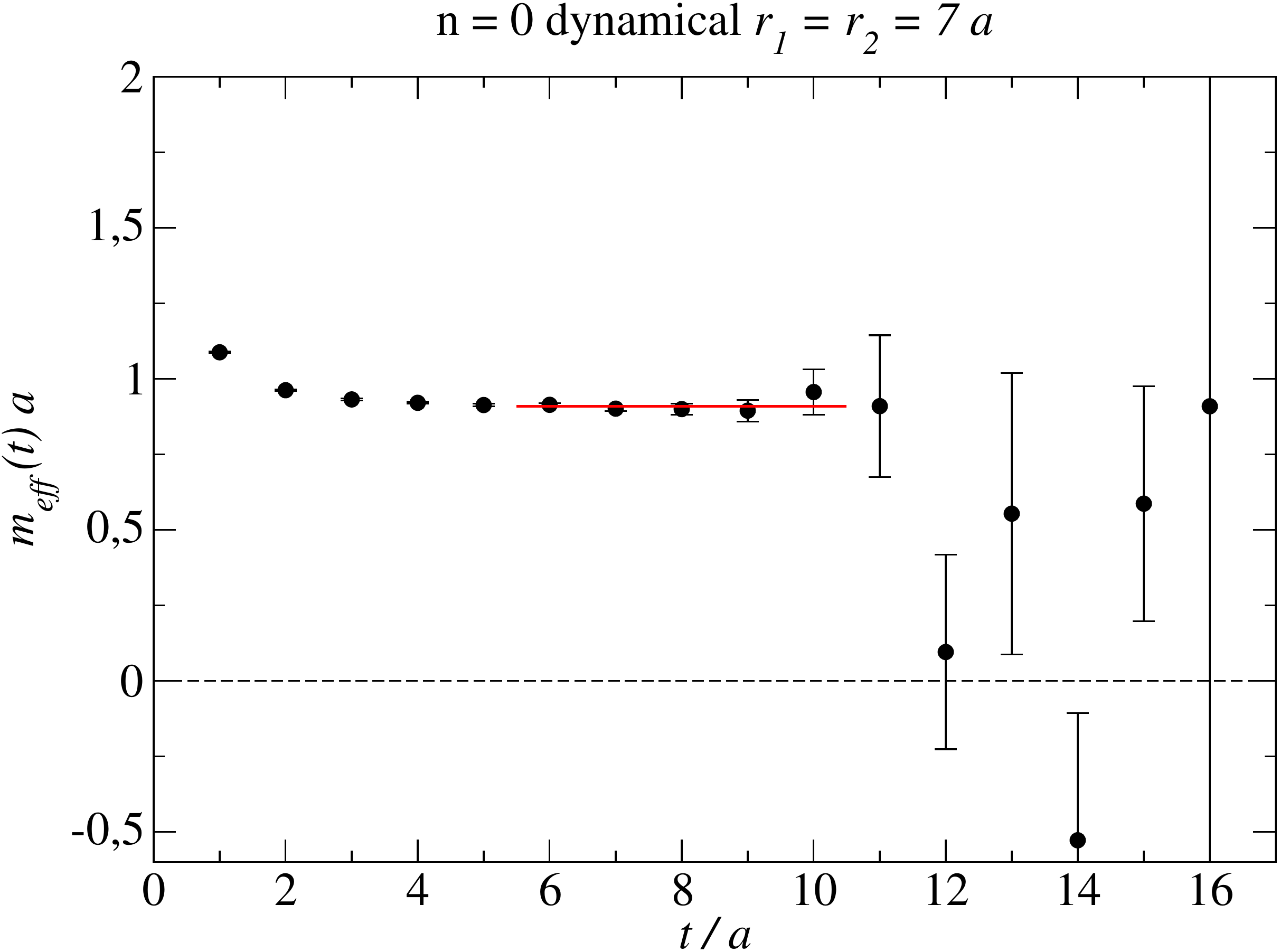}
\\
\includegraphics[scale=0.28]{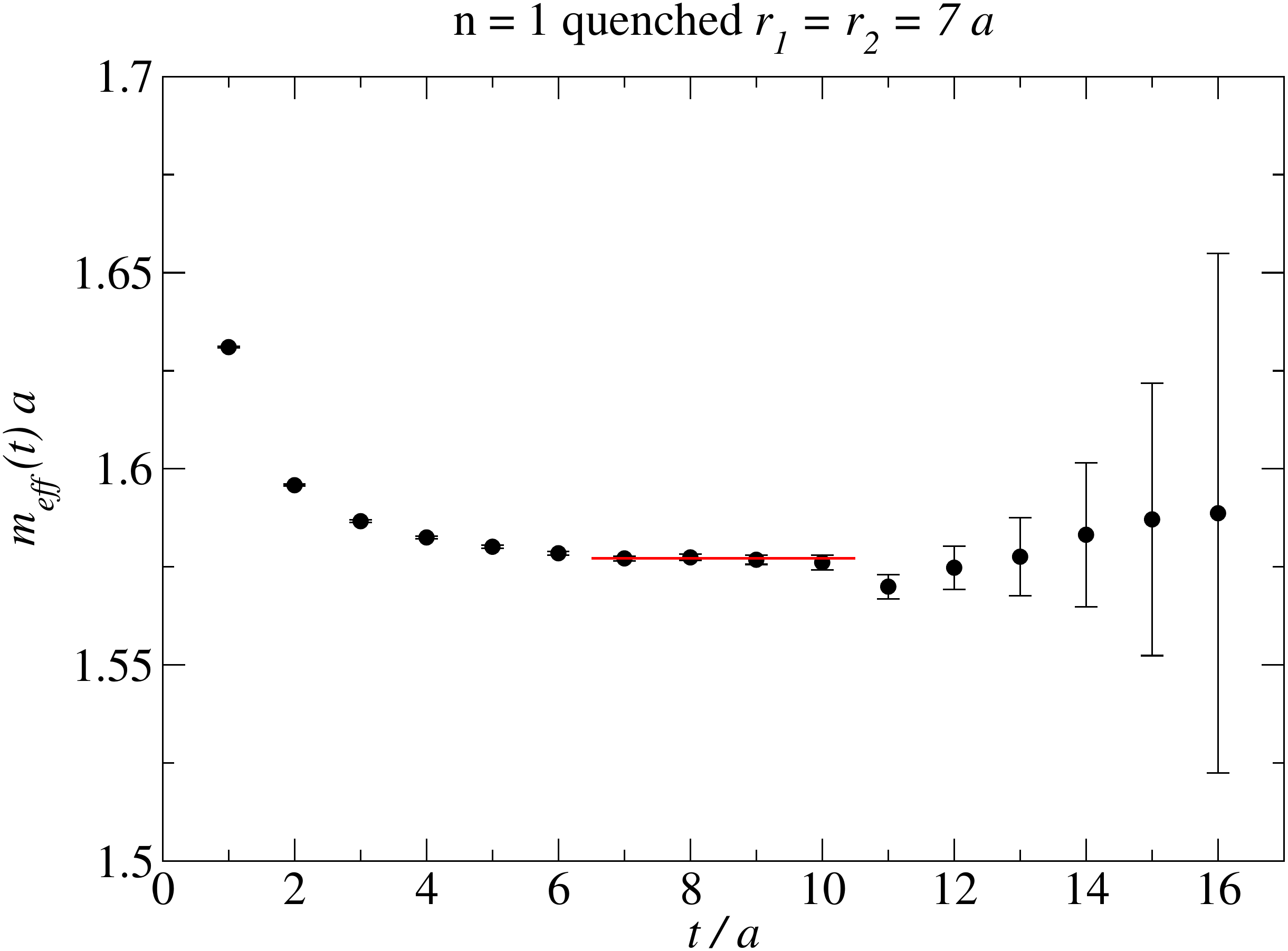}
~
\includegraphics[scale=0.28]{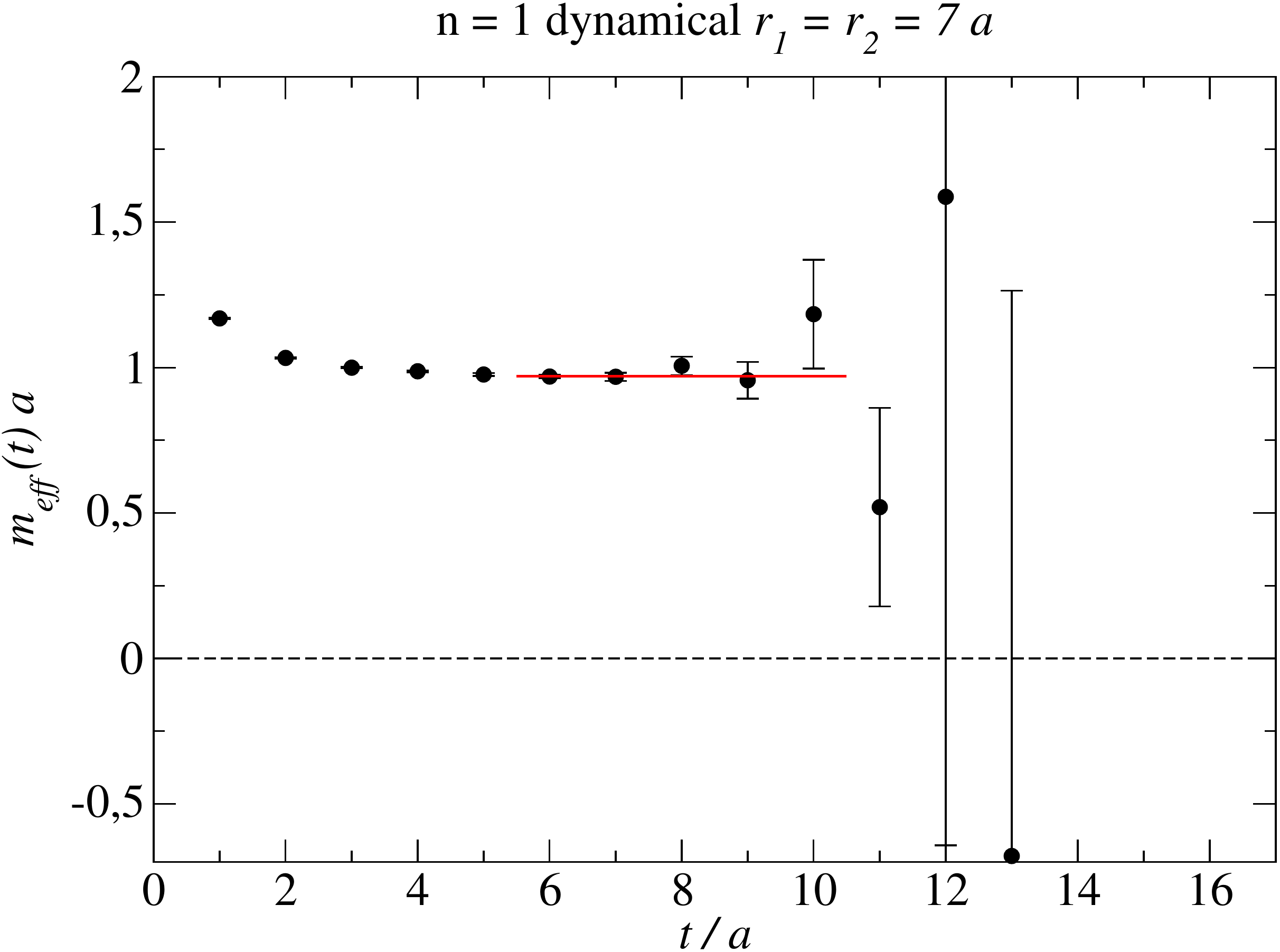}
\par\end{centering}
\caption{ (Colour online.) Effective mass plots for the quenched simulation (left) and full QCD simulation (right), 
in the anti-parallel $Q Q \bar Q \bar Q$ geometry with distances $r_1 = r_2 = 7 a$. for the ground state (top) and the first excited state (bottom). 
The red lines are the plateau fits which measure the static potential \label{fig:effective_mass}}
\end{figure*}

In Fig.~\ref{fig:effective_mass}, as an example, we show effective mass plots for the pure gauge simulation (left), full QCD simulation (right) and for the ground
state (top) and first excited state (bottom) for a $QQ \bar{Q} \bar{Q}$ system in the antiparallel geometry. 
The red curves are the results of fitting the lattice data to measure the static potential. See the appendix for further details on the numerics. 
We consider the maximum number of points aligned in a horizontal line with acceptable $\chi^2 / \text{d. o. f.}$. 
Because the noise reduction technique in the quenched simulation rejects the cases with source distances smaller than $4a$, we end up by accepting a few more results in the full QCD case than in the quenched case.

\subsection{The anti-parallel alignment}

\begin{figure*}[!t]
\includegraphics[width=0.4\paperwidth]{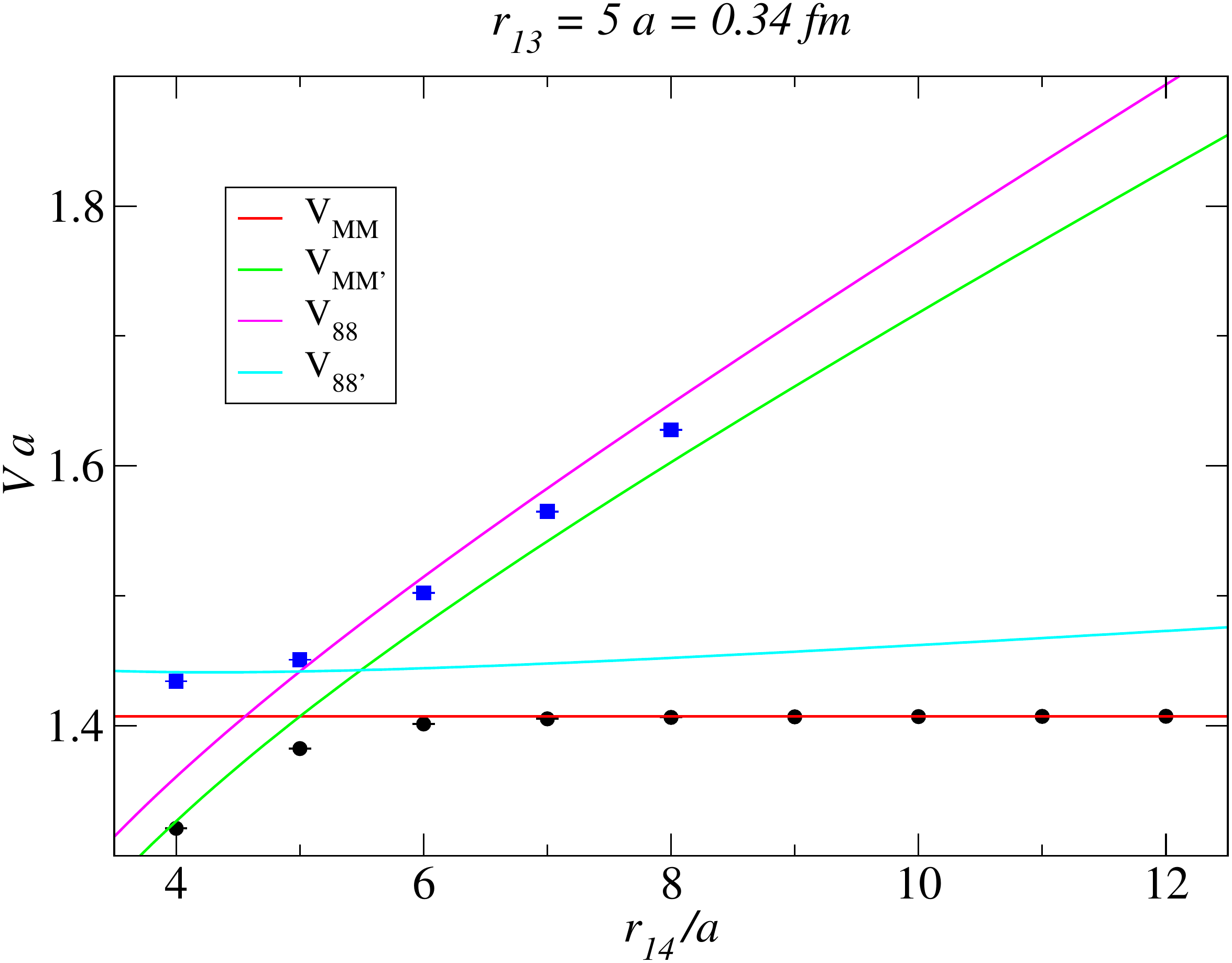}\includegraphics[width=0.4\paperwidth]{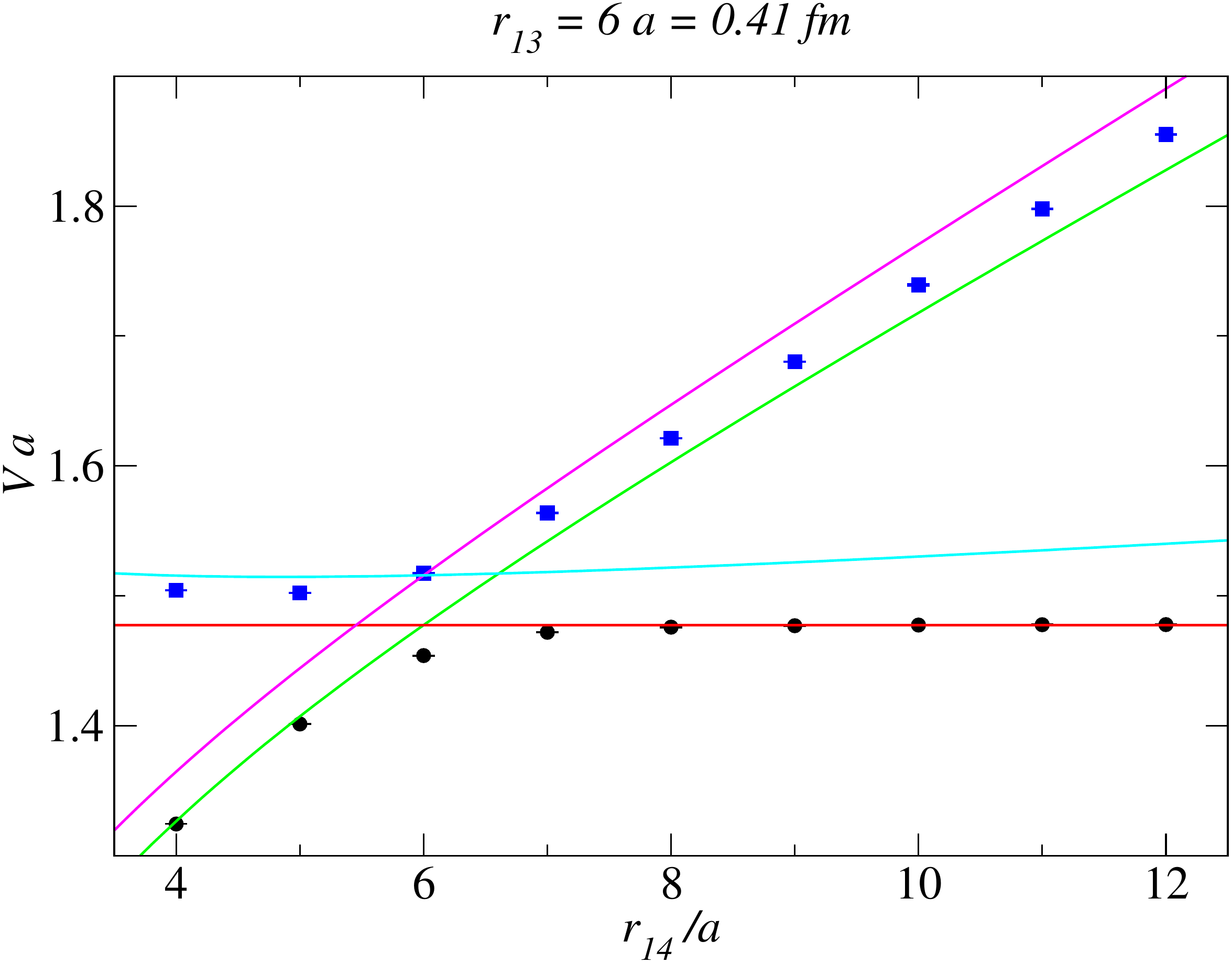}
\includegraphics[width=0.4\paperwidth]{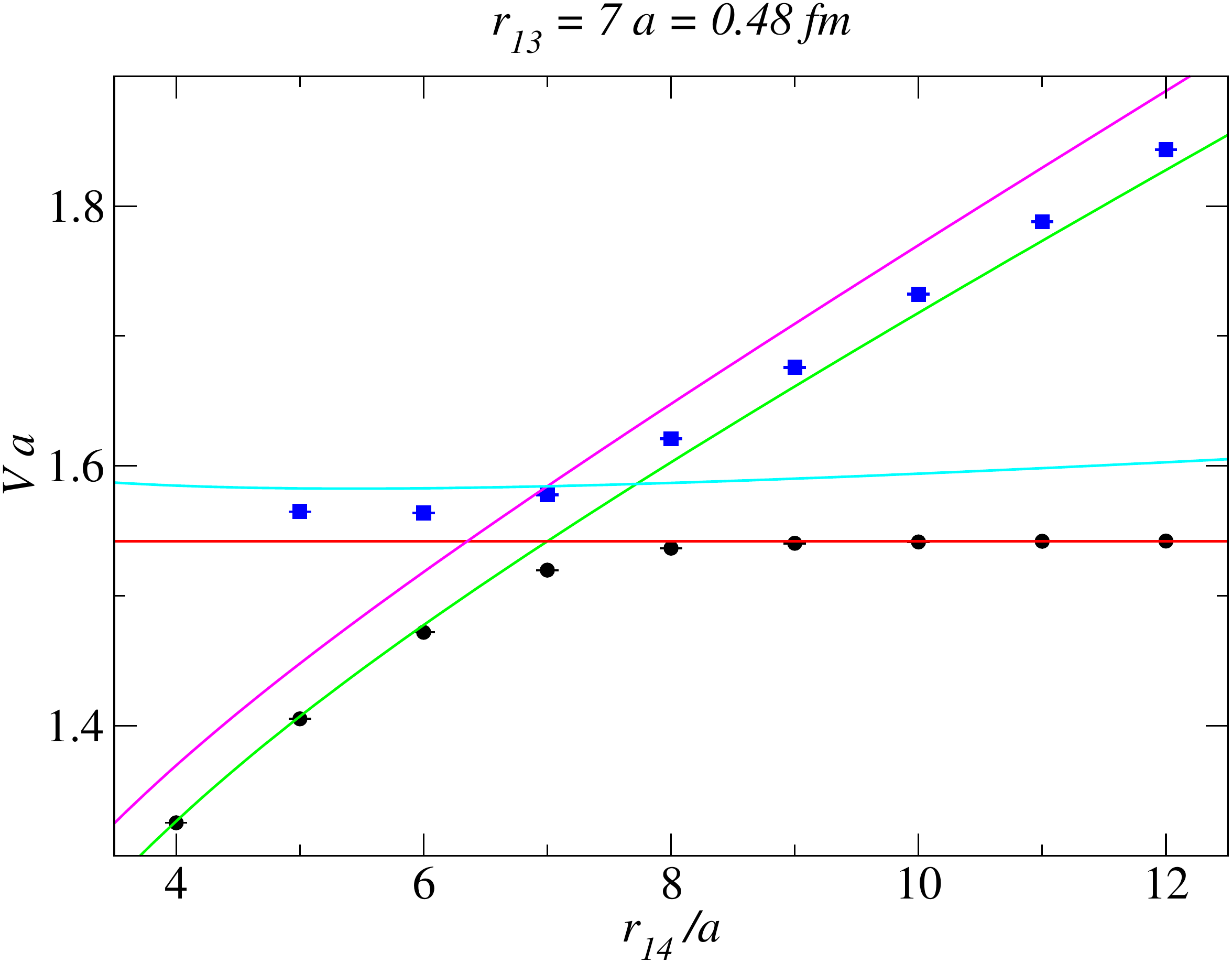}\includegraphics[width=0.4\paperwidth]{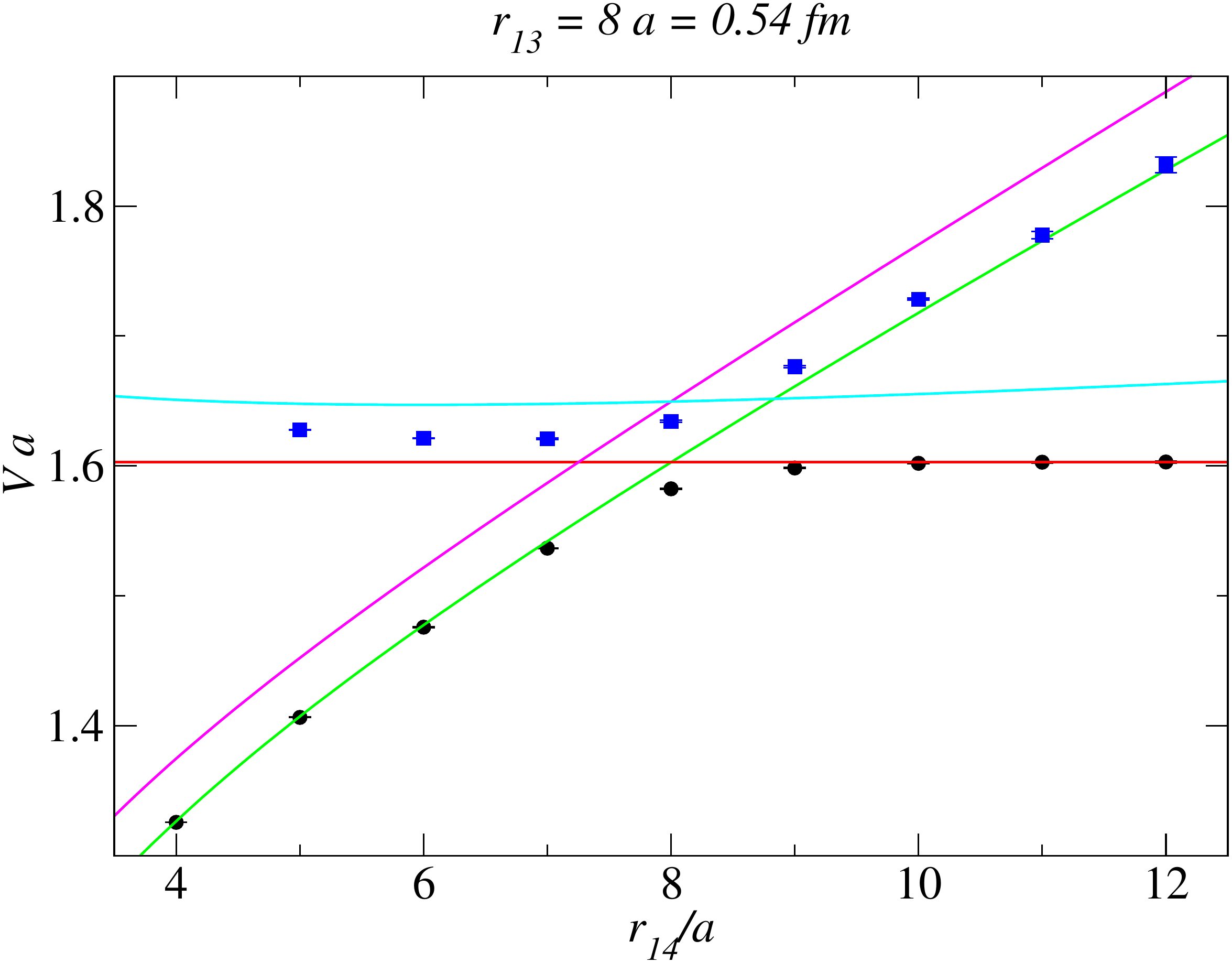}
\caption{ (Colour online.) Ground state and first excited state $Q Q \bar Q \bar Q$,computed with the quenched ensemble, for the anti-parallel alignment.
Results are compared with both two-meson potential and octet-octet
potentials. \label{fig:anti-parallel_q_res}}
\end{figure*}

\begin{figure*}[!t]
\includegraphics[width=0.4\paperwidth]{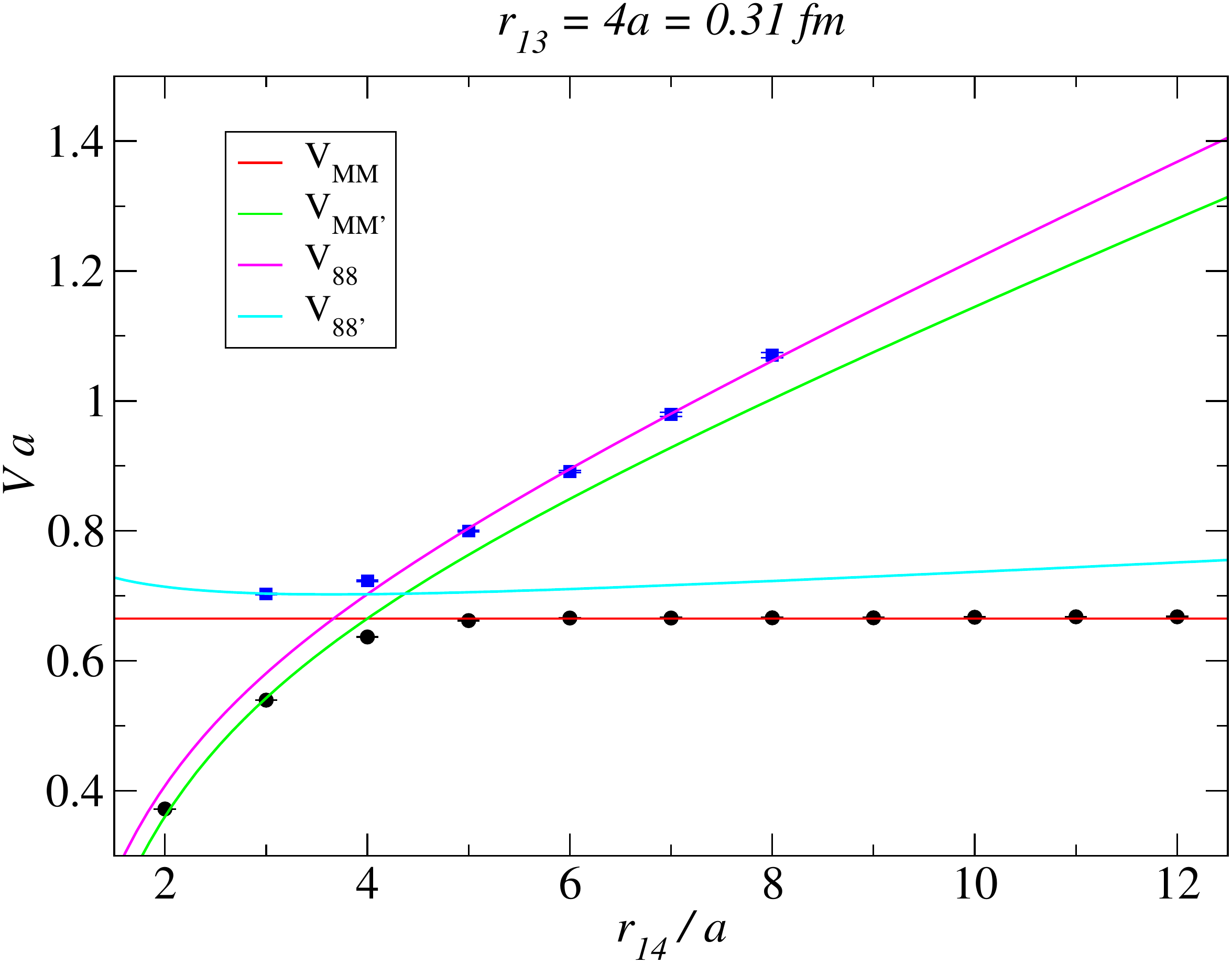}\includegraphics[width=0.4\paperwidth]{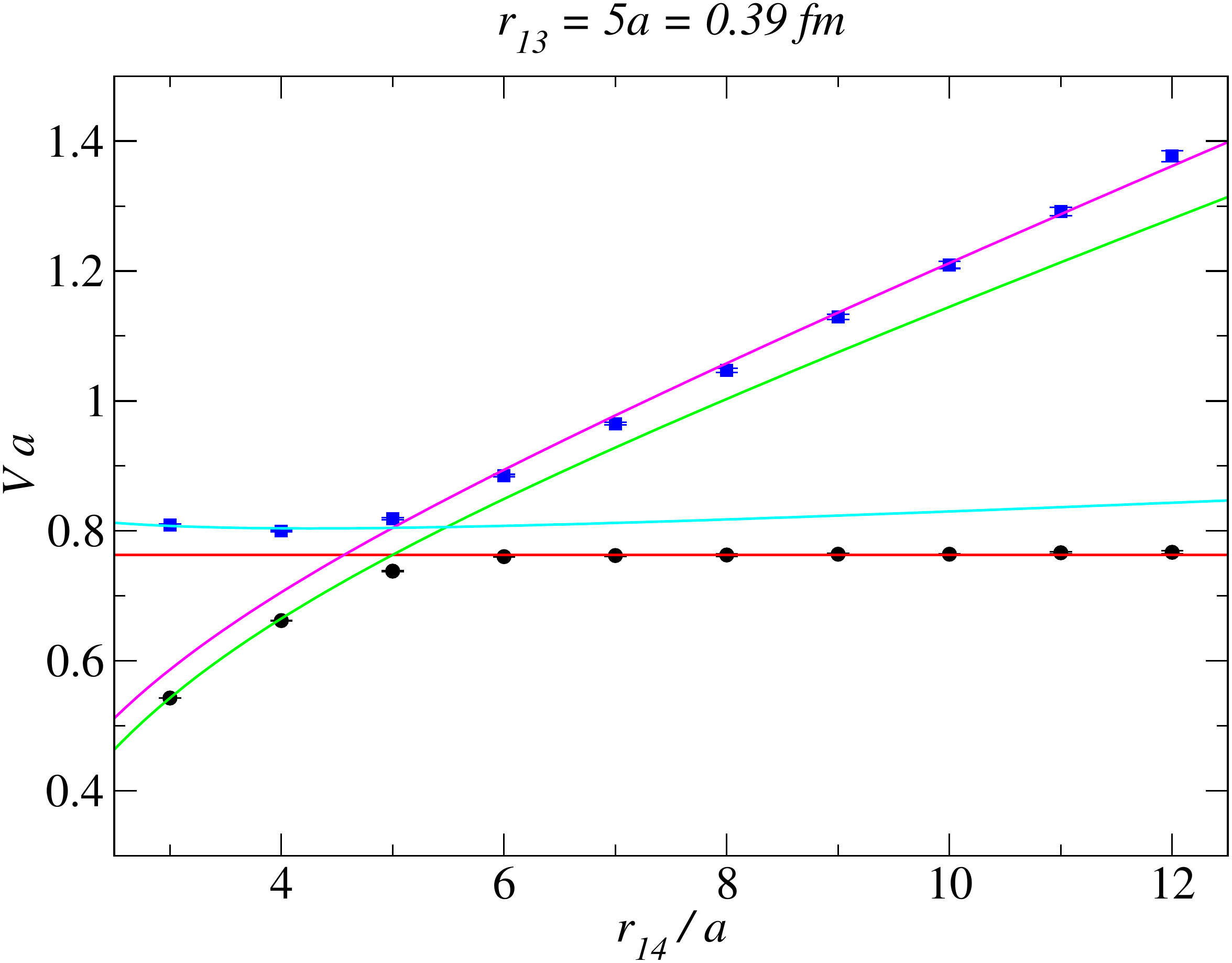}
\includegraphics[width=0.4\paperwidth]{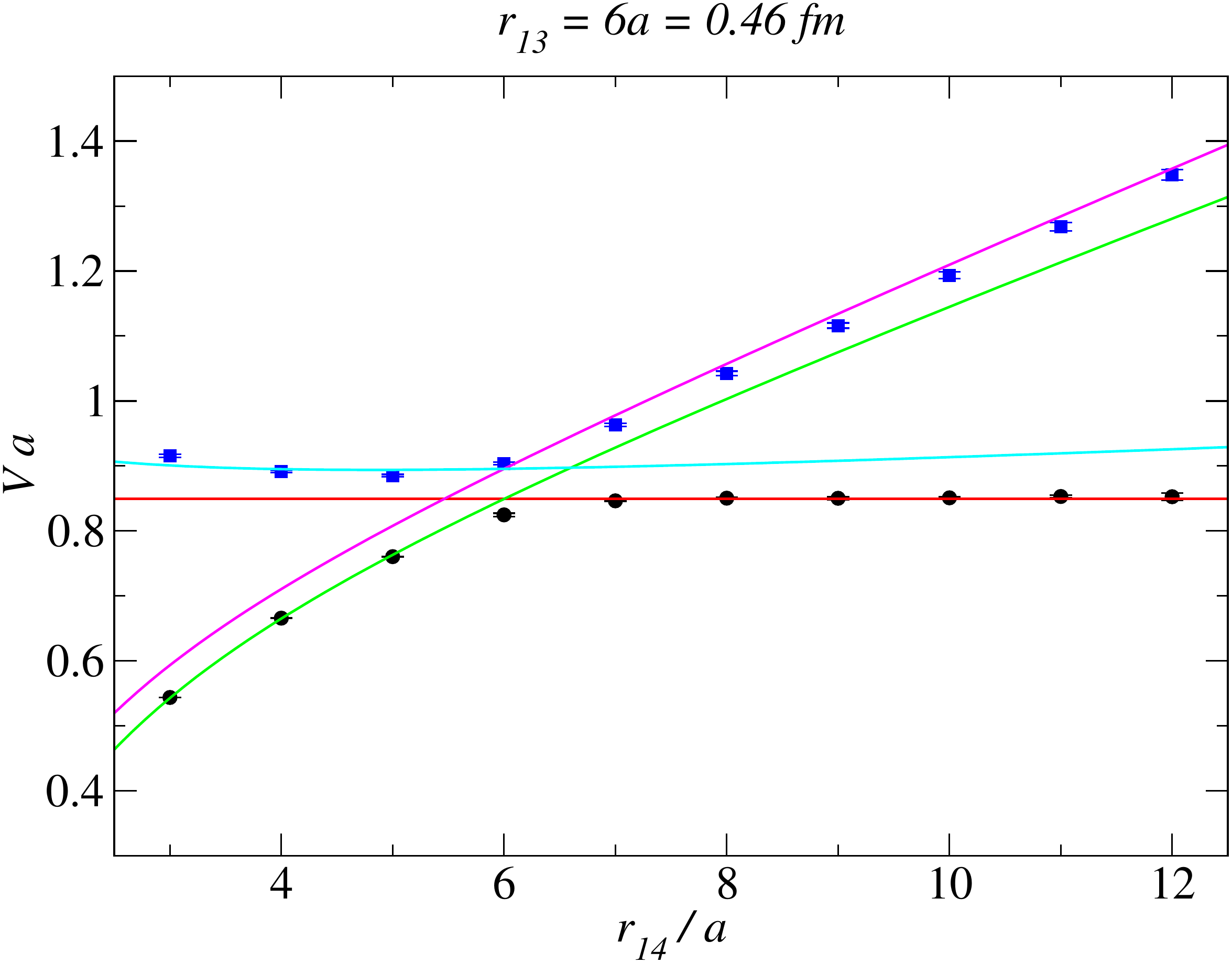}\includegraphics[width=0.4\paperwidth]{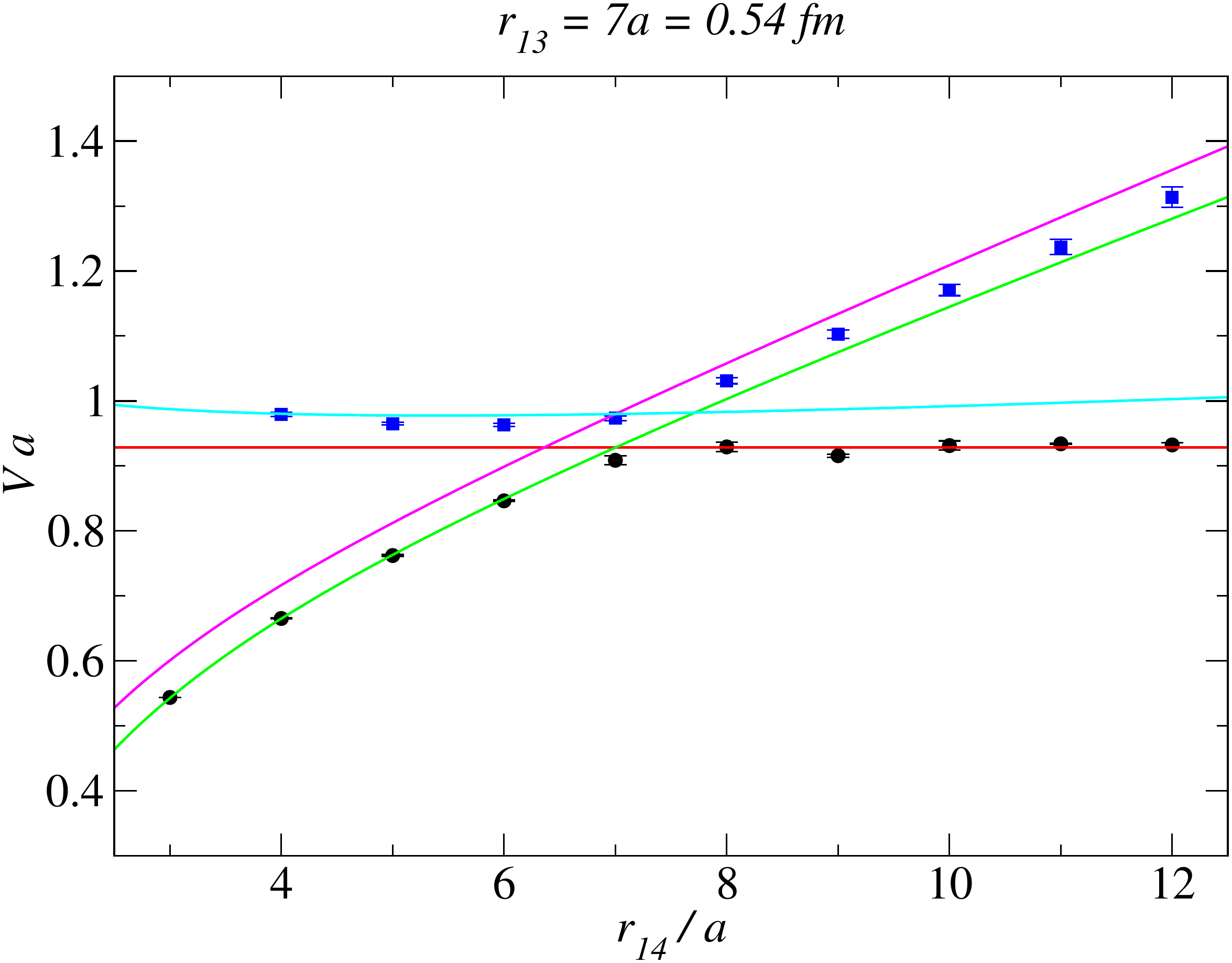}
\caption{ (Colour online.) The same in Fig. \ref{fig:anti-parallel_q_res} but for the dynamical ensemble.\label{fig:anti-parallel_dyn_res}}
\end{figure*}

We start by analyzing the simpler case of the anti-parallel geometry, where the meson-meson systems are expected to have lower energies than the tetraquark system.  Our results are plotted in Figs. \ref{fig:anti-parallel_q_res} and \ref{fig:anti-parallel_dyn_res}. Clearly there are two different trends for $r_{13} < r_{14}$ and for $r_{13}>r_{14}$ and a transition, with mixing, at the point $r_{13}=r_{24}$. Moreover we compare in detail our results with different ansatze.

From the string flip-flop paradigm of Fig. \ref{fig:tripleflipflop} we expect  the ground state of the system to be that of a two meson system 
when the distance between a quark and an anti-quark, i. e. $r_{13}$ or $r_{14}$, is much smaller than the quark-quark distance, i.e. $r_{12}$.
Then, for sufficiently small $r_{13}$ and/or $r_{14}$ the potential of the ground state of the $QQ\bar{Q}\bar{Q}$ should reproduce the string flip-flop potential,
\begin{eqnarray}
V_{0} &\simeq& V_{ff} = \min\left[ V_{MM},V_{MM'} \right]\ ,
\label{eq:VAntishort}
\end{eqnarray} 
where the two different meson -meson potentials are
\begin{eqnarray}
V_{MM} &=& 2 V_{M}(r_{13}) \ ,
\nonumber
\\
\nonumber
V_{MM'} &=& 2  V_{M}(r_{14}) \ ,
\label{eq:2vmm}
\end{eqnarray}
and $V_M$ is the ground state potential of a meson in Eq. (\ref{eq:cornell}).
Previous lattice simulations~\cite{Alexandrou:2004ak,Bornyakov:2005kn,Okiharu:2004ve} confirm that $V_0$ is compatible with
such a result. Deviations from Eq. (\ref{eq:VAntishort}) are expected at intermediate distances together with a smooth transition
from one picture to the other, i.e. from the two meson state with valence content $Q_1\bar{Q}_3$ and
$Q_2 \bar{Q}_4$ to the two meson with valence content $Q_1\bar{Q}_4$ and
$Q_2 \bar{Q}_3$.

On the other hand, for the excited state, we have two possible scenarios. From the string-flip-flop, we would again expect, when the distance between quark and anti-quark  a quark and an anti-quark, i. e. $r_{13}$ or $r_{14}$, is much smaller than the quark-quark distance, i.e. $r_{12}$, the system to be that of the next two meson system,
\begin{equation}
V_{1} 
\stackrel{?}{\simeq}
\max \left[ V_{MM},V_{MM'} \right] \ .
\label{eq:exff}
\end{equation}
However, given that the colour wavefunctions of the two mesonic states are not orthogonal, see Eq. (\ref{eq:mm_nonortho}), and Section II,
possibly the excited state is not another mesonic state and, but instead is an octet state, 
\begin{equation}
V_{1} 
\stackrel{?}{\simeq}
\max \left[ V_{88},V_{88'}) \right] \ ,
\label{eq:exoctet}
\end{equation}
where we estimate the colour octet potential assuming Casimir scaling, i.e. using the decomposition  in Eq. (\ref{eq:VCasimir}) and the values reported on 
Tab.~\ref{tab:NormCas},
\begin{eqnarray}
V_{88} &  = & \frac{1}{2}V_M(\sqrt{r_{13}^{2}+r_{14}^{2}})+\frac{7}{4}V_M(r_{14})-\frac{1}{4}V_M(r_{13})  
\ ,
\nonumber
\\
V_{88'} & = & \frac{1}{2}V_M(\sqrt{r_{13}^{2}+r_{14}^{2}})+\frac{7}{4}V_M(r_{13})-\frac{1}{4}V_M(r_{14}) \ , 
\nonumber
\\
\label{eq:flipoctet}
\end{eqnarray}

\begin{table*}[!t]
\begin{tabular}{|c|c|c|c|c|c|c|c|}
\hline 
$r_{1}$ & $r_\text{min}$ & $r_\text{max}$ & $\chi^{2}/\mbox{d.o.f.}$ & $Ca$ & $\gamma$ & $\sigma a^{2}$ & $\sigma/\sigma_{meson}$\tabularnewline
\hline 
\hline 
6 & 8 & 12 & 1.30 & 1.183(60) & 0.14(30) & 0.0570(30) & 2.28(12)\tabularnewline
\hline 
 & 9 & 12 & 0.11 & 1.252(54) & 0.50(28) & 0.0537(26) & 2.16(10)\tabularnewline
\hline 
7 & 8 & 12 & 1.00 & 1.124(77) & 0.24(37) & 0.0584(39) & 2.35(16)\tabularnewline
\hline 
 & 9 & 12 & 0.03 & 1.218(32) & 0.23(17) & 0.0537(15) & 2.16(6)\tabularnewline
\hline 
\end{tabular}
\caption{Fits of the quenched $Q Q \bar Q \bar Q$ anti-parallel alignment excited state potential to a Cornell
ansatz. }
\label{tab:mesonmesoncornell}
\end{table*}

Thus we have two different simple anzatse to interpret our results. 
The ground state potential $V_0$ and the first excited state potential $V_1$ for the quenched and dynamical ensembles are 
reported in Figs. \ref{fig:anti-parallel_q_res} and \ref{fig:anti-parallel_dyn_res}, respectively, together with $V_{MM}$,
$V_{MM'}$ and the octet potentials $V_{88}$, $V_{88'}$.

As the figures show, the ground state static potential $V_0$ as a function of $r_{14}$ is compatible with two two meson potentials
for small and large values of $r_{14}$. 
Indeed, for all $r_{13}$, at small values of $r_{14}$ the static potential is compatible with $V_{MM}$, while for large $r_{14}$ 
$V_0$ becomes compatible with $V_{MM'}$. 
We show in Table \ref{tab:mesonmesoncornell} good fits with the meson-meson potentials. 
This type of behaviour is well described by the string flip-flop potential $V_{ff}$.

\begin{figure}[!t]
\includegraphics[width=0.9\columnwidth]{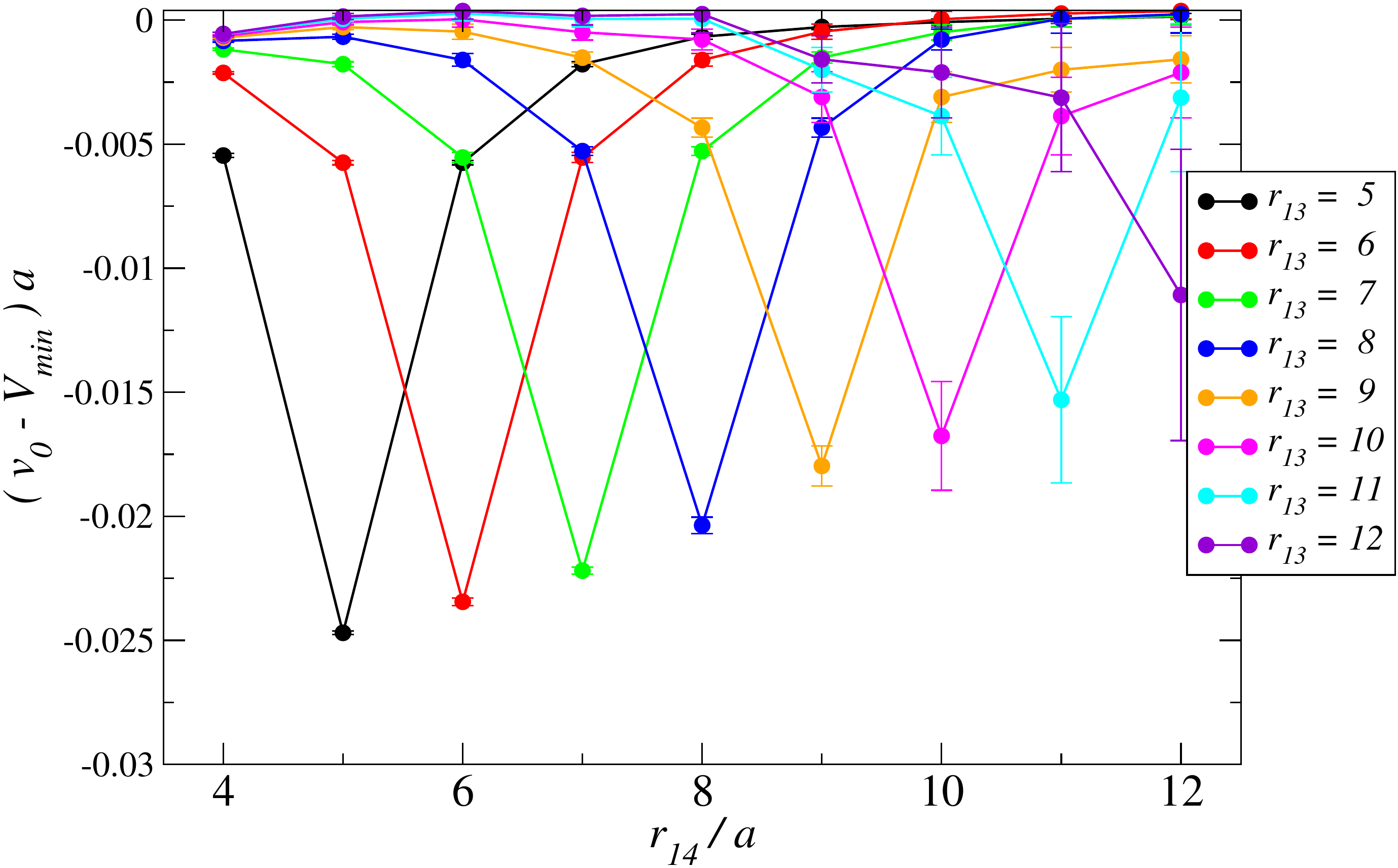}
\caption{ (Colour online.) Difference of the quenched ground state in the anti-parallel $Q Q \bar Q \bar Q$ geometry from $V_{min}=\min(V_{13}+V_{24},V_{14}+V_{23})$. The maximum difference at $r_{13}=r_{24}$ is due to the mixing between the tetraquark strings and the meson-meson strings.}
\label{Fig:dif_quenched}
\end{figure}

\begin{figure}[!t]
\includegraphics[width=0.8\columnwidth]{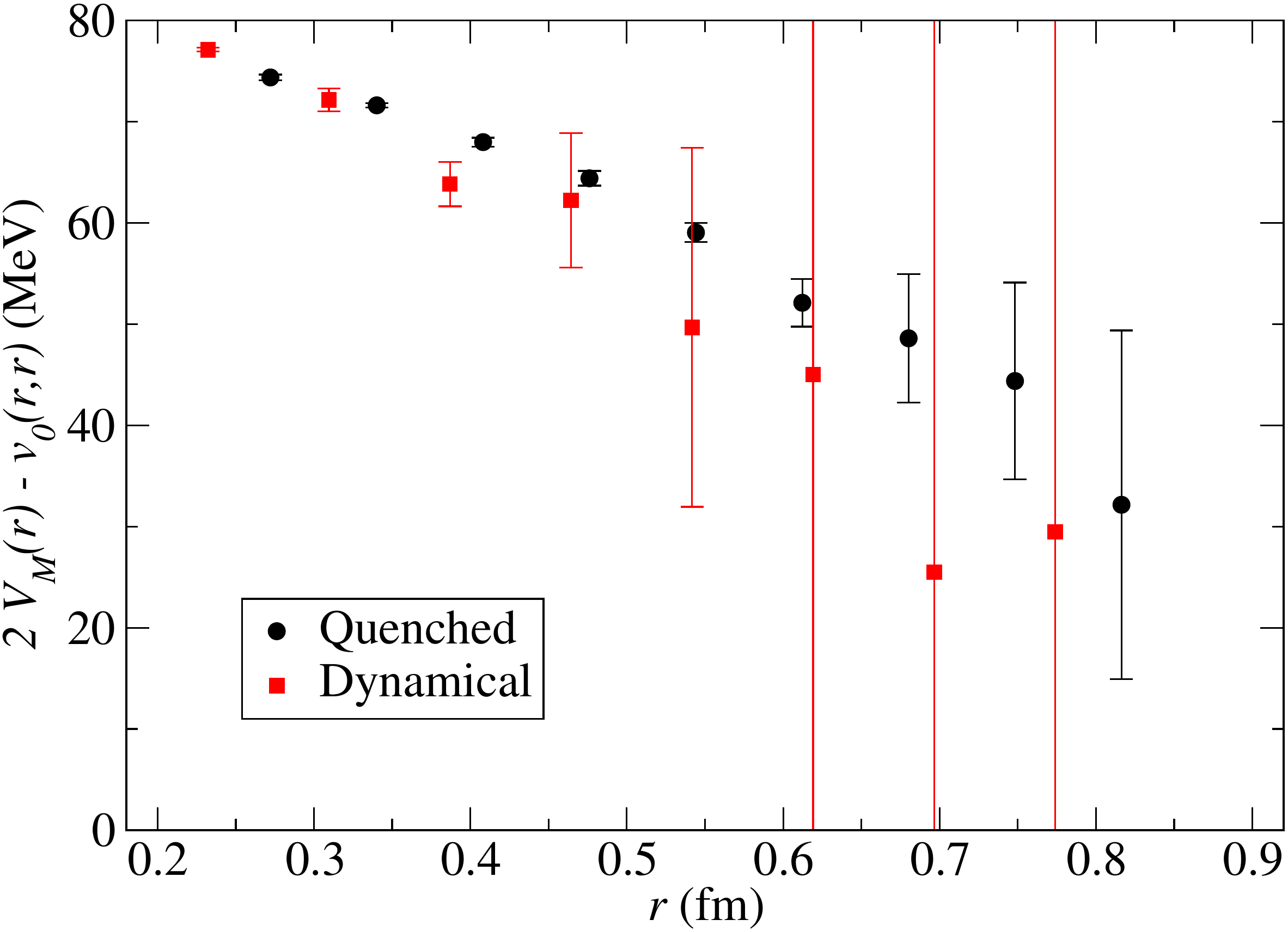}
\caption{ (Colour online.) The difference between the ground state energy $V_{0}$ and $V_{min}=\min(V_{13}+V_{24},V_{14}+V_{23})$
for $r=r_{13}=r_{14}$  in the anti-parallel $Q Q \bar Q \bar Q$ geometry. For a typical distance of $0.5$ fm the difference is of the order of $60$ MeV 
for both data sets. For $r=0.2$ fm the difference increases to $80$ MeV.}
\label{Fig:dif_quenched2}
\end{figure}

In the transition region $r_{13}\sim r_{14}$ where also $V_{MM} \sim V_{MM'}$, deviations of $V_0$ from $V_{MM}$ or $V_{MM'}$ can be seen. 
The difference between the ground state potential and the sum of the two meson potentials 
in physical units is detailed in Fig.~\ref{Fig:dif_quenched}, and in particular the transition point 
$r_{12} = r_{13}$ is analysed in Fig.~\ref{Fig:dif_quenched2}.
The results for the quenched simulation are well described assuming an off diagonal term $\Delta$ in the correlation matrix, leading to the functional form,
\begin{eqnarray}
\label{eq:antip_gs_model_gen}
V_{0}(r_{13},r_{14})&=&\frac{V_{MM}+V_{MM'}}{2}
\\ \nonumber 
&&
-\sqrt{\left(\frac{V_{MM}-V_{MM'}}{2}\right)^{2}+\Delta^{2}} \ , 
\end{eqnarray}
where we may have either,
\begin{equation}
\Delta(r_{1},r_{2})=\frac{\Delta_{0}e^{-\lambda(r_{1}+r_{2})}}{1+c(r_{1}-r_{2})^{2}}\label{eq:antip_gs_model_a}
\end{equation}
or,
\begin{equation}
\Delta(r_{1},r_{2})=\frac{\Delta_{0}}{1+c(r_{1}-r_{2})^{2}+d(r_{1}+r_{2})^{2}} \ . \label{eq:antip_gs_model_b} 
\end{equation}
Eq. (\ref{eq:antip_gs_model_gen}) interpolates between the two potentials in flip-flop picture of a meson-meson.

\begin{table}[!t]
\begin{tabular}{|c|c|c|c|c|c|}
\hline 
$r_\text{min}$ & $r_\text{max}$ & $\chi^{2}/\mbox{d.o.f.}$ & $\Delta_{0}a$ & $ca^{2}$ & $\lambda a$\tabularnewline
\hline 
\hline 
5 & 11 & 1.02 & 0.0335(16) & 0.1547(89) & 0.0309(37)\tabularnewline
\hline 
5 & 12 & 1.15 & 0.0335(16) & 0.1548(90) & 0.0309(38)\tabularnewline
\hline 
6 & 11 & 0.75 & 0.0362(49) & 0.1644(127) & 0.0363(86)\tabularnewline
\hline 
6 & 12 & 0.79 & 0.0363(49) & 0.1643(127) & 0.0364(86)\tabularnewline
\hline 
\end{tabular}
\caption{Fits of the quenched data anti-parallel alignment ground state potential
to the transition ansatz of Eqs. (\ref{eq:antip_gs_model_gen}) and (\ref{eq:antip_gs_model_a}).} \label{Tab.Fit1}
\end{table}

The fits for the functional forms in Eqs. (\ref{eq:antip_gs_model_a}) and (\ref{eq:antip_gs_model_b}) are reported in Tables \ref{Tab.Fit1} and \ref{Tab.Fit2}. 
In order to quantify the deviation from the two limits where the system behaves as a two meson system, 
we refer that the fits give a $\Delta(0.5\,\mbox{fm},0.5\,\mbox{fm})\simeq60\,\mbox{MeV}$, a number to be compared
with typical values for the meson potential which are of the order of GeV (see Fig.~\ref{fig:meson_staticpot}).
This results shows that the corrections due to $\Delta$ to the flip-flop picture are small when the quarks and anti-quarks are in an
anti-parallel geometry.

\begin{table}[!t]
\begin{tabular}{|c|c|c|c|c|c|}
\hline 
$r_\text{min}$ & $r_\text{max}$ & $\chi^{2}/\mbox{d.o.f.}$ & $\Delta_{0}a$ & $ca^{2}$ & $da^{2}$\tabularnewline
\hline 
\hline 
5 & 11 & 1.07 & 0.0285(10) & 0.1934(144) & 0.0016(3)\tabularnewline
\hline 
5 & 12 & 1.19 & 0.0285(10) & 0.1938(147) & 0.0016(3)\tabularnewline
\hline 
6 & 11 & 0.77 & 0.0294(33) & 0.2275(330) & 0.0018(7)\tabularnewline
\hline 
6 & 12 & 0.80 & 0.0295(33) & 0.2278(332) & 0.0018(7)\tabularnewline
\hline 
\end{tabular}
\caption{Fits of the quenched data anti-parallel alignment ground state potential
to the transition ansatz of Eqs. (\ref{eq:antip_gs_model_gen}) and (\ref{eq:antip_gs_model_b}).}\label{Tab.Fit2}
\end{table}

\begin{figure}[!t]
\includegraphics[width=0.8\columnwidth]{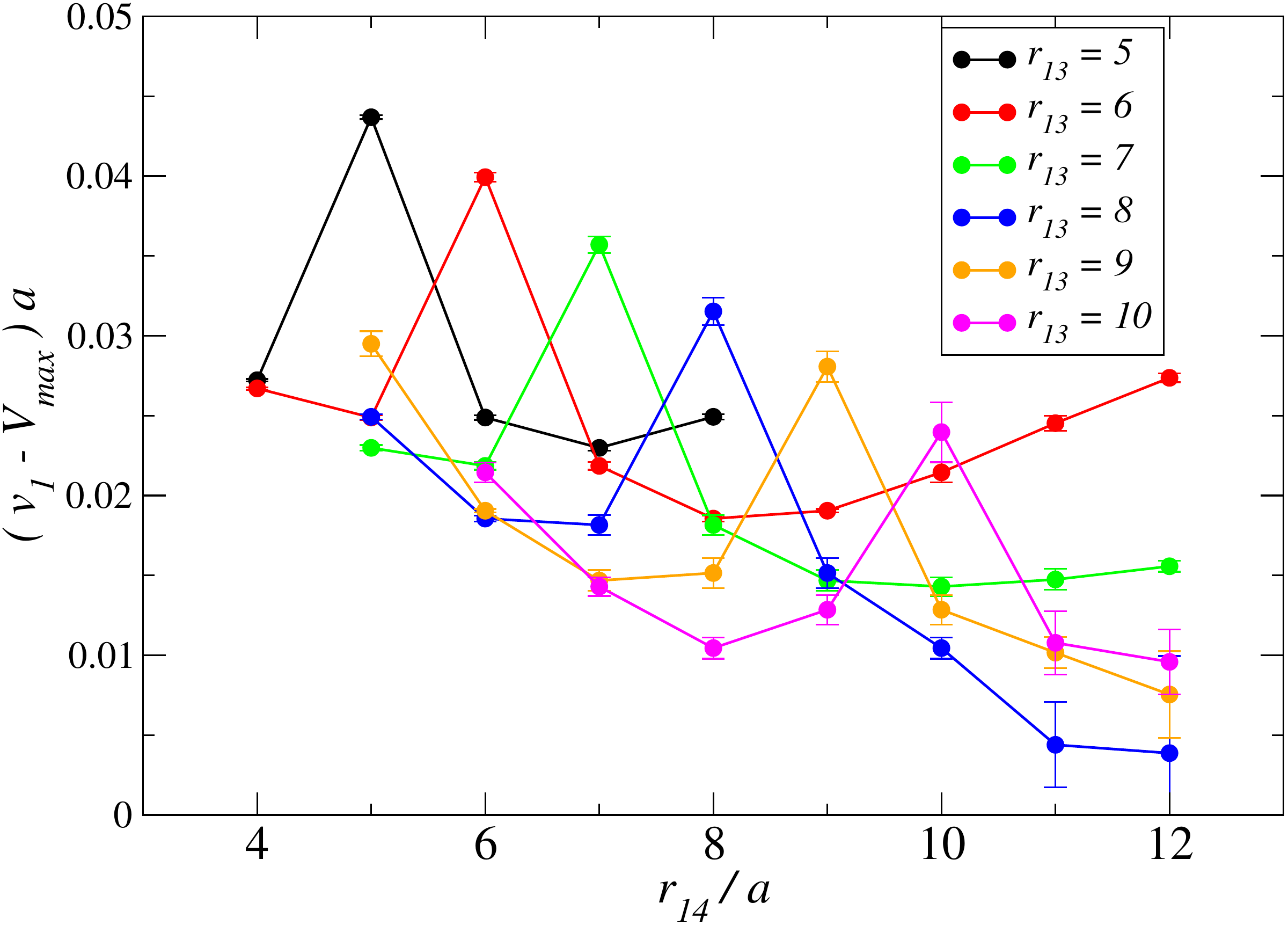}
\caption{ (Colour online.) Difference of the quenched first excited state in the anti-parallel $Q Q \bar Q \bar Q$ geometry  from $V_{max}=\max(V_{13}+V_{24},V_{14}+V_{23})$.
The maximum difference at $r_{13}=r_{24}$ is due to the mixing between the two different meson-meson strings.
Note there is also a significant difference at other distances for $r_{12}=6a$ and $r_{12}=7a$. }
\label{fig:quenchedmixingregion}
\end{figure}

The full QCD simulation shows similar results to the quenched QCD simulation. 
However, the results for $V_0$ for the full QCD configurations are not described by the same
type of functional form  given in Eq. (\ref{eq:antip_gs_model_gen})
which reproduces the flip-flop potential at large distances. We found no window where the fits are stable and, therefore, conclude that the dynamical $V_0$ is not reproduced by 
Eq. (\ref{eq:antip_gs_model_gen}) with the deviations parametrised by either Eq. (\ref{eq:antip_gs_model_a}) or Eq. (\ref{eq:antip_gs_model_b}).

In what concerns the excited state potential $V_1$ there are clearly two different regimes for $r_{13}$ very different from $r_{14}$, but we are not able to find an
analytic form compatible with the lattice data, neither for the quenched simulations nor for the full QCD simulations. 
In both  Figs. \ref{fig:anti-parallel_q_res} 
and \ref{fig:anti-parallel_dyn_res}, it is clear the static potential $V_1$ lies between the functional forms of Eq. (\ref{eq:exff}) and Eq. (\ref{eq:exoctet}).
There are subtle differences between Fig. \ref{fig:anti-parallel_q_res} 
and \ref{fig:anti-parallel_dyn_res}. In general, the full QCD case is closer to the octet expression of Eq. (\ref{eq:exoctet}) than the quenched QCD case.

A fortiori, we are not able as well to find a good ansatze to fit $V_1$ in the transition region. For a detailed view of the differences for the quenched simulation in this region, see Fig. \ref{fig:quenchedmixingregion}.

This observed behaviour for $V_1$ can be understood in terms of adjoint strings.
When the quark-anti-quark inside the octets are close to each other, they can be seen externally
as a gluon. Therefore, we have a single adjoint string with a tension of $\sigma_{A}=\frac{9}{4}\sigma$. 
On the other hand, when the quark and the anti-quark are pulled apart, the adjoint string tends to split into 
two fundamental strings, with a total string tension of $2\sigma$. The splitting of the adjoint string, gives a 
repulsive interaction between the quark-anti-quark pairs that form octets in the excited state. This is qualitatively
consistent with the behaviour predicted by Casimir scaling, where the potential for a quark and an anti-quark in an octet corresponds
to a repulsive interaction.

\subsubsection{Mixing angle}

\begin{figure}[!t]
\includegraphics[width=0.85\columnwidth]{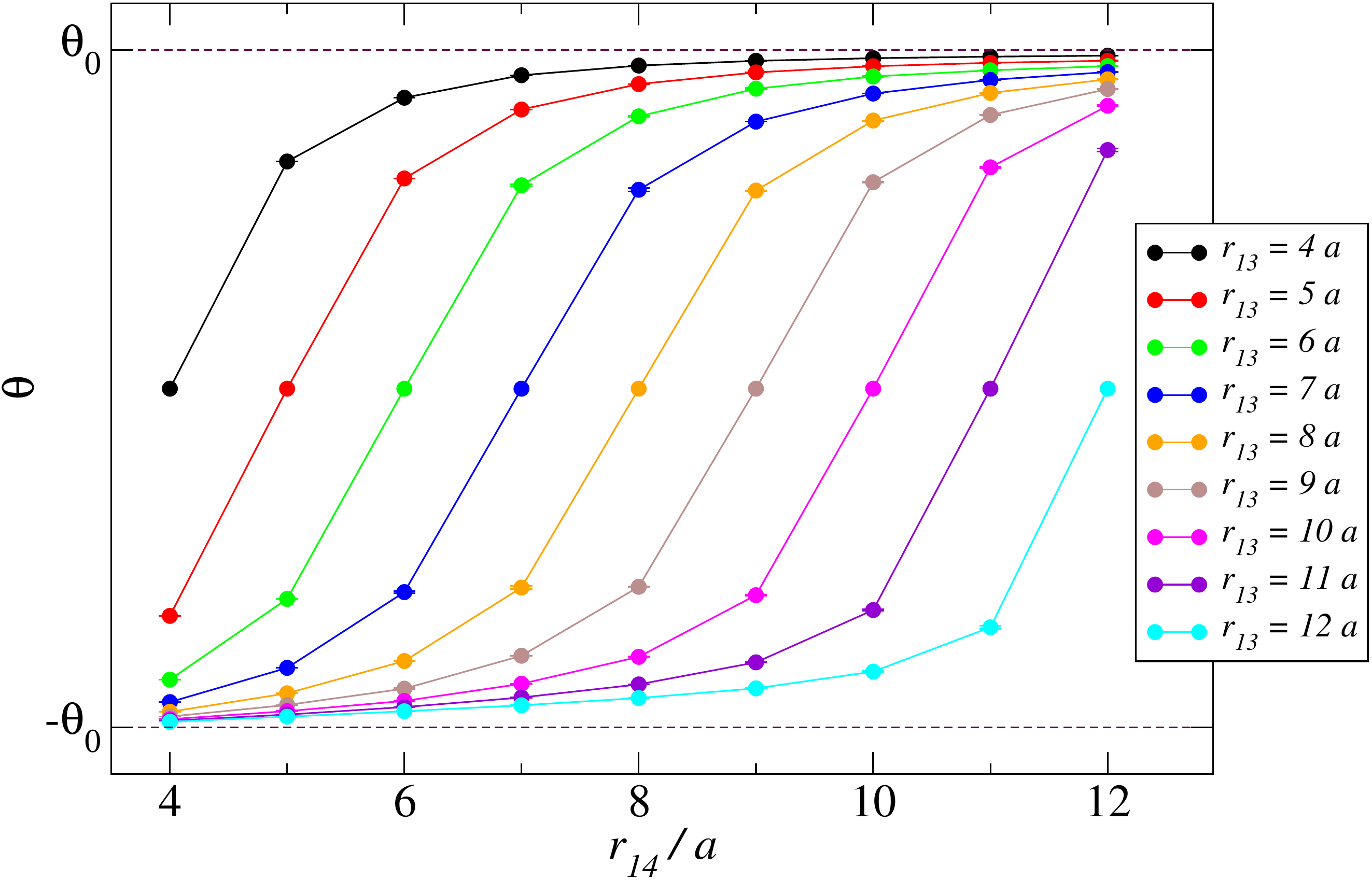}
\caption{ (Colour online.) Mixing angle $\theta$ for different values of $r_{13}$ and $r_{14}$
for quenched data, anti-parallel $Q Q \bar Q \bar Q$ geometry. 
\label{fig:mixing_angle_q}}
\end{figure}

\begin{figure}[!t]
\includegraphics[width=0.85\columnwidth]{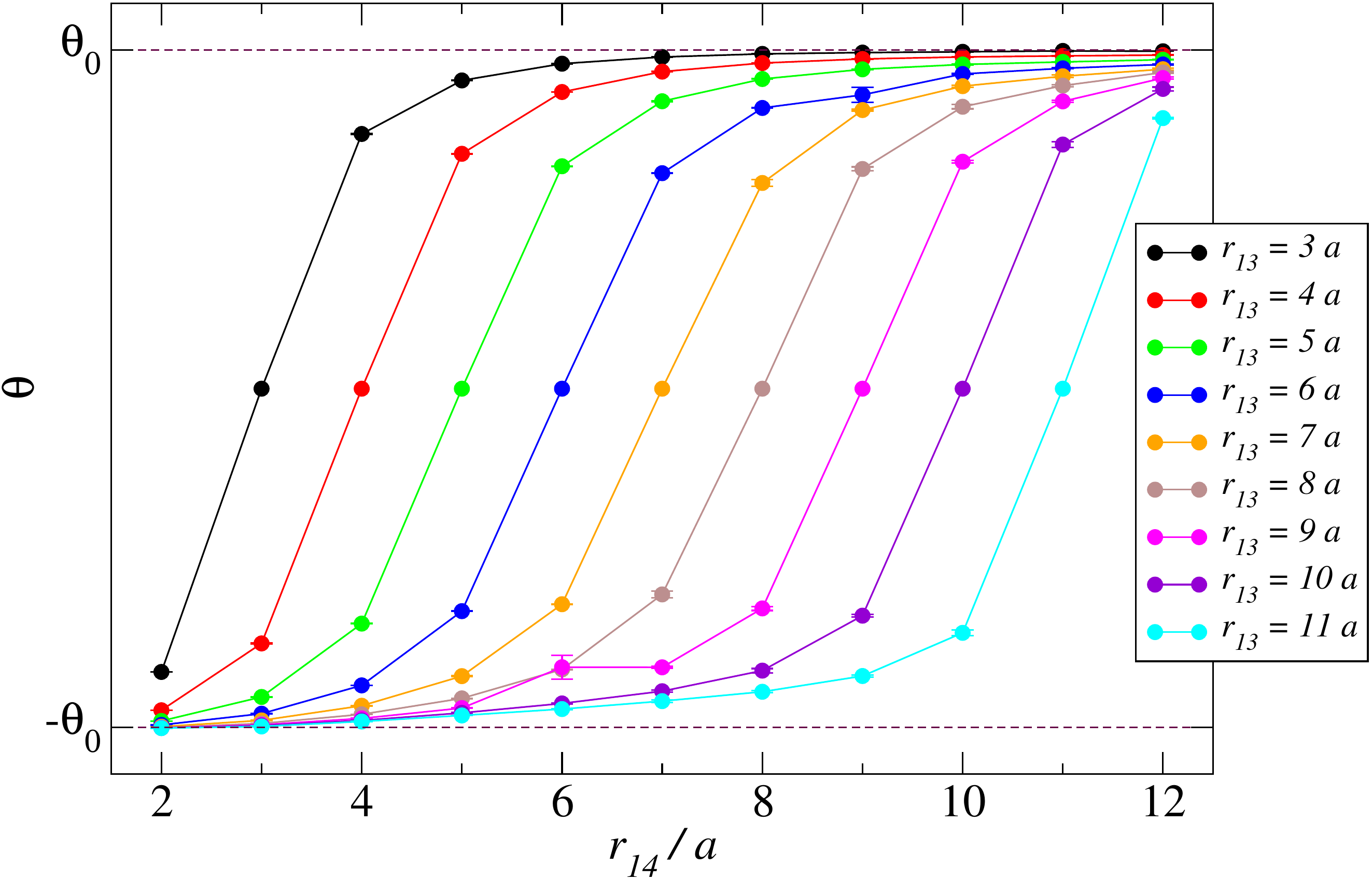}
\caption{ (Colour online.) Mixing angle $\theta$ for different values of $r_{13}$ and $r_{14}$ for
the full QCD simulation, anti-parallel $Q Q \bar Q \bar Q$ geometry.
\label{fig:mixing_angle_dyn}}
\end{figure}

For the anti-parallel geometry and for the ground state potential the lattice results show that the tetraquark is essentially a two meson state. Therefore,
one can write the most general ket describing the ground state $|u_{0}\rangle$ of a $QQ\bar{Q}\bar{Q}$ system as a linear combination of the available
colourless states
\begin{eqnarray}
|u_{0}\rangle & = & \cos\theta ~ |\mathbf{6}_{12}\bar{\mathbf{6}}_{34}\rangle+
              \sin\theta  ~ |\bar{\mathbf{3}}_{12}\mathbf{3}_{34} \rangle 
\nonumber \\
               & = & \sqrt{\frac{3}{4}} \bigg\{   \left( \frac{\cos\theta}{\sqrt{2}} + \sin\theta \right)  |\mathbf{1}_{13} {\mathbf{1}}_{24}\rangle    
\nonumber \\
               &&
                            \qquad + \left( \frac{\cos\theta}{\sqrt{2}} - \sin\theta \right)  |\mathbf{1}_{14} {\mathbf{1}}_{23}\rangle \bigg\} \ .
\end{eqnarray}
For a pure two-meson state, the mixing angle is either $\theta = \theta_{0}$,
for $|\mathbf{1}_{13}\mathbf{1}_{24}\rangle$, or $\theta = -\theta_{0}$, for  $|\mathbf{1}_{14}\mathbf{1}_{23}\rangle$,
with $\theta_{0} = \tan^{-1} ( 1 / \sqrt{2} )$. For the general case, the angle $\theta$ can be estimated using the generalized
eigenvectors obtained solving Eq. (\ref{eq:gen_eigen_wilson}) with the following operators,
\begin{eqnarray}
\mathcal{O}_{S}  &=&  \sqrt{\frac{3}{8}}\Big(\mathcal{O}_{13,24}+\mathcal{O}_{14,23}\Big)  \ ,
\nonumber
\\
\mathcal{O}_{A}  &=&  \sqrt{\frac{3}{4}}\Big(\mathcal{O}_{13,24}-\mathcal{O}_{14,23}\Big) \ .
\end{eqnarray}
The results for $\theta$ for the quenched simulation can be seen in Fig. \ref{fig:mixing_angle_q}.
From the lattice data one can estimate a typical length, or broadness, associated to the transition between the two two-meson states. In the region when $|r_{13}-r_{14}| \lesssim d_\text{trans}$, the transition occurs and the groundstate is a mixing of the $MM$ and $MM'$ states.  We estimate the typical transition length from,
\begin{equation}
d^{-1}_\text{trans} \sim \frac{d\theta(r_{13},r_{14})}{dr_{14}}\Big|_{r_{14}=r_{13}}  \ .
\end{equation}
For the quenched data, see Fig. \ref{fig:mixing_angle_q}, the derivative stays within $0.36/a$ and $0.42/a$
and, therefore, $d_{trans} \sim 0.16 - 0.19$ fm. 
For the dynamical simulation, see Fig. \ref{fig:mixing_angle_dyn}, the typical transition length is essentially the same and we find
$d_{trans} \sim 0.16 -  0.20$ fm.


The lattice data for the mixing angle gives a vanishing angle for $r_{13}=r_{14}$.
This means that the ground state for the anti-parallel alignment is given only by $|\mathbf{6}_{12}\bar{\mathbf{6}}_{34}\rangle$
and has no $|\bar{\mathbf{3}}_{12}\mathbf{3}_{34}\rangle$ component.

The results reported in Figs. \ref{fig:mixing_angle_q} and \ref{fig:mixing_angle_dyn} show that, in general, a $QQ\bar{Q}\bar{Q}$ system
is in a mixture of two possible colour meson states and it approaches meson states as the distance between the pairs of quark-anti-quark is much
smaller than the distance between quarks or anti-quarks.

\subsection{The parallel alignment}

For this particular geometry, the static potential was investigated with lattice methods in~\cite{Alexandrou:2004ak,Okiharu:2004ve}. 
For the ground state and in the limit where $r_{12}\ll r_{13}$, the authors found that the lattice data is compatible with the double-Y (or butterfly) potential,
%
\begin{eqnarray}
V_{YY} &=& 2K - \gamma\Biggr(
    \frac{1}{2r_{12}} + \frac{1}{2r_{34}}+\frac{1}{4r_{13}}
    \nonumber \\
    &&+\frac{1}{4r_{24}}+\frac{1}{4r_{14}}+\frac{1}{4r_{23}}\Biggl)+\sigma L_{min} \ ,
 \label{eq:VAligPprevious}
\end{eqnarray}
where $\gamma$ and $K$ are the estimates of the static meson potential and $\sigma$ is the fundamental string tension.
For the geometry described on the right hand side of Fig. \ref{Fig:geometrias} and for $r_{13} > r_{12} / \sqrt{3}$ the butterfly potential simplifies into,
\begin{eqnarray}
V_{YY} &=&2K-\gamma\Biggl(\frac{1}{r_{12}}+\frac{1}{2r_{13}}
\nonumber \\
&& +\frac{1}{2\sqrt{r_{12}^{2}+r_{13}^{2}}}\Biggr)+\sigma(\sqrt{3}r_{12}+r_{13}) \ ,
 \label{eq:VAligsimplified}
\end{eqnarray}

\begin{figure*}[!t]
\includegraphics[width=0.9\columnwidth]{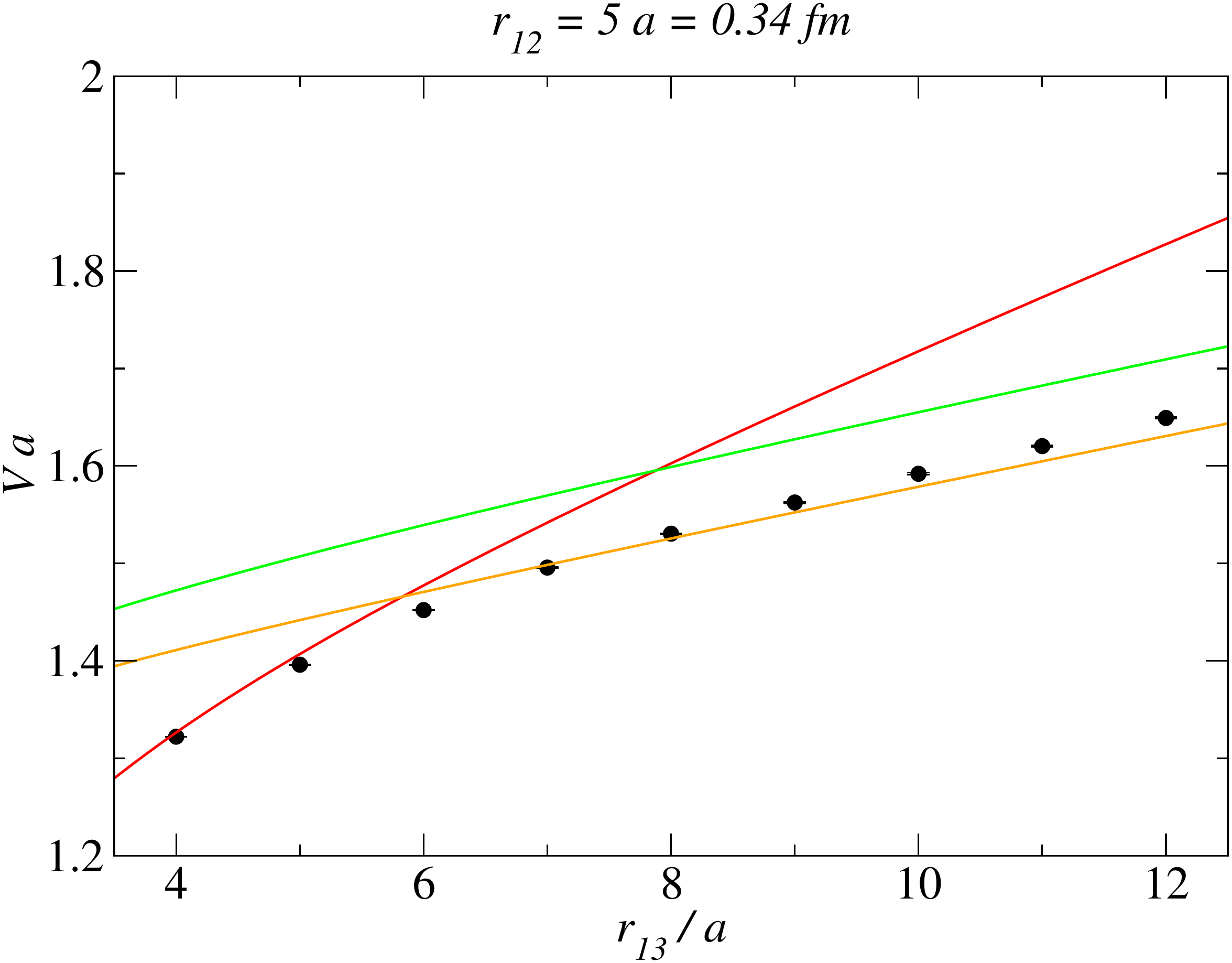}
\qquad
\includegraphics[width=0.9\columnwidth]{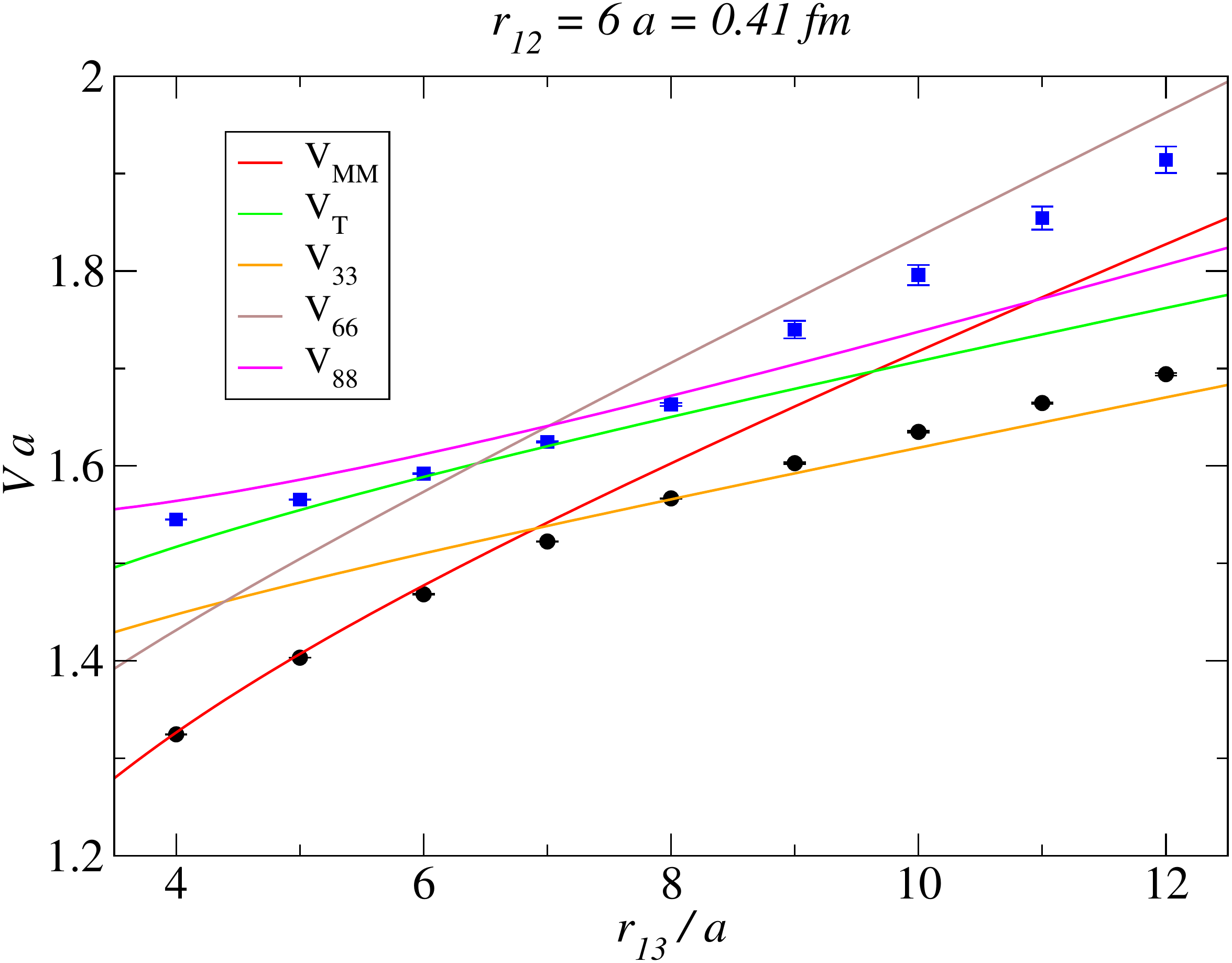}
\\ \vspace{10pt}
\includegraphics[width=0.9\columnwidth]{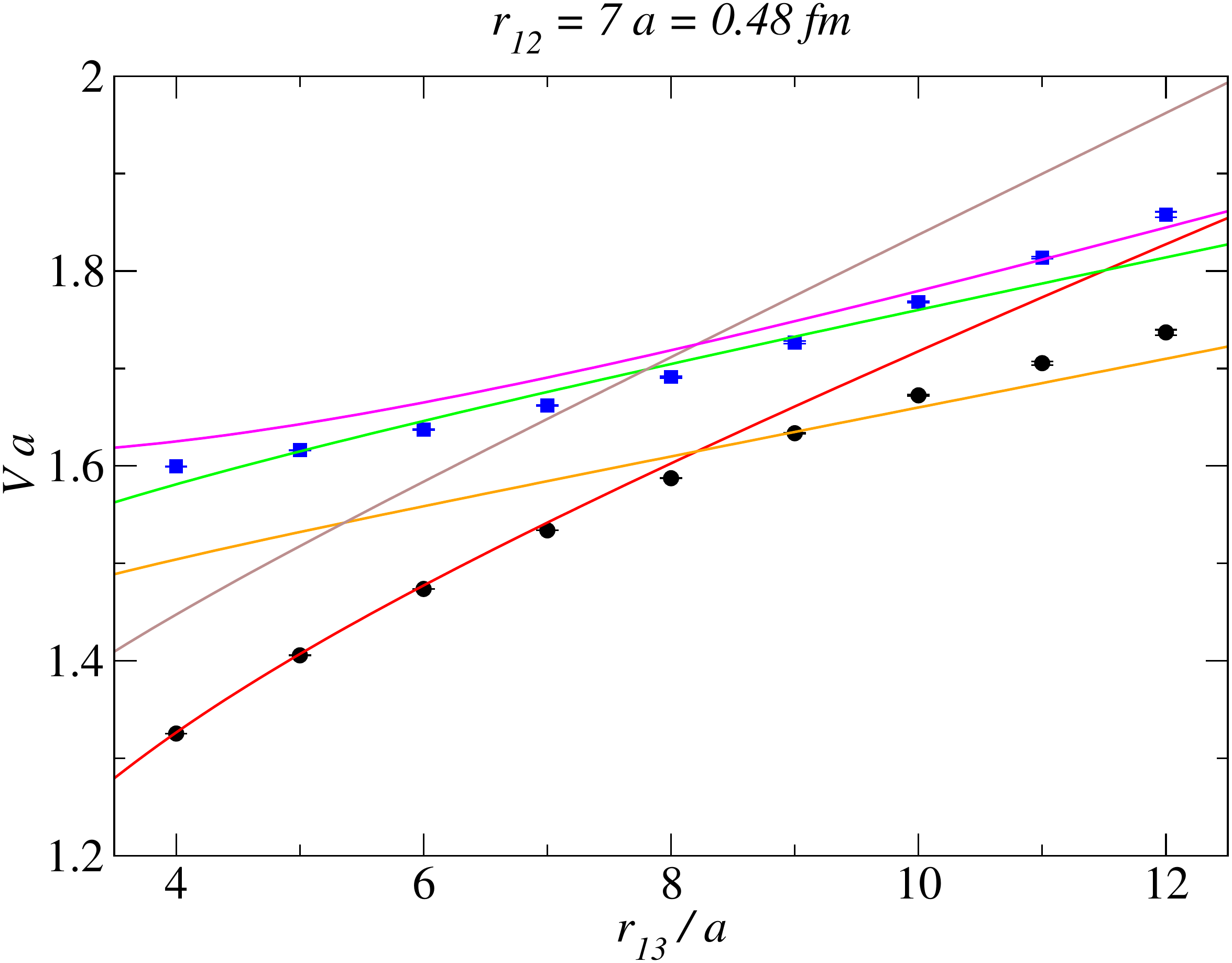}
\qquad
\includegraphics[width=0.9\columnwidth]{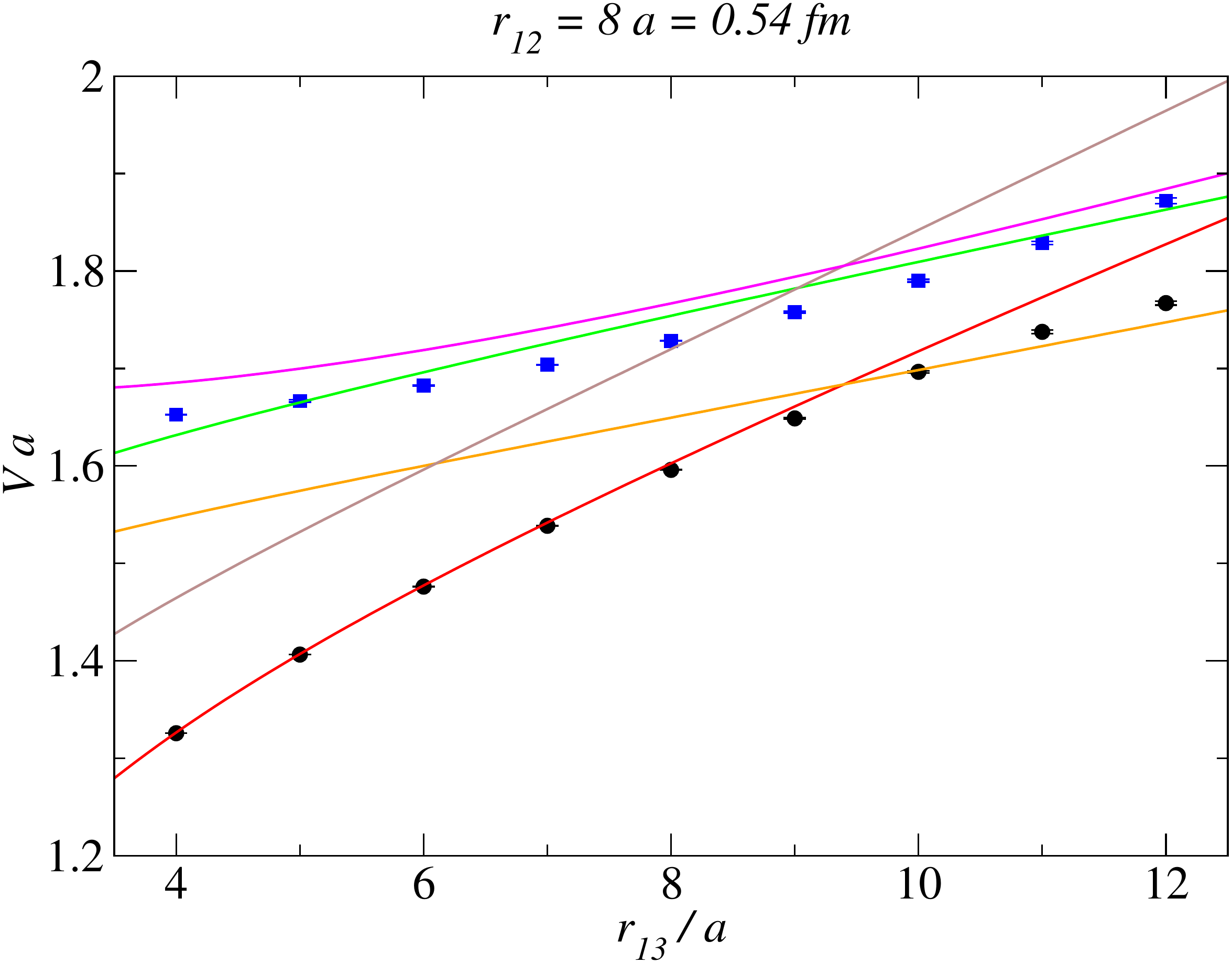}
\caption{ (Colour online.) Ground state (black dots) and first excited state (blue dots, where possible) quenched lattice estimation of the static potentials in the $QQ \bar Q \bar Q$ parallel geometry.
The figures also include fits with various potential models, namely the two-meson potential, the double Y potential and the Casimir 
anti-triplet-triplet, sextet-anti-sextet and octet-octet potentials.
\label{fig:tetra_quenched_res}}
\end{figure*}

\begin{figure*}[!t]
\includegraphics[width=0.9\columnwidth]{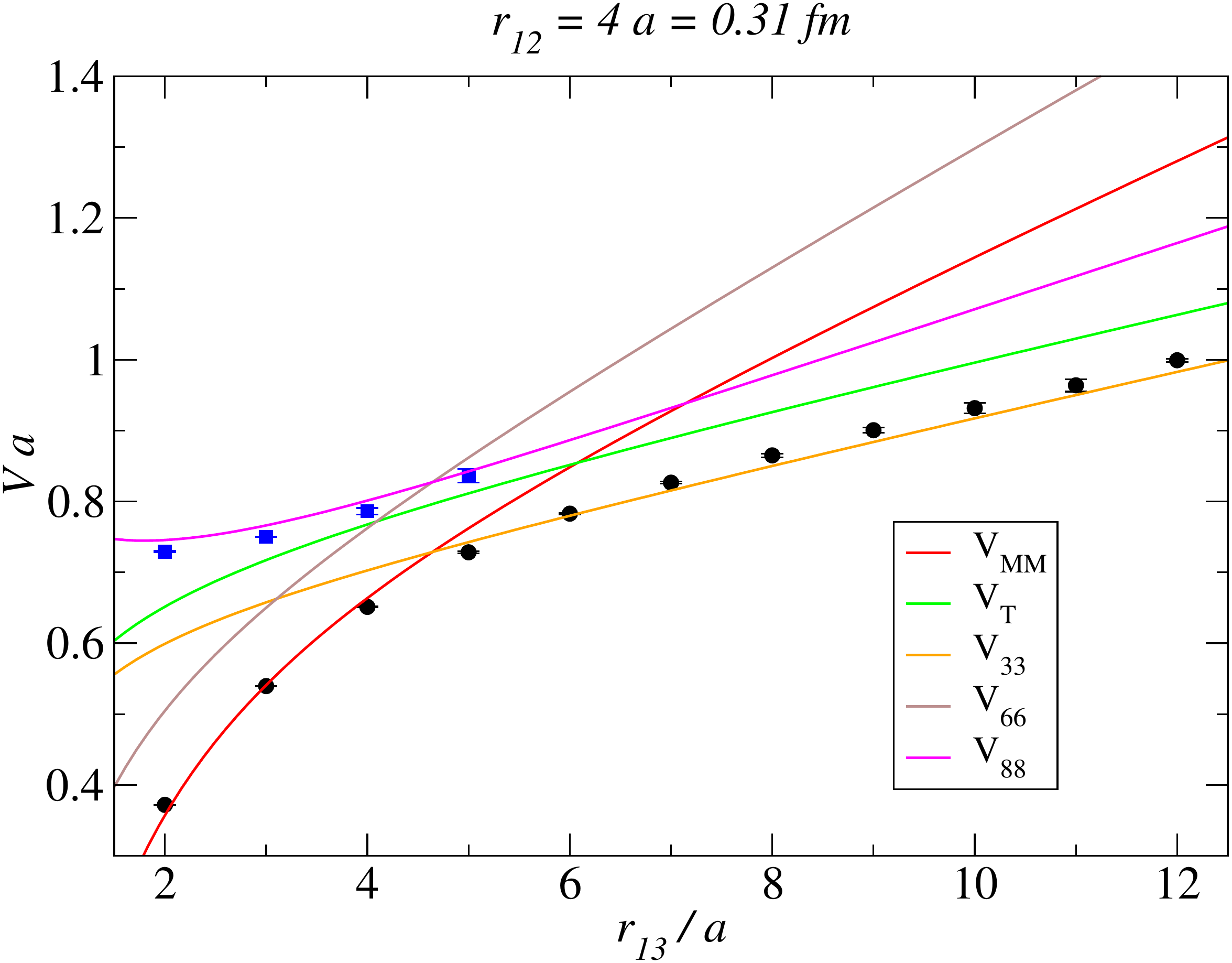}
\qquad
\includegraphics[width=0.9\columnwidth]{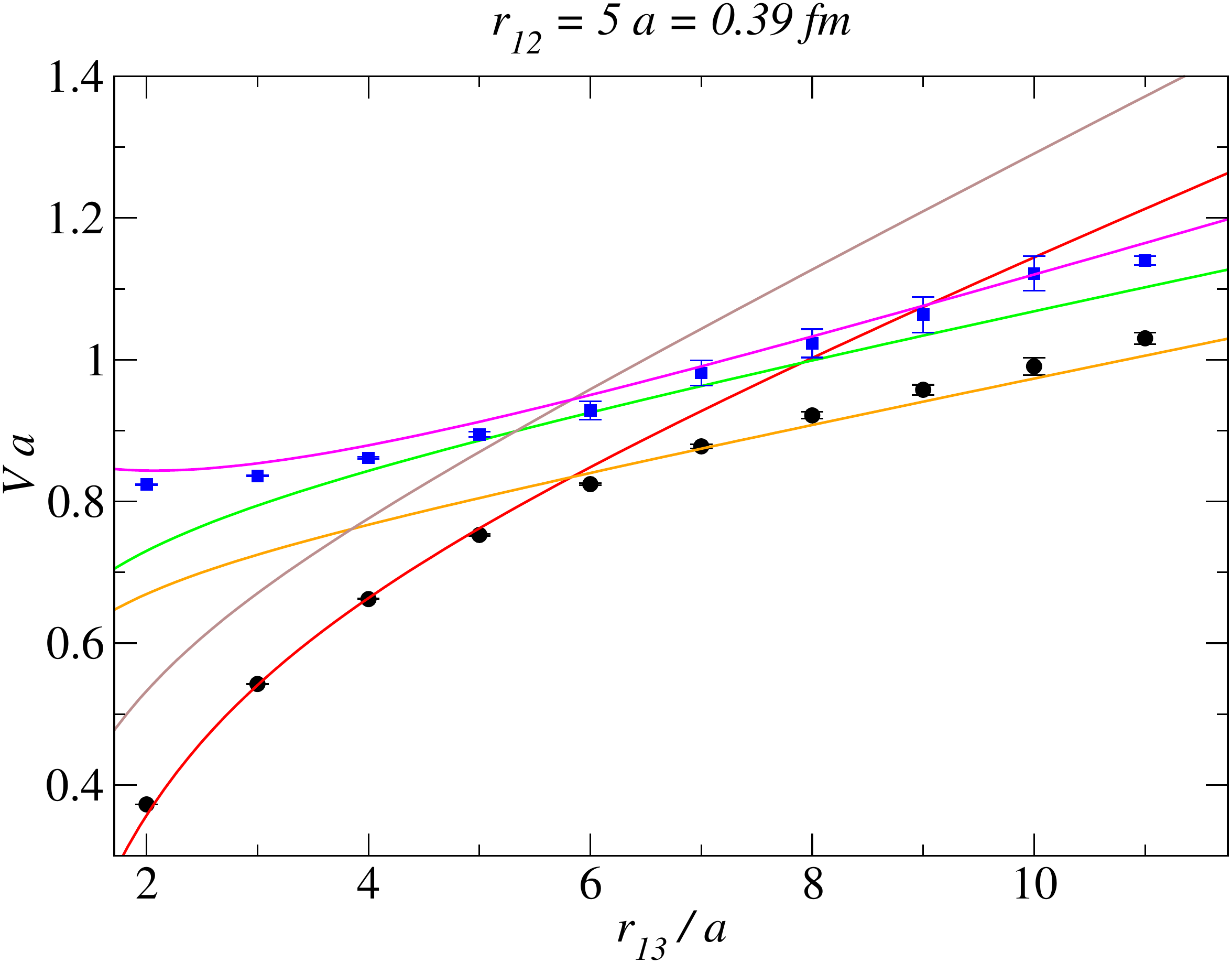}
\\ \vspace{10pt}
\includegraphics[width=0.9\columnwidth]{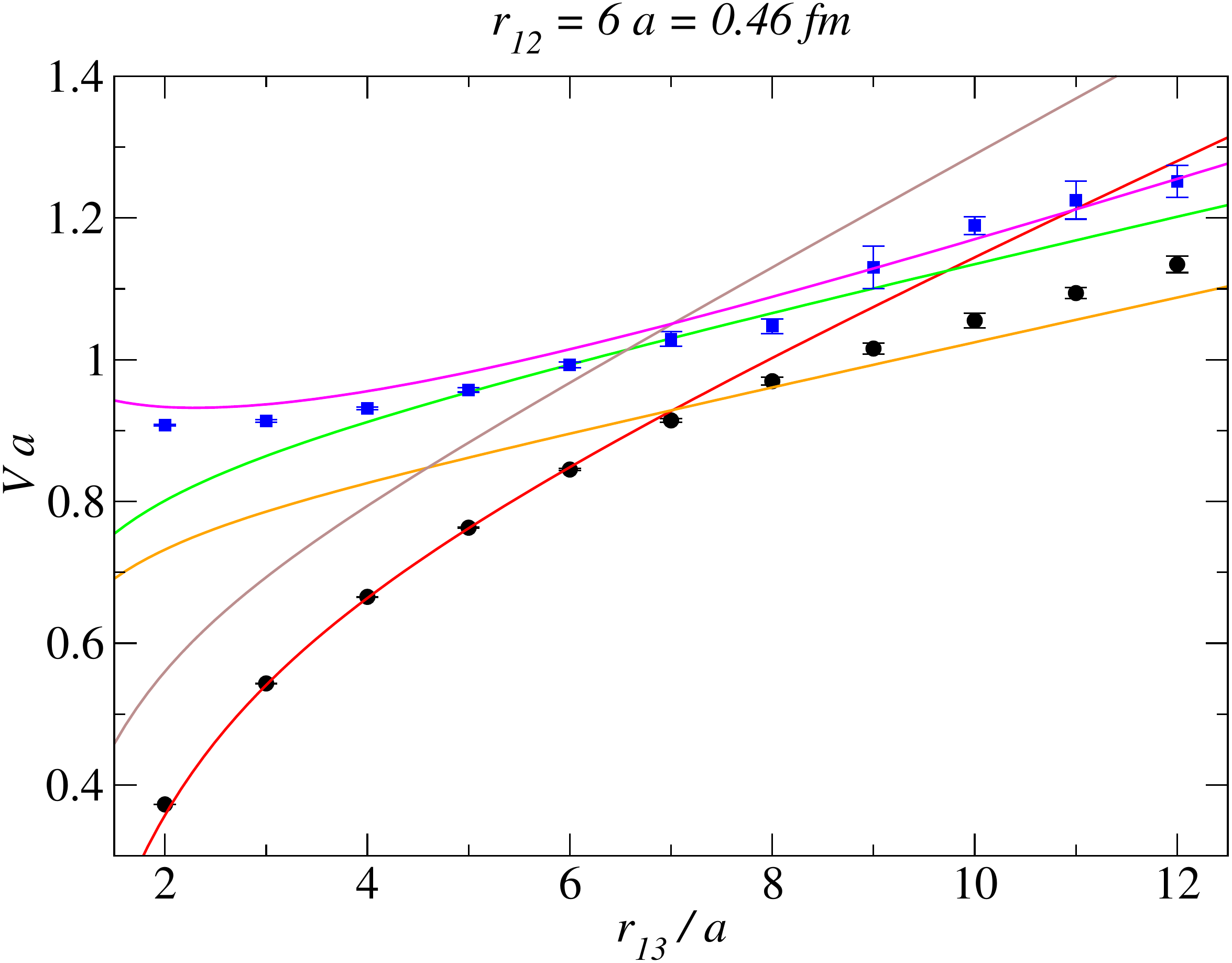}
\qquad
\includegraphics[width=0.9\columnwidth]{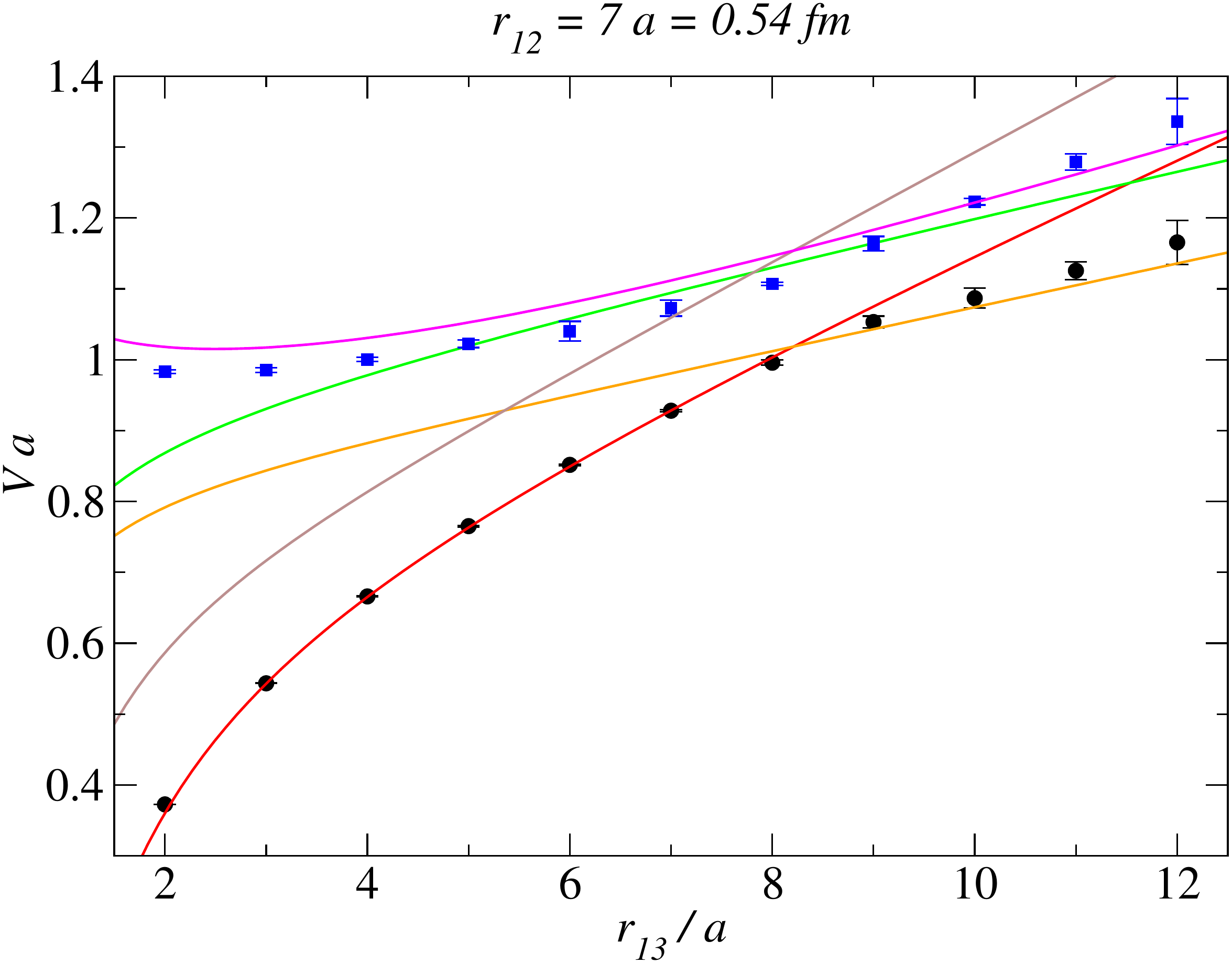}
\caption{ (Colour online.) Ground state (black dots) and first excited state (blue dots, where possible) full QCD lattice estimation of the static potentials in the $QQ \bar Q \bar Q$ parallel geometry.
The figures also include fits with various potential models, namely the two-meson potential, the double Y potential and the Casimir 
anti-triplet-triplet, sextet-anti-sextet and octet-octet potentials. 
\label{fig:tetra_dyn_res}}
\end{figure*}

Moreover, from the expression for the Casimir scaling potential given in (\ref{eq:VCasimir}) and using the results reported on Tab. \ref{tab:NormCas}
it is possible to define various types of potentials to be compared with the static potential computed from the lattice simulations.

The potential associated to the state where the quarks and anti-quarks are in triplet states leads to the so-called triplet-antitriplet or diquark-antidiquark potential,
\begin{equation}
V_{33}=\sum_{i<j}\langle\bar{\mathbf{3}}_{12}\mathbf{3}_{34}|C_{ij}|\bar{\mathbf{3}}_{12}\mathbf{3}_{34}\rangle V_{M}(r_{ij}) \ ,
\end{equation}
or in a form similar to (\ref{eq:VAligsimplified}),
\begin{eqnarray}
V_{33} &=&2K-\gamma\left(\frac{1}{r_{12}}+\frac{1}{2r_{13}}+\frac{1}{2\sqrt{r_{12}^{2}+r_{13}^{2}}}\right)
\nonumber \\
&& +\sigma\left(r_{12}+\frac{1}{2}r_{13}+\frac{1}{2}\sqrt{r_{12}^{2}+r_{13}^{2}}\right)\,.
\end{eqnarray}
Similarly, the anti-sextet-sextet potential is given by
\begin{eqnarray}
V_{66} &=& \sum_{i<j}\langle\mathbf{6}_{12}\mathbf{\bar{6}}_{34}|C_{ij}|\mathbf{6}_{12}\mathbf{\bar{6}}_{34}\rangle V_{M}(r_{ij})
\\
\nonumber 
 &=& \frac{5}{4}V_{M}(r_{13})+\frac{5}{4}V_{M}(\sqrt{r_{12}^{2}+r_{13}^{2}})-\frac{1}{2}V_{M}(r_{12}) \ , 
\end{eqnarray}
and the octet-octet potential reads
\begin{equation}
V_{88}=\frac{1}{2}V_{M}(r_{12})-\frac{1}{4}V_{M}(r_{13})+\frac{7}{4}V(\sqrt{r_{12}^{2}+r_{13}^{2}}) \ .
\label{eq:V88Align}
\end{equation}

The lattice estimates for the ground state and first excited (whenever possible) potentials can been in Figs. \ref{fig:tetra_quenched_res} 
and \ref{fig:tetra_dyn_res} for the quenched and for the dynamical simulation, respectively. 
The data shows that for large quark-anti-quark distances, i.e. for large $r_{13}$, the static potentials are compatible with a linearly 
rising function of $r_{13}$.  
This result can be viewed has an indication that the fermions on a tetraquark system are confined particles.

For both the pure gauge and dynamical simulations and for small quark-anti-quark distances, i.e. for small $r_{13}$,
and up to $r_{13} \leq  r_{12}$ the ground state potential reproduces that of a two meson state $V_{MM}$.
In this sense, one can claim that for sufficiently small quark-anti-quark distances the ground state of a $QQ\bar{Q}\bar{Q}$ system 
is a two meson state.
For the excited potential, the pure gauge results are among the double-Y potential (\ref{eq:VAligPprevious})
and the octet-octet potential (\ref{eq:V88Align}). However, for the dynamical results, the static
potential seems to be closer to $V_{88}$ at smaller and large $r_{13}$ and closer to $V_{YY}$ as $r_{13}$ approaches $ r_{12}$.

On the other hand for sufficiently large $r_{13}$, the ground state potential is essentially that of a diquark-antidiquark system
$V_{33}$ and the system enters its tetraquark phase. 
Indeed, the ground potential is given by $2 V_M$ for quark-anti-quarks distances up to $r_{13} = r_{12}$ and
is just above $V_{YY}$ for distances $r_{13} \geq r_{12} + 1$ in lattice units. These results suggests that, for this geometrical setup,
the transition of a two meson state towards a tetraquark state occurs at $r_{13} \sim r_{12} + 1$ (in lattice units).

In what concerns the dependence of $V_0$ on $r_{12}$, the lattice data suggests that the potential increases with the 
quark-quark distance and favours a $V_0 \sim V_{YY}$ for sufficiently large $r_{12}$ as was also observed in~\cite{Okiharu:2004ve,Alexandrou:2004ak}.

For the quark models with four-body tetraquark potentials, in particular the string flip-flop potential illustrated in Fig. \ref{fig:tripleflipflop} it is very important to quantify the deviation of $V_0$ from the $V_{YY}$ ansatz; and we have studied several ansatze for this difference. Clearly $V_0$ is more attractive than the tetraquark potential $V_{YY}$ of Eq. (\ref{eq:VAligPprevious}) reported by previous authors, and this favours the existence of tetraquarks.

Adding a negative constant (attractive) to the double-Y potential is not sufficient for a good fit of the lattice data for any of the sets of configurations.
Adding a correction to the double-Y potential which is linear in the
quark-quark distance,
\begin{equation}
V_{YY}^{B}=V_{YY}(r_{12},r_{34}) + \delta K + \delta\sigma_{12} ~ r_{12} \ ,
\end{equation}
describes quite well the dynamical simulation data and a fit gives $\delta K = -0.12(3) \, \sqrt{\sigma}$, $\delta\sigma_{12}=-0.34(5) \,  \sigma$,
where $\sigma$ is the fundamental string tension, for a $\chi^2 / d.o.f =0.46$ (see the tables on the appendix for details on the fits). 
The dynamical data for the deviations from $V_{YY}$ are also compatible with a Coulomb like correction
\begin{equation}
V_{YY}^{C} = V_{YY}(r_{12},r_{34}) + \delta K + \frac{\delta\gamma_{12}}{r_{12}} \ ,
\end{equation}
for a $\delta K=-0.67(4) \, \sqrt{\sigma}$, $\delta\gamma_{12}=0.22(3)$ 
with a $\chi^{2}/d.o.f. = 0.62$ (see appendix for details).
Such a functional form is not compatible with the lattice data for the pure gauge case. A possible explanation could come from
the difference in the statistics of both ensembles. Recall that the number of configurations for the pure gauge
ensemble is about ten times larger than for the dynamical simulation and, therefore, the associated statistical errors are much smaller.

In what concerns the first excited potential $V_1$, the data for the pure gauge and for the dynamical fermion simulation follows slightly 
different patterns. In the quenched simulation and for $r_{13} < r_{12}$, the potential is close to $V_T$ and the behaviour for
larger values of $r_{13}$ does not reproduces any of the potentials considered here. On the other hand, in the dynamical simulation
$V_1$ for small and large values of $r_{13}$ is just below the data for anti-sextet-sextet potential
\begin{equation}
V_{66}=\sum_{i<j}\langle\mathbf{6}_{12}\mathbf{\bar{6}}_{34}|C_{ij}|\mathbf{6}_{12}\mathbf{\bar{6}}_{34}\rangle \,  V_{M}(r_{ij}) 
\end{equation}
which, for this geometry, is given by
\begin{equation}
V_{66}=\frac{5}{4}V_{M}(r_{13})+\frac{5}{4}V_{M}(\sqrt{r_{12}^{2}+r_{13}^{2}})-\frac{1}{2}V_{M}(r_{12})
\end{equation}
and, at intermediate distances where $r_{13} \sim r_{12}$, is compatible with the octet-octet potential, 
\begin{equation}
V_{88}=\frac{1}{2}V_{M}(r_{12})-\frac{1}{4}V_{M}(r_{13})+\frac{7}{4}V(\sqrt{r_{12}^{2}+r_{13}^{2}}) \ .
\end{equation}
Further, at very small distances the potential seems to flatten for full QCD and the data also suggests a flattening or a small repulsive core. 
Note, for the quenched simulation smaller distances than 4 are not accessible, this short distance effect is not visible. 

\section{Summary and Discussion \label{Sec:final}}

In this work the static potential for a $Q Q \bar{Q} \bar{Q}$ system was investigated using both quenched and Wilson Fermion full QCD simulations for two
different geometric setups. The two geometries are designed to investigate sectors where dominantly meson-meson or tetraquark static potentials are expected.

The simulations show that whenever one distance is much larger than the other, the ground state potential and the first excited state potential are compatible
with a linearly rising function of the distance between constituents, suggesting that quarks and anti-quarks are confined particles. For the distances studied, the quenched and full QCD results are qualitatively similar, and their subtle differences only become clearer when we compare the lattice data with anzatse inspired in the string flip-flop potential and in Casimir scaling.

For the anti-parallel geometry setup, the groundstate potential $V_0$ is approximately described by a sum of two two meson potentials, i.e. it is compatible with the string flip-flop type of potential. We take this result as an indication that the $Q Q \bar{Q} \bar{Q}$ wave function is given by a superposition of two meson
states and we compute the mixing angle, as a function of the quark-anti-quark distances, which caracterize such a quantum state. 
The mixing angle shows that the tetraquark system undergoes a transition from one of the meson states to the other configuration as the
quark-antiquark distance increases, and the broadness of this transition has a typical length scale of $0.16 - 0.20$ fm.
Moreover, for the quenched simulation, we found an analytical expression which describes well the lattice groundstate.
The analytical expression is essentially a flip-flop type of potential with corrections, parametrized by $\Delta (r_1, r_2)$, which are typically 
$\lesssim 10$\% than the sum of two two mesons potentials. 

In what concerns the first excited potential $V_1$ in the anti-parallel geometry, the results show that for small enough
quark-anti-quark distances the potential is just below one of the possible octet-octet potentials and approaches a two meson potential from above
from large quark-anti-quark distances. This results for the excited potential can be interpreted in terms of and excited state including a combination of meson-meson and octet-octet states.

For the parallel geometry setup, the groundstate potential $V_0$ is compatible with  a diquark-antidiquark potential for large quark-antiquark distances and a sum of two meson potentials for small separations. Moreover, the lattice data for the full QCD simulation is 
compatible with a butterfly type of potential with corrections that we are able to parametrize. For the quenched simulation we found no analytical
expressions that are able to describe the data, but the trend is the same. 

The interpretation of the first excited potential $V_1$ for the parallet geometry, in terms of possible colour configurations is not as compliant with models as in the anti-parallel geometry. It seems that $V_1$ for the full QCD simulation
is just below the octet-octet from small quark-anti-quark distances and approaches again the octet-octet potential for at large distances. For the quenched simulation, the interpretation of $V_1$ in terms of colour components is not so clear, as the lattice data seems to point for a combination of different colour potentials.

Importantly for quark models with four-body tetraquark potentials, in particular for the string flip-flop potential illustrated in Fig. \ref{fig:tripleflipflop}, we obtain a groundstate potential $V_0$ more attractive, by a difference of $-300$ to $-500$ MeV, than the butterfly potential reported by previous authors
\cite{Alexandrou:2004ak,Bornyakov:2005kn,Okiharu:2004ve,Okiharu:2004wy}, and this favours the existence of tetraquarks.

As an outlook, it would be interesting to measure the static $Q Q \bar Q \bar Q$ potentials for larger distances and for different geometries. We leave this for future studies.

\begin{acknowledgements}
The authors are extremely grateful to Nuno Cardoso \cite{Cardoso:2011xu,Cardoso:2010di} for generating the ensemble of quenched configurations utilized in this work. The authors also acknowledge both the use of CPU and GPU servers of the collaboration PtQCD \cite{PtQCD}, supported by NVIDIA, CFTP and FCT grant UID/FIS/00777/2013, and the Laboratory for Advanced Computing at University of Coimbra  \cite{LCA} for providing HPC resources that have contributed to the research results reported within this paper. 
M. C. is supported by FCT under the contract SFRH/BPD/73140/2010.
P. J. S. acknowledges support
by FCT under contracts SFRH/BPD/40998/2007 and SFRH/BPD/109971/2015.
\end{acknowledgements}

\bibliographystyle{ieeetr}
\bibliography{bib}

\begin{thebibliography}{10}

\bibitem{Olive:2016xmw}
C.~Patrignani {\em et~al.}, ``{Review of Particle Physics},'' {\em Chin.
  Phys.}, vol.~C40, no.~10, p.~100001, 2016.

\bibitem{Briceno:2015rlt}
R.~A. Briceno {\em et~al.}, ``{Issues and Opportunities in Exotic Hadrons},''
  {\em Chin. Phys.}, vol.~C40, no.~4, p.~042001, 2016.

\bibitem{Lebed:2016hpi}
R.~F. Lebed, R.~E. Mitchell, and E.~S. Swanson, ``{Heavy-Quark QCD Exotica},''
  2016.

\bibitem{Esposito20171}
A.~Esposito, A.~Pilloni, and A.~Polosa, ``Multiquark resonances,'' {\em Physics
  Reports}, vol.~668, pp.~1 -- 97, 2017.
\newblock Multiquark Resonances.

\bibitem{Belle:2011aa}
A.~Bondar {\em et~al.}, ``{Observation of two charged bottomonium-like
  resonances in Y(5S) decays},'' {\em Phys. Rev. Lett.}, vol.~108, p.~122001,
  2012.

\bibitem{Choi:2007wga}
S.~Choi {\em et~al.}, ``{Observation of a resonance-like structure in the pi+-
  psi-prime mass distribution in exclusive B -> K pi+- psi-prime decays},''
  {\em Phys.Rev.Lett.}, vol.~100, p.~142001, 2008.

\bibitem{Liu:2013dau}
Z.~Q. Liu {\em et~al.}, ``{Study of $e^+e^- → π^+ π^- J/ψ$ and Observation
  of a Charged Charmoniumlike State at Belle},'' {\em Phys. Rev. Lett.},
  vol.~110, p.~252002, 2013.

\bibitem{Chilikin:2014bkk}
K.~Chilikin {\em et~al.}, ``{Observation of a new charged charmoniumlike state
  in $\bar{B}^0 → J/ψK^-π^+$ decays},'' {\em Phys. Rev.}, vol.~D90, no.~11,
  p.~112009, 2014.

\bibitem{Xiao:2013iha}
T.~Xiao, S.~Dobbs, A.~Tomaradze, and K.~K. Seth, ``{Observation of the Charged
  Hadron $Z\_c^{\pm}(3900)$ and Evidence for the Neutral $Z\_c^0(3900)$ in
  $e^+e^-\to \pi\pi J/\psi$ at $\sqrt{s}=4170$ MeV},'' {\em Phys. Lett.},
  vol.~B727, pp.~366--370, 2013.

\bibitem{Ablikim:2013mio}
M.~Ablikim {\em et~al.}, ``{Observation of a Charged Charmoniumlike Structure
  in $e^+e^-$ → $π^+π^-$ J/ψ at $\sqrt{s}$ =4.26  GeV},'' {\em Phys.
  Rev. Lett.}, vol.~110, p.~252001, 2013.

\bibitem{Ablikim:2013emm}
M.~Ablikim {\em et~al.}, ``{Observation of a charged charmoniumlike structure
  in $e^+e^- \to (D^{*} \bar{D}^{*})^{\pm} \pi^\mp$ at $\sqrt{s}=4.26$GeV},''
  {\em Phys. Rev. Lett.}, vol.~112, no.~13, p.~132001, 2014.

\bibitem{Ablikim:2013wzq}
M.~Ablikim {\em et~al.}, ``{Observation of a Charged Charmoniumlike Structure
  $Z\_c$(4020) and Search for the $Z\_c$(3900) in $e^+e^- \to π^+π^-h\_c$},''
  {\em Phys. Rev. Lett.}, vol.~111, no.~24, p.~242001, 2013.

\bibitem{Ablikim:2013xfr}
M.~Ablikim {\em et~al.}, ``{Observation of a charged $(D\bar{D}^{*})^\pm$ mass
  peak in $e^{+}e^{-} \to \pi D\bar{D}^{*}$ at $\sqrt{s} =$ 4.26 GeV},'' {\em
  Phys. Rev. Lett.}, vol.~112, no.~2, p.~022001, 2014.

\bibitem{Ablikim:2014dxl}
M.~Ablikim {\em et~al.}, ``{Observation of $e^+e^- → π^0π^0h\_c$ and a
  Neutral Charmoniumlike Structure $Z\_c(4020)^0$},'' {\em Phys. Rev. Lett.},
  vol.~113, no.~21, p.~212002, 2014.

\bibitem{Aaij:2014jqa}
R.~Aaij {\em et~al.}, ``{Observation of the resonant character of the
  $Z(4430)^-$ state},'' {\em Phys.Rev.Lett.}, vol.~112, no.~22, p.~222002,
  2014.

\bibitem{Aaij:2015tga}
R.~Aaij {\em et~al.}, ``{Observation of J/ψp Resonances Consistent with
  Pentaquark States in Λ$_b^0$ → J/ψK$^-$p Decays},'' {\em Phys. Rev.
  Lett.}, vol.~115, p.~072001, 2015.

\bibitem{Cardoso:2008dd}
M.~Cardoso and P.~Bicudo, ``{Microscopic calculation of the decay of
  Jaffe-Wilczek tetraquarks, and the Z(4433)},'' {\em AIP Conf.Proc.},
  vol.~1030, p.~352, 2008.

\bibitem{Prelovsek:2014swa}
S.~Prelovsek, C.~Lang, L.~Leskovec, and D.~Mohler, ``{Study of the $Z_c^+$
  channel using lattice QCD},'' 2014.

\bibitem{Leskovec:2014gxa}
L.~Leskovec, S.~Prelovsek, C.~Lang, and D.~Mohler, ``{Study of the Zc+ channel
  in lattice QCD},'' 2014.

\bibitem{Ikeda:2013vwa}
Y.~Ikeda, B.~Charron, S.~Aoki, T.~Doi, T.~Hatsuda, T.~Inoue, N.~Ishii,
  K.~Murano, H.~Nemura, and K.~Sasaki, ``{Charmed tetraquarks $T\_{cc}$ and
  $T\_{cs}$ from dynamical lattice QCD simulations},'' {\em Phys. Lett.},
  vol.~B729, pp.~85--90, 2014.

\bibitem{Guerrieri:2014nxa}
A.~L. Guerrieri, M.~Papinutto, A.~Pilloni, A.~D. Polosa, and N.~Tantalo,
  ``{Flavored tetraquark spectroscopy},'' {\em PoS}, vol.~LATTICE2014, p.~106,
  2015.

\bibitem{Wagner:2010ad}
M.~Wagner, ``{Forces between static-light mesons},'' {\em PoS},
  vol.~LATTICE2010, p.~162, 2010.

\bibitem{Wagner:2011ev}
M.~Wagner, ``{Static-static-light-light tetraquarks in lattice QCD},'' {\em
  Acta Phys. Polon. Supp.}, vol.~4, pp.~747--752, 2011.

\bibitem{Born:1927}
M.~Born and R.~Oppenheimer, ``{Zur Quantentheorie der Molekeln},'' {\em Annalen
  der Physik}, vol.~389, p.~457, 1927.

\bibitem{Bicudo:2012qt}
P.~Bicudo and M.~Wagner, ``{Lattice QCD signal for a bottom-bottom
  tetraquark},'' {\em Phys. Rev.}, vol.~D87, no.~11, p.~114511, 2013.

\bibitem{Brown:2012tm}
Z.~S. Brown and K.~Orginos, ``{Tetraquark bound states in the heavy-light
  heavy-light system},'' {\em Phys.Rev.}, vol.~D86, p.~114506, 2012.

\bibitem{Bicudo:2015kna}
P.~Bicudo, K.~Cichy, A.~Peters, and M.~Wagner, ``{BB interactions with static
  bottom quarks from Lattice QCD},'' {\em Phys. Rev.}, vol.~D93, no.~3,
  p.~034501, 2016.

\bibitem{Peters:2015tra}
A.~Peters, P.~Bicudo, K.~Cichy, B.~Wagenbach, and M.~Wagner, ``{Exploring
  possibly existing $q q \bar b \bar b$ tetraquark states with $q q = ud, ss,
  cc$},'' {\em PoS}, vol.~LATTICE2015, p.~095, 2016.

\bibitem{Bicudo:2015vta}
P.~Bicudo, K.~Cichy, A.~Peters, B.~Wagenbach, and M.~Wagner, ``{Evidence for
  the existence of $u d \bar{b} \bar{b}$ and the non-existence of $s s \bar{b}
  \bar{b}$ and $c c \bar{b} \bar{b}$ tetraquarks from lattice QCD},'' {\em
  Phys. Rev.}, vol.~D92, no.~1, p.~014507, 2015.

\bibitem{Bicudo:2016ooe}
P.~Bicudo, J.~Scheunert, and M.~Wagner, ``{Including heavy spin effects in the
  prediction of a $\bar{b} \bar{b} u d$ tetraquark with lattice QCD
  potentials},'' {\em Phys. Rev.}, vol.~D95, no.~3, p.~034502, 2017.

\bibitem{Bicudo:2016jwl}
P.~Bicudo, J.~Scheunert, and M.~Wagner, ``{Including heavy spin effects in a
  lattice QCD study of static-static-light-light tetraquarks},'' in {\em
  {Proceedings, 34th International Symposium on Lattice Field Theory (Lattice
  2016): Southampton, UK, July 24-30, 2016}}, 2016.

\bibitem{Peters:2016isf}
A.~Peters, P.~Bicudo, L.~Leskovec, S.~Meinel, and M.~Wagner, ``{Lattice QCD
  study of heavy-heavy-light-light tetraquark candidates},'' in {\em
  {Proceedings, 34th International Symposium on Lattice Field Theory (Lattice
  2016): Southampton, UK, July 24-30, 2016}}, 2016.

\bibitem{Fishbane:1977ay}
P.~M. Fishbane and M.~T. Grisaru, ``{Consequences of a Color Induced Van Der
  Waals' Force Between Hadrons},'' {\em Phys.Lett.}, vol.~B74, p.~98, 1978.

\bibitem{Appelquist:1978rt}
T.~Appelquist and W.~Fischler, ``{Some Remarks on Van Der Waals Forces in
  {QCD}},'' {\em Phys.Lett.}, vol.~B77, p.~405, 1978.

\bibitem{Willey:1978fm}
R.~Willey, ``{ON QUARK VAN DER WAALS FORCES},'' {\em Phys.Rev.}, vol.~D18,
  p.~270, 1978.

\bibitem{Matsuyama:1978hf}
S.~Matsuyama and H.~Miyazawa, ``{LONG RANGE FORCES BETWEEN HADRONS},'' {\em
  Prog.Theor.Phys.}, vol.~61, p.~942, 1979.

\bibitem{Gavela:1979zu}
M.~Gavela, A.~Le~Yaouanc, L.~Oliver, O.~Pene, J.~Raynal, {\em et~al.}, ``{VAN
  DER WAALS LIKE FORCES BETWEEN HADRONS INDUCED BY COLOR CONFINING
  POTENTIALS},'' {\em Phys.Lett.}, vol.~B82, p.~431, 1979.

\bibitem{Feinberg:1983zz}
G.~Feinberg and J.~Sucher, ``{Long-range forces between a charged and neutral
  system},'' {\em Phys.Rev.}, vol.~A27, pp.~1958--1967, 1983.

\bibitem{Miyazawa:1979vx}
H.~Miyazawa, ``{String Flipflop and Quark Matter},'' {\em Phys.Rev.}, vol.~D20,
  p.~2953, 1979.

\bibitem{Miyazawa:1980ft}
H.~Miyazawa, ``{String Flipflop, Quark Matter and Multi - Baryon Resonances.
  (Talk)},'' {\em {\it In *Hakone 1980, Proceedings, High-energy Nuclear
  Interactions and Properties Of Dense Nuclear Matter, Vol. 2*, Iii.224-226}},
  1980.

\bibitem{Oka:1984yx}
M.~Oka, ``{HADRON HADRON INTERACTION IN THE STRING FLIPFLOP MODEL OF QUARK
  CONFINEMENT WITH COLOR, SPIN AND FLAVOR DEGREES OF FREEDOM. 1. MESON MESON
  INTERACTION},'' {\em Phys.Rev.}, vol.~D31, pp.~2274--2287, 1985.

\bibitem{Oka:1985vg}
M.~Oka and C.~Horowitz, ``{HADRON HADRON INTERACTION IN THE STRING FLIPFLOP
  MODEL OF QUARK CONFINEMENT WITH COLOR, SPIN AND FLAVOR DEGREES OF FREEDOM. 2.
  NUCLEON NUCLEON INTERACTION},'' {\em Phys.Rev.}, vol.~D31, pp.~2773--2779,
  1985.

\bibitem{Karliner:2003dt}
M.~Karliner and H.~J. Lipkin, ``{A Diquark - triquark model for the K N
  pentaquark},'' {\em Phys.Lett.}, vol.~B575, pp.~249--255, 2003.

\bibitem{Carlson:1991zt}
J.~Carlson and V.~R. Pandharipande, ``{Absence of exotic hadrons in flux tube
  quark models},'' {\em Phys. Rev.}, vol.~D43, pp.~1652--1658, 1991.

\bibitem{Vijande:2007ix}
J.~Vijande, A.~Valcarce, and J.-M. Richard, ``{Stability of multiquarks in a
  simple string model},'' {\em Phys.Rev.}, vol.~D76, p.~114013, 2007.

\bibitem{Vijande:2009xx}
J.~Vijande, A.~Valcarce, J.-M. Richard, and N.~Barnea, ``{Four-quark
  stability},'' {\em Few Body Syst.}, vol.~45, pp.~99--103, 2009.

\bibitem{Beinker:1995qe}
M.~Beinker, B.~Metsch, and H.~Petry, ``{Bound q**2 - anti-q**2 states in a
  constituent quark model},'' {\em J.Phys.}, vol.~G22, pp.~1151--1160, 1996.

\bibitem{Zouzou:1986qh}
S.~Zouzou, B.~Silvestre-Brac, C.~Gignoux, and J.~Richard, ``{FOUR QUARK BOUND
  STATES},'' {\em Z.Phys.}, vol.~C30, p.~457, 1986.

\bibitem{Bicudo:2010mv}
P.~Bicudo and M.~Cardoso, ``{Decays of tetraquark resonances in a two-variable
  approximation to the triple flip-flop potential},'' {\em Phys.Rev.},
  vol.~D83, p.~094010, 2011.

\bibitem{Bicudo:2015bra}
P.~Bicudo and M.~Cardoso, ``{Tetraquark bound states and resonances in the
  unitary and microscopic triple string flip-flop quark model, the
  light-light-antiheavy-antiheavy $q q \bar Q\bar Q$ case study},'' {\em Phys.
  Rev.}, vol.~D94, no.~9, p.~094032, 2016.

\bibitem{Lenz:1985jk}
F.~Lenz, J.~Londergan, E.~Moniz, R.~Rosenfelder, M.~Stingl, {\em et~al.},
  ``{Quark Confinement and Hadronic Interactions},'' {\em Annals Phys.},
  vol.~170, p.~65, 1986.

\bibitem{Heupel:2012ua}
W.~Heupel, G.~Eichmann, and C.~S. Fischer, ``{Tetraquark Bound States in a
  Bethe-Salpeter Approach},'' {\em Phys. Lett.}, vol.~B718, pp.~545--549, 2012.

\bibitem{Bali:2000gf}
G.~S. Bali, ``{QCD forces and heavy quark bound states},'' {\em Phys. Rept.},
  vol.~343, pp.~1--136, 2001.

\bibitem{Alexandrou:2004ak}
C.~Alexandrou and G.~Koutsou, ``{The Static tetraquark and pentaquark
  potentials},'' {\em Phys. Rev.}, vol.~D71, p.~014504, 2005.

\bibitem{Bornyakov:2005kn}
V.~G. Bornyakov, P.~{\relax Yu}. Boyko, M.~N. Chernodub, and M.~I. Polikarpov,
  ``{Interactions of confining strings in SU(3) gluodynamics},'' 2005.

\bibitem{Okiharu:2004ve}
F.~Okiharu, H.~Suganuma, and T.~T. Takahashi, ``{Detailed analysis of the
  tetraquark potential and flip-flop in SU(3) lattice QCD},'' {\em Phys. Rev.},
  vol.~D72, p.~014505, 2005.

\bibitem{Okiharu:2004wy}
F.~Okiharu, H.~Suganuma, and T.~T. Takahashi, ``{First study for the pentaquark
  potential in SU(3) lattice QCD},'' {\em Phys. Rev. Lett.}, vol.~94,
  p.~192001, 2005.

\bibitem{Bicudo:2007xp}
P.~Bicudo, M.~Cardoso, and O.~Oliveira, ``{Study of the gluon-quark-antiquark
  static potential in SU(3) lattice QCD},'' {\em Phys. Rev.}, vol.~D77,
  p.~091504, 2008.

\bibitem{Cardoso:2008sb}
M.~Cardoso and P.~Bicudo, ``{First study of the three-gluon static potential in
  Lattice QCD},'' {\em Phys. Rev.}, vol.~D78, p.~074508, 2008.

\bibitem{Cardoso:2011xu}
N.~Cardoso and P.~Bicudo, ``{Generating SU(Nc) pure gauge lattice QCD
  configurations on GPUs with CUDA and OpenMP},'' {\em Comput. Phys. Commun.},
  vol.~184, pp.~509--518, 2013.

\bibitem{Cardoso:2010di}
N.~Cardoso and P.~Bicudo, ``{SU(2) Lattice Gauge Theory Simulations on Fermi
  GPUs},'' {\em J. Comput. Phys.}, vol.~230, pp.~3998--4010, 2011.

\bibitem{PtQCD}
http://nemea.ist.utl.pt/ ptqcd

\bibitem{LCA}
http://www.uc.pt/lca

\bibitem{Edwards:2004sx}
R.~G. Edwards and B.~Joo, ``{The Chroma software system for lattice QCD},''
  {\em Nucl. Phys. Proc. Suppl.}, vol.~140, p.~832, 2005.
\newblock [,832(2004)].

\bibitem{Clark:2011ir}
M.~A. Clark, B.~Joo, A.~D. Kennedy, and P.~J. Silva, ``{Improving dynamical
  lattice QCD simulations through integrator tuning using Poisson brackets and
  a force-gradient integrator},'' {\em Phys. Rev.}, vol.~D84, p.~071502, 2011.

\bibitem{Kennedy:2012gk}
A.~D. Kennedy, P.~J. Silva, and M.~A. Clark, ``{Shadow Hamiltonians, Poisson
  Brackets, and Gauge Theories},'' {\em Phys. Rev.}, vol.~D87, no.~3,
  p.~034511, 2013.

\bibitem{Albanese:1987ds}
M.~Albanese {\em et~al.}, ``{Glueball Masses and String Tension in Lattice
  QCD},'' {\em Phys. Lett.}, vol.~B192, pp.~163--169, 1987.

\bibitem{Cardoso:2013lla}
N.~Cardoso, M.~Cardoso, and P.~Bicudo, ``{Inside the SU(3) quark-antiquark QCD
  flux tube: screening versus quantum widening},'' {\em Phys. Rev.}, vol.~D88,
  p.~054504, 2013.

\bibitem{Parisi:1983hm}
G.~Parisi, R.~Petronzio, and F.~Rapuano, ``{A Measurement of the String Tension
  Near the Continuum Limit},'' {\em Phys. Lett.}, vol.~B128, pp.~418--420,
  1983.

\bibitem{Hasenfratz:2001hp}
A.~Hasenfratz and F.~Knechtli, ``{Flavor symmetry and the static potential with
  hypercubic blocking},'' {\em Phys. Rev.}, vol.~D64, p.~034504, 2001.

\bibitem{Sommer:1993ce}
R.~Sommer, ``{A New way to set the energy scale in lattice gauge theories and
  its applications to the static force and alpha-s in SU(2) Yang-Mills
  theory},'' {\em Nucl. Phys.}, vol.~B411, pp.~839--854, 1994.

\end{thebibliography}

\appendix

\section{Tables of Results \label{App:tabelas}}

\clearpage

\begin{table*}[!t]
\begin{tabular}{|c|c|ccc||c|c|ccc|}
\hline 
$r_{13}$ & $r_{14}$ & $V_{0}$ & $t_{i}-t_{f}$ & $\chi_{r}^{2}$ & $r_{13}$ & $r_{14}$ & $V_{0}$ & $t_{i}-t_{f}$ & $\chi_{r}^{2}$\tabularnewline
\hline 
\hline 
4 & 4 & 1.3010(1) & 9-16 & 0.94 & 7 & 7 & 1.51984(14) & 6-10 & 0.52\tabularnewline
\hline 
 & 5 & 1.32119(8) & 8-16 & 0.82 &  & 8 & 1.53675(17) & 6-10 & 0.65\tabularnewline
\hline 
 & 6 & 1.32452(4) & 5-16 & 0.62 &  & 9 & 1.54052(23) & 6-16 & 0.48\tabularnewline
\hline 
 & 7 & 1.32547(5) & 5-16 & 0.69 &  & 10 & 1.54154(31) & 6-16 & 0.88\tabularnewline
\hline 
 & 8 & 1.32581(5) & 5-16 & 0.53 &  & 11 & 1.54208(32) & 6-16 & 1.06\tabularnewline
\hline 
 & 9 & 1.32594(5) & 5-16 & 0.82 &  & 12 & 1.54220(36) & 6-16 & 0.49\tabularnewline
\hline 
 & 10 & 1.32601(5) & 5-16 & 0.76 & 8 & 8 & 1.58247(33) & 6-16 & 0.96\tabularnewline
\hline 
 & 11 & 1.32608(6) & 5-16 & 1.40 &  & 9 & 1.59850(38) & 6-13 & 1.01\tabularnewline
\hline 
 & 12 & 1.32604(6) & 5-10 & 0.85 &  & 10 & 1.60205(42) & 6-16 & 0.38\tabularnewline
\hline 
5 & 5 & 1.38254(8) & 5-11 & 0.92 &  & 11 & 1.60289(59) & 6-16 & 1.09\tabularnewline
\hline 
 & 6 & 1.40149(9) & 5-11 & 0.63 &  & 12 & 1.60232(13) & 7-12 & 0.52\tabularnewline
\hline 
 & 7 & 1.40546(10) & 5-16 & 0.70 & 9 & 9 & 1.64322(81) & 6-16 & 1.02\tabularnewline
\hline 
 & 8 & 1.40657(11) & 5-16 & 0.87 &  & 10 & 1.65809(101) & 6-15 & 0.94\tabularnewline
\hline 
 & 9 & 1.40657(12) & 5-16 & 1.13 &  & 11 & 1.65919(90) & 7-16 & 0.76\tabularnewline
\hline 
 & 10 & 1.40716(12) & 5-16 & 0.86 &  & 12 & 1.65961(95) & 7-10 & 0.75\tabularnewline
\hline 
 & 11 & 1.40729(11) & 5-16 & 0.64 & 10 & 10 & 1.70108(219) & 7-16 & 1.00\tabularnewline
\hline 
 & 12 & 1.40729(12) & 6-13 & 1.06 &  & 11 & 1.71397(156) & 7-10 & 1.03\tabularnewline
\hline 
6 & 6 & 1.45412(15) & 5-10 & 0.88 &  & 12 & 1.71573(183) & 7-10 & 1.06\tabularnewline
\hline 
 & 7 & 1.47203(21) & 5-16 & 1.00 & 11 & 11 & 1.75794(335) & 7-10 & 1.26\tabularnewline
\hline 
 & 8 & 1.47596(26) & 5-16 & 1.21 &  & 12 & 1.77037(216) & 7-10 & 0.70\tabularnewline
\hline 
 & 9 & 1.47710(31) & 5-16 & 1.34 & 12 & 12 & 1.81682(432) & 7-10 & 0.74\tabularnewline
\hline 
 & 10 & 1.47760(31) & 5-16 & 1.37 & \multicolumn{1}{c}{} & \multicolumn{1}{c}{} &  &  & \multicolumn{1}{c}{}\tabularnewline
\cline{1-5} 
 & 11 & 1.47783(31) & 5-16 & 1.20 & \multicolumn{1}{c}{} & \multicolumn{1}{c}{} &  &  & \multicolumn{1}{c}{}\tabularnewline
\cline{1-5} 
 & 12 & 1.47763(13) & 6-13 & 0.38 & \multicolumn{1}{c}{} & \multicolumn{1}{c}{} &  &  & \multicolumn{1}{c}{}\tabularnewline
\cline{1-5} 
\end{tabular}\caption{Data for the ground state potential in the anti-parallel geometry for the pure-gauge simulation. 
All values are in lattice units.}
\end{table*}

\begin{table*}[!t]
\begin{tabular}{|c|c|ccc||c|c|ccc|}
\hline 
$r_{13}$ & $r_{14}$ & $V_{1}$ & $t_{i}-t_{f}$ & $\chi_{r}^{2}$ & $r_{13}$ & $r_{14}$ & $V_{1}$ & $t_{i}-t_{f}$ & $\chi_{r}^{2}$\tabularnewline
\hline 
\hline 
4 & 4 & 1.37342(8) & 4-7 & 1.11 & 7 & 7 & 1.57714(10) & 7-10 & 0.17\tabularnewline
\hline 
 & 5 & 1.43446(6) & 4-6 & 0.41 &  & 8 & 1.62034(21) & 7-16 & 0.23\tabularnewline
\hline 
 & 6 & 1.50427(8) & 4-5 & 0.32 &  & 9 & 1.67517(27) & 7-13 & 0.39\tabularnewline
\hline 
 & 7 & - & - & - &  & 10 & 1.73150(34) & 7-13 & 0.26\tabularnewline
\hline 
 & 8 & - & - & - &  & 11 & 1.78729(32) & 7-14 & 0.34\tabularnewline
\hline 
 & 9 & - & - & - &  & 12 & 1.84313(215) & 7-15 & 0.23\tabularnewline
\hline 
 & 10 & - & - & - & 8 & 8 & 1.63344(67) & 7-15 & 0.39\tabularnewline
\hline 
 & 11 & - & - & - &  & 9 & 1.67532(85) & 7-14 & 0.48\tabularnewline
\hline 
 & 12 & - & - & - &  & 10 & 1.72029(29) & 9-10 & 0.01\tabularnewline
\hline 
5 & 5 & 1.45092(12) & 5-7 & 0.66 &  & 11 & 1.77764(267) & 8-10 & 0.51\tabularnewline
\hline 
 & 6 & 1.50245(14) & 5-7 & 0.57 &  & 12 & 1.83159(607) & 8-15 & 0.51\tabularnewline
\hline 
 & 7 & 1.56502(18) & 5-7 & 1.02 & 9 & 9 & 1.68826(48) & 7-16 & 0.24\tabularnewline
\hline 
 & 8 & 1.62775(18) & 5-6 & 0.40 &  & 10 & 1.72195(153) & 9-11 & 0.20\tabularnewline
\hline 
 & 9 & - & - & - &  & 11 & 1.78195(74) & 7-8 & 0.37\tabularnewline
\hline 
 & 10 & - & - & - &  & 12 & 1.83526(272) & 7-12 & 0.57\tabularnewline
\hline 
 & 11 & - & - & - & 10 & 10 & 1.73977(76) & 7-14 & 0.22\tabularnewline
\hline 
 & 12 & - & - & - &  & 11 & 1.78172(233) & 7-12 & 0.72\tabularnewline
\hline 
6 & 6 & 1.51750(20) & 6-14 & 1.12 &  & 12 & 1.83504(281) & 7-14 & 0.41\tabularnewline
\hline 
 & 7 & 1.56389(24) & 6-16 & 0.44 & 11 & 11 & 1.79142(126) & 7-12 & 0.40\tabularnewline
\hline 
 & 8 & 1.62139(19) & 6-8 & 0.64 &  & 12 & 1.83291(374) & 7-12 & 0.49\tabularnewline
\hline 
 & 9 & 1.68025(11) & 6-7 & 0.07 & 12 & 12 & 1.84085(34) & 7-13 & 0.16\tabularnewline
\hline 
 & 10 & 1.73928(61) & 6-7 & 1.03 & \multicolumn{1}{c}{} & \multicolumn{1}{c}{} &  &  & \multicolumn{1}{c}{}\tabularnewline
\cline{1-5} 
 & 11 & 1.79777(48) & 5-6 & 0.79 & \multicolumn{1}{c}{} & \multicolumn{1}{c}{} &  &  & \multicolumn{1}{c}{}\tabularnewline
\cline{1-5} 
 & 12 & 1.85509(28) & 5-6 & 0.17 & \multicolumn{1}{c}{} & \multicolumn{1}{c}{} &  &  & \multicolumn{1}{c}{}\tabularnewline
\cline{1-5} 
\end{tabular}\caption{Data for the first excited state potential for the anti-parallel geometry and for the pure-gauge simulation. All values are in lattice
units.}
\end{table*}

\begin{table*}[!t]
\begin{tabular}{|c|c|ccc||c|c|ccc|}
\hline 
$r_{13}$ & $r_{14}$ & $V_{0}$ & $t_{i}-t_{f}$ & $\chi_{r}^{2}$ & $r_{13}$ & $r_{14}$ & $V_{0}$ & $t_{i}-t_{f}$ & $\chi_{r}^{2}$\tabularnewline
\hline 
\hline 
3 & 3 & 0.5126(1) & 6-10 & 0.07 & 6 & 6 & 0.8248(26) & 6-8 & 0.62\tabularnewline
\hline 
 & 4 & 0.5397(2) & 6-10 & 0.07 &  & 7 & 0.8464(12) & 6-7 & 0.26\tabularnewline
\hline 
 & 5 & 0.5431(2) & 6-10 & 0.24 &  & 8 & 0.8504(18) & 6-8 & 0.79\tabularnewline
\hline 
 & 6 & 0.5438(1) & 6-10 & 0.12 &  & 9 & 0.8503(25) & 6-8 & 0.82\tabularnewline
\hline 
 & 7 & 0.5437(2) & 6-10 & 0.07 &  & 10 & 0.8520(10) & 6-7 & 0.18\tabularnewline
\hline 
 & 8 & 0.5438(2) & 6-10 & 0.20 &  & 11 & 0.8531(18) & 6-11 & 0.38\tabularnewline
\hline 
 & 9 & 0.5433(3) & 6-10 & 0.34 & 7 & 7 & 0.9088(67) & 6-10 & 0.60\tabularnewline
\hline 
 & 10 & 0.5434(3) & 6-10 & 0.32 &  & 8 & 0.9293(72) & 6-10 & 1.07\tabularnewline
\hline 
 & 11 & 0.5440(2) & 6-10 & 0.39 &  & 9 & 0.9157(23) & 7-8 & 0.12\tabularnewline
\hline 
4 & 4 & 0.6364(6) & 6-10 & 0.50 &  & 10 & 0.9314(71) & 6-8 & 1.12\tabularnewline
\hline 
 & 5 & 0.6620(3) & 6-10 & 0.08 &  & 11 & 0.9342(10) & 6-8 & 0.19\tabularnewline
\hline 
 & 6 & 0.6658(1) & 6-10 & 0.02 & 8 & 8 & 0.9854(214) & 6-12 & 0.83\tabularnewline
\hline 
 & 7 & 0.6654(9) & 6-10 & 0.80 &  & 9 & 1.0023(159) & 6-11 & 1.25\tabularnewline
\hline 
 & 8 & 0.6655(10) & 6-10 & 0.66 &  & 10 & 1.0048(96) & 6-10 & 0.70\tabularnewline
\hline 
 & 9 & 0.6654(9) & 6-10 & 0.58 &  & 11 & 1.0004(82) & 6-10 & 0.40\tabularnewline
\hline 
 & 10 & 0.6667(5) & 6-10 & 0.32 & 9 & 9 & 1.0650(277) & 6-10 & 1.01\tabularnewline
\hline 
 & 11 & 0.6677(4) & 6-10 & 0.14 &  & 10 & 1.0734(178) & 6-10 & 0.55\tabularnewline
\hline 
5 & 5 & 0.7382(9) & 6-10 & 0.26 &  & 11 & 1.0805(26) & 5-6 & 0.22\tabularnewline
\hline 
 & 6 & 0.7605(5) & 6-10 & 0.29 & 10 & 10 & 1.1377(56) & 5-6 & 0.64\tabularnewline
\hline 
 & 7 & 0.7621(14) & 6-10 & 0.66 &  & 11 & 1.1491(18) & 5-6 & 0.06\tabularnewline
\hline 
 & 8 & 0.7630(20) & 6-10 & 1.09 & 11 & 11 & 1.1630(298) & 6-9 & 0.14\tabularnewline
\hline 
 & 9 & 0.7642(11) & 6-10 & 0.54 & \multicolumn{1}{c}{} & \multicolumn{1}{c}{} &  &  & \multicolumn{1}{c}{}\tabularnewline
\cline{1-5} 
 & 10 & 0.7643(8) & 6-10 & 0.28 & \multicolumn{1}{c}{} & \multicolumn{1}{c}{} &  &  & \multicolumn{1}{c}{}\tabularnewline
\cline{1-5} 
 & 11 & 0.7664(16) & 6-10 & 0.20 & \multicolumn{1}{c}{} & \multicolumn{1}{c}{} &  &  & \multicolumn{1}{c}{}\tabularnewline
\cline{1-5} 
\end{tabular}\caption{Data for the ground state potential for the anti-parallel geometry and for the full QCD simulation. All values are in lattice units.}
\end{table*}

\begin{table*}[!t]
\begin{tabular}{|c|c|ccc||c|c|ccc|}
\hline 
$r_{13}$ & $r_{14}$ & $V_{1}$ & $t_{i}-t_{f}$ & $\chi_{r}^{2}$ & $r_{13}$ & $r_{14}$ & $V_{1}$ & $t_{i}-t_{f}$ & $\chi_{r}^{2}$\tabularnewline
\hline 
\hline 
3 & 3 & 0.6039(2) & 6-7 & 0.12 & 6 & 6 & 0.8988(38) & 6-12 & 0.16\tabularnewline
\hline 
 & 4 & 0.7062(10) & 6-7 & 0.91 &  & 7 & 0.9549(19) & 6-15 & 0.69\tabularnewline
\hline 
 & 5 & - & - & - &  & 8 & 1.0233(101) & 6-10 & 1.26\tabularnewline
\hline 
 & 6 & - & - & - &  & 9 & 1.0933(184) & 6-11 & 0.34\tabularnewline
\hline 
 & 7 & - & - & - &  & 10 & 1.1690(8) & 6-10 & 0.10\tabularnewline
\hline 
 & 8 & - & - & - &  & 11 & 1.2341(136) & 6-9 & 0.25\tabularnewline
\hline 
 & 9 & - & - & - & 7 & 7 & 0.9704(33) & 6-10 & 0.67\tabularnewline
\hline 
 & 10 & - & - & - &  & 8 & 1.0235(96) & 6-10 & 0.81\tabularnewline
\hline 
 & 11 & - & - & - &  & 9 & 1.0952(79) & 6-10 & 0.34\tabularnewline
\hline 
4 & 4 & 0.7234(3) & 6-10 & 0.19 &  & 10 & 1.1683(170) & 6-9 & 1.26\tabularnewline
\hline 
 & 5 & 0.7990(8) & 6-13 & 0.12 &  & 11 & 1.2274(408) & 6-9 & 0.31\tabularnewline
\hline 
 & 6 & 0.8960(22) & 6-12 & 0.78 & 8 & 8 & 1.0482(35) & 6-12 & 0.42\tabularnewline
\hline 
 & 7 & 0.9974(47) & 6-7 & 1.20 &  & 9 & 1.1122(9) & 6-8 & 0.11\tabularnewline
\hline 
 & 8 & - & - & - &  & 10 & 1.1956(30) & 6-9 & 0.42\tabularnewline
\hline 
 & 9 & - & - & - &  & 11 & 1.2612(89) & 6-9 & 0.06\tabularnewline
\hline 
 & 10 & - & - & - & 9 & 9 & 1.1219(15) & 6-10 & 0.13\tabularnewline
\hline 
 & 11 & - & - & - &  & 10 & 1.1852(360) & 6-9 & 0.92\tabularnewline
\hline 
5 & 5 & 0.8160(12) & 6-14 & 0.27 &  & 11 & 1.2526(507) & 6-11 & 0.26\tabularnewline
\hline 
 & 6 & 0.8821(16) & 6-12 & 0.27 & 10 & 10 & 1.1621(476) & 6-9 & 0.79\tabularnewline
\hline 
 & 7 & 0.9617(25) & 6-12 & 0.63 &  & 11 & 1.2069(751) & 6-9 & 1.46\tabularnewline
\hline 
 & 8 & 1.0409(16) & 6-12 & 0.63 & 11 & 11 & 1.2052(998) & 6-9 & 1.30\tabularnewline
\hline 
 & 9 & 1.1287(136) & 6-8 & 0.49 & \multicolumn{1}{c}{} & \multicolumn{1}{c}{} &  &  & \multicolumn{1}{c}{}\tabularnewline
\cline{1-5} 
 & 10 & 1.1992(362) & 6-9 & 0.71 & \multicolumn{1}{c}{} & \multicolumn{1}{c}{} &  &  & \multicolumn{1}{c}{}\tabularnewline
\cline{1-5} 
 & 11 & 1.2929(606) & 6-8 & 0.88 & \multicolumn{1}{c}{} & \multicolumn{1}{c}{} &  &  & \multicolumn{1}{c}{}\tabularnewline
\cline{1-5} 
\end{tabular}\caption{Data for the first excited state potential for the anti-parallel geometry and for the full QCD simulation. All values are in lattice
units.}
\end{table*}

\begin{table*}[!t]
\begin{tabular}{|cc|ccc||cc|ccc||cc|ccc|}
\hline 
$r_{12}$ & $r_{13}$ & $V_{0}$ & $t_{i}-t_{f}$ & $\chi_{r}^{2}$ & $r_{12}$ & $r_{13}$ & $V_{0}$ & $t_{i}-t_{f}$ & $\chi_{r}^{2}$ & $r_{12}$ & $r_{13}$ & $V_{0}$ & $t_{i}-t_{f}$ & $\chi_{r}^{2}$\tabularnewline
\hline 
\hline 
4 & 4 & 1.31360(7) & 9-16 & 0.90 & 7 & 4 & 1.32551(5) & 7-16 & 0.64 & 10 & 4 & 1.32607(8) & 7-14 & 1.09\tabularnewline
\hline 
 & 5 & 1.37514(19) & 9-16 & 0.90 &  & 5 & 1.40571(11) & 7-16 & 0.88 &  & 5 & 1.40713(10) & 7-15 & 0.75\tabularnewline
\hline 
 & 6 & 1.41847(24) & 9-14 & 0.78 &  & 6 & 1.47380(8) & 7-16 & 0.18 &  & 6 & 1.47735(10) & 6-16 & 0.23\tabularnewline
\hline 
 & 7 & 1.45346(28) & 9-15 & 0.76 &  & 7 & 1.53399(6) & 7-15 & 0.41 &  & 7 & 1.54129(20) & 7-14 & 0.54\tabularnewline
\hline 
 & 8 & 1.48445(25) & 9-15 & 0.28 &  & 8 & 1.57844(29) & 7-15 & 1.10 &  & 8 & 1.60145(57) & 7-15 & 0.99\tabularnewline
\hline 
 & 9 & 1.51359(33) & 9-15 & 0.49 &  & 9 & 1.63355(32) & 7-15 & 0.68 &  & 9 & 1.65887(79) & 7-10 & 0.86\tabularnewline
\hline 
 & 10 & 1.54153(63) & 9-12 & 0.49 &  & 10 & 1.67247(85) & 7-14 & 0.66 &  & 10 & 1.71108(252) & 8-10 & 1.37\tabularnewline
\hline 
 & 11 & 1.57062(87) & 8-12 & 0.91 &  & 11 & 1.70545(208) & 7-14 & 0.81 &  & 11 & 1.76163(373) & 8-10 & 1.18\tabularnewline
\hline 
 & 12 & 1.59793(83) & 8-12 & 0.84 &  & 12 & 1.73700(298) & 7-13 & 1.03 &  & 12 & 1.81737(460) & 7-13 & 0.67\tabularnewline
\hline 
5 & 4 & 1.32224(5) & 5-16 & 0.38 & 8 & 4 & 1.32583(5) & 6-16 & 0.67 & 11 & 4 & 1.32618(9) & 8-12 & 1.02\tabularnewline
\hline 
 & 5 & 1.39610(12) & 6-16 & 0.88 &  & 5 & 1.40657(8) & 7-16 & 0.53 &  & 5 & 1.40724(10) & 6-16 & 0.85\tabularnewline
\hline 
 & 6 & 1.45207(32) & 8-16 & 1.07 &  & 6 & 1.47618(10) & 7-16 & 0.34 &  & 6 & 1.47750(12) & 6-16 & 0.19\tabularnewline
\hline 
 & 7 & 1.49576(37) & 8-16 & 0.94 &  & 7 & 1.53861(10) & 7-14 & 0.25 &  & 7 & 1.54185(33) & 6-16 & 0.92\tabularnewline
\hline 
 & 8 & 1.53054(45) & 9-13 & 1.10 &  & 8 & 1.59605(24) & 7-12 & 0.18 &  & 8 & 1.60217(25) & 7-10 & 0.32\tabularnewline
\hline 
 & 9 & 1.56251(81) & 8-15 & 1.00 &  & 9 & 1.64883(59) & 7-15 & 0.63 &  & 9 & 1.65978(75) & 7-10 & 1.01\tabularnewline
\hline 
 & 10 & 1.59199(100) & 8-15 & 1.05 &  & 10 & 1.69660(123) & 7-14 & 0.89 &  & 10 & 1.71562(138) & 7-10 & 0.98\tabularnewline
\hline 
 & 11 & 1.62032(89) & 8-13 & 0.72 &  & 11 & 1.73760(192) & 7-12 & 0.70 &  & 11 & 1.76992(238) & 7-10 & 0.88\tabularnewline
\hline 
 & 12 & 1.64931(87) & 7-15 & 0.46 &  & 12 & 1.76709(231) & 8-12 & 0.21 &  & 12 & 1.82503(171) & 7-15 & 0.67\tabularnewline
\hline 
6 & 4 & 1.32467(4) & 5-16 & 0.39 & 9 & 4 & 1.32597(6) & 6-15 & 0.85 & 12 & 4 & 1.32661(17) & 10-16 & 0.85\tabularnewline
\hline 
 & 5 & 1.40329(12) & 5-16 & 0.74 &  & 5 & 1.40698(9) & 6-14 & 1.02 &  & 5 & 1.40731(12) & 6-16 & 1.07\tabularnewline
\hline 
 & 6 & 1.46806(17) & 6-16 & 0.49 &  & 6 & 1.47693(11) & 6-14 & 0.17 &  & 6 & 1.47769(11) & 6-14 & 0.26\tabularnewline
\hline 
 & 7 & 1.52250(34) & 6-16 & 0.57 &  & 7 & 1.54083(22) & 6-14 & 0.28 &  & 7 & 1.54211(33) & 6-16 & 0.52\tabularnewline
\hline 
 & 8 & 1.56670(24) & 7-16 & 0.32 &  & 8 & 1.59999(25) & 7-16 & 0.72 &  & 8 & 1.60188(67) & 7-16 & 1.11\tabularnewline
\hline 
 & 9 & 1.60289(89) & 8-14 & 1.01 &  & 9 & 1.65665(13) & 7-15 & 0.26 &  & 9 & 1.65957(131) & 7-16 & 1.21\tabularnewline
\hline 
 & 10 & 1.63500(109) & 8-11 & 0.58 &  & 10 & 1.70918(103) & 7-14 & 0.84 &  & 10 & 1.71533(282) & 7-16 & 1.08\tabularnewline
\hline 
 & 11 & 1.66453(89) & 8-16 & 0.19 &  & 11 & 1.75771(215) & 7-14 & 0.89 &  & 11 & 1.77108(299) & 7-10 & 1.34\tabularnewline
\hline 
 & 12 & 1.69415(147) & 8-15 & 0.74 &  & 12 & 1.80102(441) & 7-15 & 0.93 &  & 12 & 1.82536(277) & 7-13 & 0.42\tabularnewline
\hline 
\end{tabular}\caption{Data for the ground state potential for the parallel geometry and for the pure-gauge simulation. All values are in lattice units.}
\end{table*}

\begin{table*}[!t]
\begin{tabular}{|cc|ccc||cc|ccc||cc|ccc|}
\hline 
$r_{12}$ & $r_{13}$ & $V_{0}$ & $t_{i}-t_{f}$ & $\chi_{r}^{2}$ & $r_{12}$ & $r_{13}$ & $V_{0}$ & $t_{i}-t_{f}$ & $\chi_{r}^{2}$ & $r_{12}$ & $r_{13}$ & $V_{0}$ & $t_{i}-t_{f}$ & $\chi_{r}^{2}$\tabularnewline
\hline 
\hline 
4 & 4 & - & - & - & 7 & 4 & 1.59960(18) & 7-13 & 0.40 & 10 & 4 & 1.74590(101) & 7-16 & 0.33\tabularnewline
\hline 
 & 5 & - & - & - &  & 5 & 1.61629(19) & 7-14 & 0.39 &  & 5 & 1.75813(40) & 7-16 & 0.32\tabularnewline
\hline 
 & 6 & - & - & - &  & 6 & 1.63726(37) & 7-15 & 0.84 &  & 6 & 1.77430(162) & 7-14 & 0.80\tabularnewline
\hline 
 & 7 & - & - & - &  & 7 & 1.66201(25) & 7-12 & 0.39 &  & 7 & 1.79224(143) & 7-13 & 0.54\tabularnewline
\hline 
 & 8 & - & - & - &  & 8 & 1.69130(104) & 7-15 & 0.67 &  & 8 & 1.81258(123) & 7-14 & 0.91\tabularnewline
\hline 
 & 9 & - & - & - &  & 9 & 1.72705(87) & 7-10 & 0.91 &  & 9 & 1.83436(263) & 7-15 & 0.91\tabularnewline
\hline 
 & 10 & - & - & - &  & 10 & 1.76824(107) & 7-10 & 0.49 &  & 10 & 1.85436(263) & 7-15 & 0.86\tabularnewline
\hline 
 & 11 & - & - & - &  & 11 & 1.81376(100) & 7-14 & 0.75 &  & 11 & 1.87964(35) & 7-16 & 0.63\tabularnewline
\hline 
 & 12 & - & - & - &  & 12 & 1.85802(299) & 7-13 & 0.55 &  & 12 & 1.90681(118) & 7-12 & 0.30\tabularnewline
\hline 
5 & 4 & 1.48851(16) & 7-11 & 0.61 & 8 & 4 & 1.65264(44) & 7-15 & 0.31 & 11 & 4 & 1.78291(307) & 7-15 & 0.96\tabularnewline
\hline 
 & 5 & - & - & - &  & 5 & 1.66678(121) & 6-16 & 0.76 &  & 5 & 1.79754(100) & 7-14 & 1.27\tabularnewline
\hline 
 & 6 & - & - & - &  & 6 & 1.68246(25) & 7-14 & 0.42 &  & 6 & 1.81565(156) & 7-11 & 1.04\tabularnewline
\hline 
 & 7 & - & - & - &  & 7 & 1.70390(12) & 7-15 & 0.38 &  & 7 & 1.83565(337) & 7-14 & 0.70\tabularnewline
\hline 
 & 8 & - & - & - &  & 8 & 1.72841(33) & 7-15 & 0.42 &  & 8 & 1.85526(365) & 7-15 & 0.74\tabularnewline
\hline 
 & 9 & - & - & - &  & 9 & 1.75774(96) & 7-15 & 0.65 &  & 9 & 1.87600(800) & 7-15 & 0.60\tabularnewline
\hline 
 & 10 & - & - & - &  & 10 & 1.79002(155) & 7-11 & 0.59 &  & 10 & 1.89616(84) & 7-12 & 0.80\tabularnewline
\hline 
 & 11 & - & - & - &  & 11 & 1.82871(186) & 7-13 & 0.40 &  & 11 & 1.91507(533) & 7-12 & 0.84\tabularnewline
\hline 
 & 12 & - & - & - &  & 12 & 1.87212(329) & 7-15 & 0.91 &  & 12 & 1.93991(618) & 7-13 & 0.67\tabularnewline
\hline 
6 & 4 & 1.54514(37) & 6-13 & 0.85 & 9 & 4 & 1.70183(27) & 7-15 & 0.68 & 12 & 4 & 1.80257(107) & 8-12 & 1.17\tabularnewline
\hline 
 & 5 & 1.56559(22) & 7-14 & 0.59 &  & 5 & 1.71365(17) & 7-16 & 0.17 &  & 5 & 1.83135(262) & 7-14 & 0.46\tabularnewline
\hline 
 & 6 & 1.59216(34) & 7-14 & 0.33 &  & 6 & 1.72868(41) & 7-15 & 0.73 &  & 6 & 1.85330(267) & 7-11 & 0.34\tabularnewline
\hline 
 & 7 & 1.62468(83) & 7-15 & 0.89 &  & 7 & 1.74737(42) & 7-15 & 0.36 &  & 7 & 1.87618(254) & 7-13 & 0.39\tabularnewline
\hline 
 & 8 & - & - & - &  & 8 & 1.76888(84) & 7-14 & 0.80 &  & 8 & 1.89751(246) & 7-13 & 0.18\tabularnewline
\hline 
 & 9 & - & - & - &  & 9 & 1.79388(113) & 7-14 & 0.86 &  & 9 & 1.91888(266) & 7-13 & 0.27\tabularnewline
\hline 
 & 10 & - & - & - &  & 10 & 1.82044(154) & 7-14 & 0.22 &  & 10 & 1.93594(431) & 7-9 & 0.45\tabularnewline
\hline 
 & 11 & - & - & - &  & 11 & 1.85045(370) & 7-14 & 0.51 &  & 11 & 1.95381(962) & 7-15 & 0.89\tabularnewline
\hline 
 & 12 & - & - & - &  & 12 & 1.88700(724) & 7-14 & 0.94 &  & 12 & 1.97689(190) & 7-13 & 0.45\tabularnewline
\hline 
\end{tabular}\caption{Data for the first excited state potential and for the parallel geometry for the pure-gauge simulation. All values are in lattice units.}
\end{table*}

\begin{table*}[!t]
\begin{tabular}{|cc|ccc||cc|ccc||cc|ccc|}
\hline 
$r_{12}$ & $r_{13}$ & $V_{0}$ & $t_{i}-t_{f}$ & $\chi_{r}^{2}$ & $r_{12}$ & $r_{13}$ & $V_{0}$ & $t_{i}-t_{f}$ & $\chi_{r}^{2}$ & $r_{12}$ & $r_{13}$ & $V_{0}$ & $t_{i}-t_{f}$ & $\chi_{r}^{2}$\tabularnewline
\hline 
\hline 
3 & 3 & 0.5280(4) & 5-15 & 1.07 & 6 & 3 & 0.5432(3) & 5-13 & 0.47 & 9 & 3 & 0.5435(3) & 5-14 & 0.74\tabularnewline
\hline 
 & 4 & 0.6163(6) & 7-16 & 0.29 &  & 4 & 0.6654(4) & 5-10 & 0.29 &  & 4 & 0.6665(6) & 5-15 & 0.59\tabularnewline
\hline 
 & 5 & 0.6718(11) & 7-12 & 0.75 &  & 5 & 0.7630(6) & 5-15 & 0.58 &  & 5 & 0.7661(13) & 5-13 & 1.03\tabularnewline
\hline 
 & 6 & 0.7162(13) & 7-12 & 0.29 &  & 6 & 0.8450(13) & 5-13 & 0.87 &  & 6 & 0.8447(57) & 7-13 & 1.01\tabularnewline
\hline 
 & 7 & 0.7553(14) & 7-12 & 0.74 &  & 7 & 0.9143(27) & 5-11 & 0.58 &  & 7 & 0.9331(30) & 5-15 & 0.59\tabularnewline
\hline 
 & 8 & 0.7925(11) & 7-10 & 0.09 &  & 8 & 0.9696(56) & 5-15 & 0.92 &  & 8 & 1.0092(39) & 5-11 & 0.66\tabularnewline
\hline 
 & 9 & 0.8259(34) & 7-10 & 0.57 &  & 9 & 1.0157(75) & 5-11 & 0.93 &  & 9 & 1.0813(52) & 5-10 & 0.80\tabularnewline
\hline 
 & 10 & 0.8546(27) & 7-9 & 0.59 &  & 10 & 1.0551(104) & 5-13 & 1.05 &  & 10 & 1.1512(14) & 5-6 & 0.12\tabularnewline
\hline 
 & 11 & 0.8882(27) & 7-8 & 0.30 &  & 11 & 1.0939(77) & 5-10 & 1.26 &  & 11 & 1.2129(148) & 5-9 & 1.60\tabularnewline
\hline 
4 & 3 & 0.5396(2) & 7-16 & 0.46 & 7 & 3 & 0.5434(3) & 5-16 & 0.92 & 10 & 3 & 0.5433(4) & 5-16 & 0.74\tabularnewline
\hline 
 & 4 & 0.6513(6) & 7-16 & 0.43 &  & 4 & 0.6660(6) & 5-13 & 0.64 &  & 4 & 0.6666(6) & 5-16 & 1.07\tabularnewline
\hline 
 & 5 & 0.7283(15) & 7-16 & 0.66 &  & 5 & 0.7652(9) & 5-14 & 0.74 &  & 5 & 0.7658(16) & 5-16 & 0.90\tabularnewline
\hline 
 & 6 & 0.7827(12) & 6-16 & 0.91 &  & 6 & 0.8516(12) & 5-13 & 0.76 &  & 6 & 0.8505(18) & 6-12 & 0.46\tabularnewline
\hline 
 & 7 & 0.8264(16) & 6-15 & 0.45 &  & 7 & 0.9281(17) & 5-10 & 1.27 &  & 7 & 0.9237(53) & 7-9 & 0.71\tabularnewline
\hline 
 & 8 & 08649(26) & 6-14 & 0.39 &  & 8 & 0.9958(36) & 5-8 & 1.07 &  & 8 & 1.0054(92) & 6-15 & 0.74\tabularnewline
\hline 
 & 9 & 0.9006(38) & 6-9 & 0.95 &  & 9 & 1.0531(84) & 5-10 & 1.27 &  & 9 & 1.0766(164) & 6-12 & 0.92\tabularnewline
\hline 
 & 10 & 0.9316(73) & 6-14 & 0.79 &  & 10 & 1.0868(140) & 6-11 & 0.76 &  & 10 & 1.0857(296) & 7-9 & 0.47\tabularnewline
\hline 
 & 11 & 0.9638(88) & 6-16 & 0.80 &  & 11 & 1.1255(125) & 6-10 & 1.22 &  & 11 & 1.2039(334) & 6-9 & 0.72\tabularnewline
\hline 
5 & 3 & 0.5425(3) & 5-16 & 0.42 & 8 & 3 & 0.5437(2) & 5-15 & 0.46 & 11 & 3 & 0.5431(4) & 5-14 & 0.80\tabularnewline
\hline 
 & 4 & 0.6622(6) & 5-16 & 0.63 &  & 4 & 0.6666(5) & 5-16 & 0.83 &  & 4 & 0.6665(5) & 5-12 & 0.63\tabularnewline
\hline 
 & 5 & 0.7528(12) & 6-11 & 0.72 &  & 5 & 0.7664(9) & 5-13 & 1.18 &  & 5 & 0.7657(2) & 6-14 & 0.47\tabularnewline
\hline 
 & 6 & 0.8244(13) & 6-12 & 0.25 &  & 6 & 0.8534(14) & 5-15 & 0.89 &  & 6 & 0.8531(11) & 6-14 & 0.45\tabularnewline
\hline 
 & 7 & 0.8778(27) & 6-16 & 0.38 &  & 7 & 0.9336(9) & 5-10 & 0.40 &  & 7 & 0.9318(13) & 6-11 & 0.42\tabularnewline
\hline 
 & 8 & 0.9214(48) & 6-16 & 0.36 &  & 8 & 1.0065(29) & 5-11 & 0.69 &  & 8 & 1.0040(63) & 6-11 & 0.31\tabularnewline
\hline 
 & 9 & 0.9575(70) & 6-11 & 0.94 &  & 9 & 1.0652(128) & 6-10 & 1.20 &  & 9 & 1.0490(319) & 7-10 & 1.18\tabularnewline
\hline 
 & 10 & 0.9905(123) & 6-9 & 2.34 &  & 10 & 1.1188(330) & 6-10 & 0.96 &  & 10 & 1.1035(584) & 7-11 & 0.81\tabularnewline
\hline 
 & 11 & 1.0301(84) & 6-11 & 1.09 &  & 11 & 1.1548(428) & 6-10 & 0.96 &  & 11 & 1.2040(147) & 6-8 & 0.52\tabularnewline
\hline 
\end{tabular}\caption{Data for the ground state potential and for the parallel geometry and for the full QCD simulation. All values are in lattice units.}
\end{table*}

\begin{table*}[!t]
\begin{tabular}{|cc|ccc||cc|ccc||cc|ccc|}
\hline 
$r_{12}$ & $r_{13}$ & $V_{1}$ & $t_{i}-t_{f}$ & $\chi_{r}^{2}$ & $r_{12}$ & $r_{13}$ & $V_{1}$ & $t_{i}-t_{f}$ & $\chi_{r}^{2}$ & $r_{12}$ & $r_{13}$ & $V_{1}$ & $t_{i}-t_{f}$ & $\chi_{r}^{2}$\tabularnewline
\hline 
\hline 
3 & 3 & 0.6637(29) & 8-12 & 1.14 & 6 & 3 & 0.9136(16) & 5-15 & 5-15 & 9 & 3 & 1.0911(229) & 6-11 & 0.99\tabularnewline
\hline 
 & 4 & - & - & - &  & 4 & 0.9313(20) & 5-9 & 0.42 &  & 4 & 1.1233(40) & 5-10 & 0.52\tabularnewline
\hline 
 & 5 & - & - & - &  & 5 & 0.9575(33) & 5-11 & 0.89 &  & 5 & 1.1457(79) & 5-11 & 0.62\tabularnewline
\hline 
 & 6 & 0.6054(299) & 10-15 & 1.05 &  & 6 & 0.9926(42) & 5-12 & 1.30 &  & 6 & 1.1692(194) & 5-11 & 1.22\tabularnewline
\hline 
 & 7 & 0.5737(83) & 9-16 & 0.66 &  & 7 & 1.0292(103) & 6-9 & 0.85 &  & 7 & 1.1897(286) & 5-11 & 2.08\tabularnewline
\hline 
 & 8 & 0.5398(41) & 9-11 & 0.37 &  & 8 & 1.0472(105) & 7-10 & 0.23 &  & 8 & 1.2112(244) & 5-10 & 1.29\tabularnewline
\hline 
 & 9 & 0.5420(79) & 9-10 & 0.41 &  & 9 & 1.1304(298) & 6-11 & 1.35 &  & 9 & 1.2476(130) & 5-10 & 0.55\tabularnewline
\hline 
 & 10 & 0.5399(22) & 8-9 & 0.08 &  & 10 & 1.1892(126) & 6-7 & 0.72 &  & 10 & 1.2876(108) & 5-9 & 0.76\tabularnewline
\hline 
 & 11 & 0.5491(149) & 8-9 & 1.33 &  & 11 & 1.2250(268) & 6-10 & 0.74 &  & 11 & 1.3220(308) & 5-9 & 0.60\tabularnewline
\hline 
4 & 3 & 0.7502(4) & 7-8 & 0.10 & 7 & 3 & 0.9855(32) & 5-12 & 0.67 & 10 & 3 & 1.1340(108) & 6-11 & 0.69\tabularnewline
\hline 
 & 4 & 0.7863(45) & 7-12 & 1.16 &  & 4 & 1.0004(30) & 5-12 & 0.75 &  & 4 & 1.1560(137) & 6-10 & 0.81\tabularnewline
\hline 
 & 5 & 0.8363(94) & 7-15 & 1.47 &  & 5 & 1.0226(55) & 5-11 & 0.62 &  & 5 & 1.1748(75) & 6-8 & 0.63\tabularnewline
\hline 
 & 6 & 0.8490(292) & 8-13 & 1.49 &  & 6 & 1.0401(140) & 6-11 & 1.00 &  & 6 & 1.1807(221) & 6-7 & 1.41\tabularnewline
\hline 
 & 7 & 0.8468(298) & 8-13 & 0.94 &  & 7 & 1.0727(113) & 6-9 & 0.50 &  & 7 & 1.1899(578) & 6-10 & 1.21\tabularnewline
\hline 
 & 8 & 0.8285(299) & 8-13 & 0.59 &  & 8 & 1.1070(23) & 6-11 & 0.45 &  & 8 & 1.2202(592) & 6-9 & 1.17\tabularnewline
\hline 
 & 9 & 0.8141(492) & 8-12 & 1.36 &  & 9 & 1.1638(102) & 5-9 & 1.01 &  & 9 & 1.3152(104) & 5-6 & 0.91\tabularnewline
\hline 
 & 10 & 0.7706(714) & 8-12 & 1.22 &  & 10 & 1.2227(44) & 5-9 & 0.66 &  & 10 & 1.3472(243) & 5-9 & 0.48\tabularnewline
\hline 
 & 11 & 0.6894(1127) & 8-14 & 1.42 &  & 11 & 1.2788(115) & 5-7 & 0.62 &  & 11 & 1.3764(32) & 5-9 & 0.18\tabularnewline
\hline 
5 & 3 & 0.8360(7) & 5-8 & 0.40 & 8 & 3 & 1.0504(73) & 5-11 & 1.20 & 11 & 3 & 1.1707(7) & 6-7 & 0.001\tabularnewline
\hline 
 & 4 & 0.8614(13) & 5-8 & 1.20 &  & 4 & 1.0642(30) & 5-12 & 0.24 &  & 4 & 1.1916(188) & 6-9 & 0.22\tabularnewline
\hline 
 & 5 & 0.8946(36) & 6-9 & 0.89 &  & 5 & 1.0850(77) & 5-10 & 0.75 &  & 5 & 1.2134(375) & 6-10 & 1.01\tabularnewline
\hline 
 & 6 & 0.9284(129) & 7-10 & 1.28 &  & 6 & 1.1121(60) & 5-7 & 0.97 &  & 6 & 1.2265(479) & 6-10 & 0.46\tabularnewline
\hline 
 & 7 & 0.9813(179) & 7-12 & 0.66 &  & 7 & 1.1154(113) & 6-12 & 1.12 &  & 7 & 1.2458(381) & 6-10 & 0.18\tabularnewline
\hline 
 & 8 & 1.0230(200) & 7-13 & 1.03 &  & 8 & 1.1610(141) & 5-10 & 0.99 &  & 8 & 1.2981(96) & 6-8 & 0.32\tabularnewline
\hline 
 & 9 & 1.0636(251) & 7-12 & 0.94 &  & 9 & 1.1988(67) & 5-9 & 0.87 &  & 9 & 1.3944(490) & 5-10 & 0.84\tabularnewline
\hline 
 & 10 & 1.1215(245) & 6-9 & 0.81 &  & 10 & 1.2453(48) & 5-10 & 0.34 &  & 10 & 1.4238(236) & 5-8 & 0.44\tabularnewline
\hline 
 & 11 & 1.1400(60) & 6-7 & 0.12 &  & 11 & 1.2913(190) & 5-12 & 0.81 &  & 11 & 1.4508(480) & 5-9 & 0.40\tabularnewline
\hline 
\end{tabular}\caption{Data for the first excited state potential for the parallel geometry and for the full QCD simulation. All values are in lattice units.}
\end{table*}

\end{document}